\documentclass[aps,prl,twocolumn,superscriptaddress]{revtex4-2}
\usepackage[T1]{fontenc}
\usepackage{amsmath,amssymb,amsfonts}
\usepackage{graphicx}
\usepackage{hyperref}
\usepackage{braket}
\usepackage{tikz}
\usetikzlibrary{decorations.pathreplacing,calc}
\definecolor{tnteal}{HTML}{4FB39A}
\definecolor{tngray}{HTML}{8C8C8C}
\definecolor{permblue}{HTML}{2F6FB0}

\usepackage{booktabs}
\usepackage{tikz}

\usepackage{blindtext}
\AtBeginDocument{\def\selectlanguage#1{}}
\begin{document}
\setcounter{secnumdepth}{3}

\title{Typical Superpositions are Magical}

\author{Joshuah T. Heath}
\email{joshuah.heath@mu.ie}
\affiliation{Department of Physics, Science Building, Maynooth University, Maynooth, Co. Kildare W23 F2H6, Ireland}
\affiliation{Hamilton Institute, Eolas Building, Maynooth University, Maynooth, Co. Kildare W23 A3HY, Ireland}

\author{Andrew Urilyon}
\email{Andrew.Urilyon@mu.ie}
\affiliation{Department of Physics, Science Building, Maynooth University, Maynooth, Co. Kildare W23 F2H6, Ireland}

\author{Damien Quinn}
\email{damien.quinn.2026@mumail.ie}
\affiliation{Department of Physics, Science Building, Maynooth University, Maynooth, Co. Kildare W23 F2H6, Ireland}

\author{Graham Kells}
\email{Graham.Kells@mu.ie}
\affiliation{Department of Physics, Science Building, Maynooth University, Maynooth, Co. Kildare W23 F2H6, Ireland}
\affiliation{Hamilton Institute, Eolas Building, Maynooth University, Maynooth, Co. Kildare W23 A3HY, Ireland}
\affiliation{Dublin Institute for Advanced Studies, School of Theoretical Physics, 10 Burlington Road, Dublin, Co. Dublin D04 C932, Ireland}

\date{\today}

\begin{abstract}
\noindent The non-stabilizerness, or "magic", of a quantum state is a common metric for the hardness of simulation on a classical computer, and is often quantified by the stabilizer R\'enyi entropy (SRE). We show how random superpositions of classically simulable states saturate the SRE to the symmetry-restricted Haar-random maximum. For any large-$N$ random pure-state ensemble satisfying the conditions of Schur's lemma, the average SRE of the $m$-component superposition $\langle M_2(m)\rangle=N-\log_2[\langle W(m)\rangle]$ is governed by an average witness $\langle W(m)\rangle$ that universally decomposes into a Haar floor and an ensemble-specific excess. As the number of components $m$ increases, the excess is suppressed, resulting in a logarithmic scaling of the SRE before plateauing at the Haar-valued maximum. We derive a general closed form for $\langle W(m)\rangle$ in the thermodynamic limit, and numerically verify our result for random stabilizer states and random Gaussian states.
\end{abstract}

\maketitle

{\it Introduction} -- Quantum computational complexity is often defined as the minimum number of non-trivial gates required to prepare a given quantum state~\cite{Nielsen2006May,Plesch2011Mar,Aaronson2016Jul,Hu2020Sep}. Within the Clifford paradigm of quantum computing, non-trivial gates correspond to non-Clifford operations, since Clifford circuits can be efficiently simulated on a classical computer~\cite{Gottesman1998Jul,AaronsonGottesman2004,bravyi2019}. The study of non-Cliffordness (also called non-stabilizerness or "magic"~\cite{Bravyi2005Feb,Howard2014Jun}) has recently attracted a large amount of interest within the quantum information community, both as a metric for non-classical behavior~\cite{Veitch2014Jan,Howard2017Mar,leone2022,Projansky2024Aug,White2024Dec} and as a tool for understanding complex many-body phases~\cite{White2021Feb,Liu2022May,Oliviero2022Oct,Haug2023Jan,Odavic2023Oct,Tarabunga2023Oct,collura2024,Hoshino2026Feb,Ding2025Aug,Russomanno2025Aug,Lamma2026Aug,Sabharwal2026Sep}. In particular, Haar random systems in the thermodynamic limit have been proposed to maximize non-stabilizerness~\cite{turkeshiPRB2025,Szombathy2025Oct}, while also exhibiting volume-law entanglement at the Page value~\cite{Page1993Aug}, circuit complexity exponential in the system size $N$~\cite{Iaconis2021Feb}, and maximized non-Gaussianity density~\cite{Sierant2026Jan,Hoshino2026Oct}. In this way, proximity to the large-$N$ Haar random ensemble can serve as a universal, gate-set-independent proxy for computational complexity~\cite{Roberts2017Apr,Iaconis2021Feb}.

While many metrics for non-stabilizerness have been proposed~\cite{Veitch2014Jan,Bravyi2016Jun,Howard2017Mar,Wang2020Mar}, the stabilizer R\'enyi entropy (SRE)~\cite{leone2022} has become popular due to its simplicity, computational accessibility~\cite{Haug2023Jan,HaugPiroli2023Jan,Liu2025Feb,huang2026,Xiao2026}, and clear operational interpretation~\cite{Bittel2026Apr}.
For some pure state $|\phi\rangle$ of $N$ qubits, the $\alpha=2$ SRE is defined as $M_2(|\phi\rangle)\equiv N-\log_2[W(|\phi\rangle)]$. The non-stabilizer witness is defined as the fourth moment $W(|\phi\rangle)=\sum_{P\in\mathbb{P}_N}\langle \phi|P|\phi\rangle^4$, with the sum taken over all Pauli strings $P$ in the Pauli group $\mathbb{P}_N$. 
The SRE vanishes if $|\phi\rangle$ is a stabilizer state, and, for a large-$N$ Haar random ensemble, saturates (on average) to a maximum of $N-2$~\cite{leone2022,turkeshiPRB2025}.
Nevertheless, a rigorous understanding of how a low-magic random state might approach the Haar ceiling of the SRE is still lacking.

\begin{figure}[t]
    \includegraphics[width=\columnwidth]{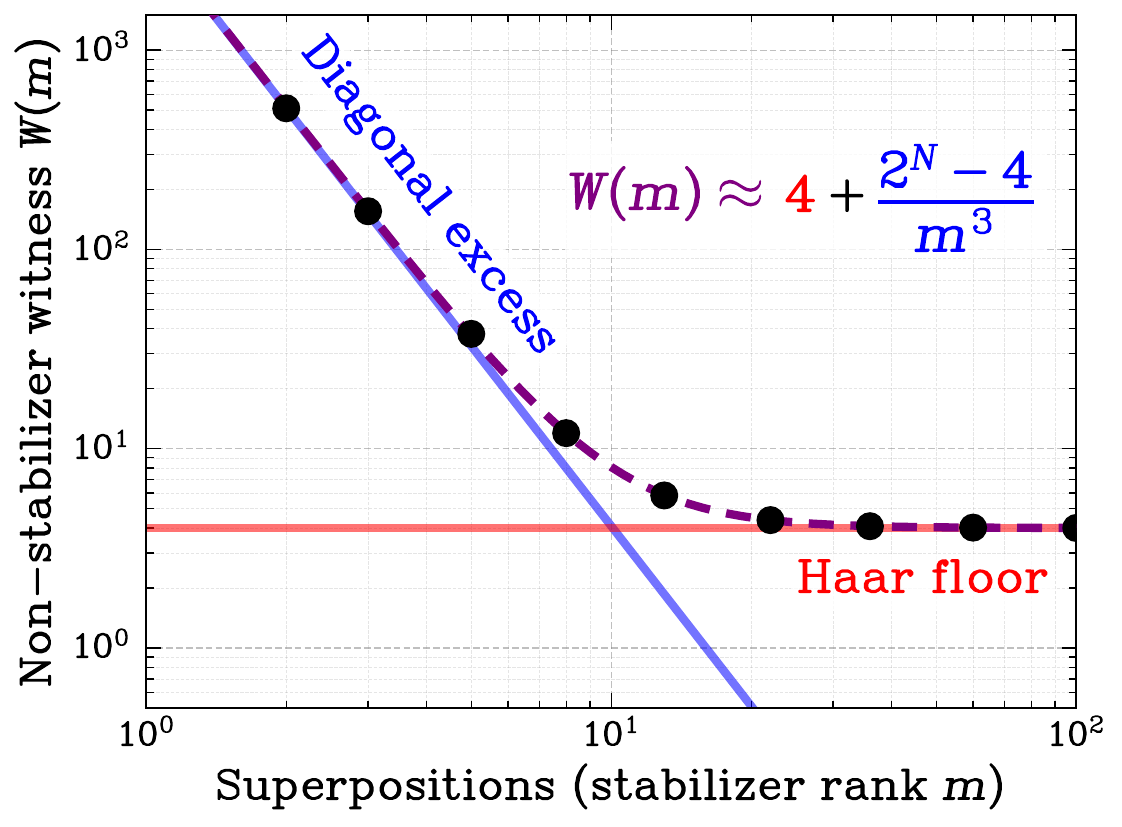}
    \caption{\small The average non-stabilizer witness versus the stabilizer rank $m$ for Haar random stabilizer states. The black dots are numerical data taken at $N=12$ and averaged over $100$ samples, while the purple dashed line is the theoretical prediction given by the in-figure equation. Increasing the stabilizer rank dilutes the diagonal excess to the Haar floor of the witness, which is a ceiling for the stabilizer R\'enyi entropy.}
    \label{fig:FrontPage}
\end{figure}

In this Letter, we study the magic inherent to typical superpositions~\cite{Goldstein2006Feb,Reimann2007Oct,Wilde2017Feb,TypicalEntanglement} of random ensembles, and derive a general law by which the SRE grows with the number of superposed states $m$ and their weights. For superpositions $|\Psi_m\rangle=\sum_{i=1}^m\lambda_i|\phi_i\rangle$ with weights $p_i=|\lambda_i|^2$~\footnote{We fix the weights by $\sum_ip_i=1$ rather than normalizing each realization of $|\Psi_m\rangle$. As the components are not orthogonal, this introduces an average error of order $2^{-N}$ in the witness.

} and pure-state components $|\phi_i\rangle$ drawn independently from an ensemble satisfying the conditions of Schur's lemma~\cite{georgi,Serre,Etingof2009Jan}, we find that for a large number of qubits $N$ the ensemble-averaged witness reduces to a universal form
\begin{align}
    \langle W_\Psi(m)\rangle=\{a-(b+c) \} \|p\|_4^4+ b\,\|p\|_2^4 +c,\label{GeneralLaw}
\end{align}
where $\|p\|_l^k\equiv\big(\sum_ip_i^l\big)^{k/l}$ is the $k$-th power of the $l$-norm of the weights and $a$, $b$, and $c$ are constants. The constant $a=\langle W_\Psi(1)\rangle$ is the average witness for a single component in the superposition; $b$ is composed of ensemble-averaged second moments of same-component Pauli string expectation values, which vanishes when the ensemble forms a state 2-design~\cite{Renes2004,KlappeneckerRoetteler2005,AmbainisEmerson2007,Dankert2009Jul}; and $c$ is constructed from cross-component Pauli string expectations, which all sum to the symmetry-restricted Haar-random witness.
Upon increasing the number of components, both $a$ and $b$ become diluted, thus pushing the witness towards the Haar floor given by $c$ (see Fig.~\ref{fig:FrontPage}). Such dilution results in a universal logarithmic ascent of the SRE towards the Haar ceiling. In addition to deriving Eqn.~\eqref{GeneralLaw}, we consider the witness and resultant SRE for three canonical ensembles: random stabilizer states~\cite{Smith2006Dec,Koenig2014Dec,Struchalin2021Jan}, random Gaussian states~\cite{Magan2016Jan,BoutinPRR2021,Bhattacharjee2021Dec,bhk,Yu2023Jan,swietek2026onebody,Turkeshi2026Sep} and, in the Supplemental Material~\cite{Supplement}, random computational-basis states. 

{\it Non-stabilizer witness for typical $m$-component superpositions} -- We define the SRE for an $m$-component superposition $|\Psi\rangle$ of pure states $|\phi_i\rangle$ to be $\langle M_2(m)\rangle=N-\log_2(\langle W_\Psi(m)\rangle)$. We'll write the $m$-component witness $W_\Psi(m)$ as
\begin{align}
     W_\Psi(m)=n_{\textrm{fix}}+\sum_{P\in \widetilde{\mathbb{P}}_N}\langle\Psi| P|\Psi \rangle^4,\label{Witness}
 \end{align}
where in the above we have separated the sum over the {\it fixed Pauli strings} (which all sum to some number $n_{\textrm{fix}}$) and the {\it filtered Pauli strings} $P\in \widetilde{\mathbb{P}}_N$~\cite{turkeshiPRB2025}. For some random ensemble, a symmetry restriction will make a subset of Pauli strings all take a "fixed" average value $\langle \phi_i|P|\phi_i\rangle=\pm 1\,\forall \,i$, thus contributing a factor $n_{\textrm{fix}}$ to the witness independent of the particular form of $|\Psi\rangle$. For example, the string of all identities is always a fixed Pauli string, while for random Gaussian states of fixed parity the fermion-parity string is also fixed. 

Our goal is to evaluate the sum over the filtered Pauli strings, given in Eqn.\eqref{Witness}, at large $N$. We split the summand into terms which are diagonal and off-diagonal in the components of the superposition:
\begin{align}
    \langle \Psi|P|\Psi\rangle&=\sum_i p_i\langle\phi_i|P|\phi_i\rangle+\sum_{i\not=j}p_{ij}\langle\phi_i|P|\phi_j\rangle\notag\\
    &\equiv\sigma_{\textrm{diag}}(P)+\sigma_{\textrm{off}}(P),\label{diag_and_off}
\end{align}
where $p_{i}=|\lambda_i|^2$ (as before) and $p_{ij}\equiv \lambda_i \lambda_j^*$ are the weights of the superposition's diagonal and off-diagonal Pauli spectra $\sigma(P)$, respectively. Taking the expected value of Eqn.~\eqref{diag_and_off} to the fourth power, 
the non-stabilizer witness is determined as the sum of five terms: $\mathbb{E}[\sigma_{\textrm{diag}}^4(P)]$, $\mathbb{E}[\sigma_{\textrm{diag}}^3(P)\sigma_{\textrm{off}}(P)]$, $\mathbb{E}[\sigma_{\textrm{diag}}^2(P)\sigma_{\textrm{off}}^2(P)]$, $\mathbb{E}[\sigma_{\textrm{diag}}(P)\sigma_{\textrm{off}}^3(P)]$, and $\mathbb{E}[\sigma_{\textrm{off}}^4(P)]$, 
with each term summed over the filtered Pauli group $ \widetilde{\mathbb{P}}_N$.

\begin{figure*}[t]
    \centering
    \includegraphics[width=\textwidth]{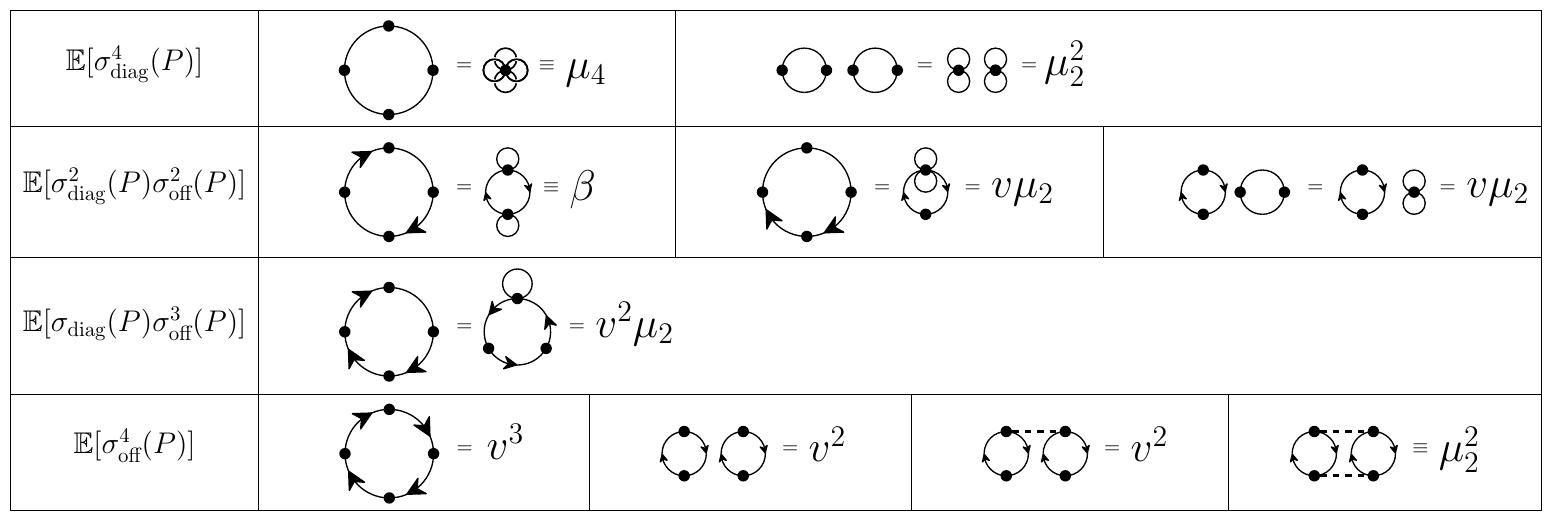}
    \caption{\small Diagrammatic representations for each non-vanishing contribution to the non-stabilizer witness of the $m$-state superposition. $\sigma_{\textrm{diag}}(P)$ and $\sigma_{\textrm{off}}(P)$ are defined in Eqn.~\eqref{diag_and_off}.   The non-stabilizer witness is a function of Pauli strings, $P$, and goes as $\langle \phi_i|P|\phi_j\rangle\langle \phi_j|P|\phi_k\rangle\langle \phi_k|P|\phi_\ell\rangle\langle \phi_\ell|P|\phi_i\rangle$, with the black nodes in the above diagrams representing the four possible indices $i,\,j,\,k,\,\ell$. A solid line connecting the nodes denotes a symmetric cycle where connected indices are equivalent; an arrow denotes a directed cycle where connected indices differ; and a dashed line denotes equality of the connected nodes whose contraction is prevented by the presence of one or more directed cycles. Symmetric cycles may be contracted, which is shown for several diagrams as a "pinching" of two or four adjacent nodes. $\beta$ is a class-dependent diagram that cannot be generically reduced to a product of symmetric or directed moments, given by $\mu_2$ and $v$ respectively.}
    \label{fig:table1}
\end{figure*}

Expanding Eqn.~\eqref{diag_and_off} to the fourth power results in products of four amplitudes $\langle\phi_i|P|\phi_j\rangle$ with indices that form a closed cycle $i\to j\to k\to\ell\to i$. The random phases of the components average to zero unless the off-diagonal factors form a closed loop, so every term of $\mathbb{E}[\sigma_{\textrm{diag}}^3(P)\sigma_{\textrm{off}}(P)]$ vanishes upon taking the expected value. In addition, any independent diagonal factor averages to $\mathbb{E}\langle\phi|P|\phi\rangle=\textrm{Tr}(P\overline{\rho})=0$. This follows if we assume the ensemble satisfies the conditions of Schur's lemma~\cite{georgi,Serre,Etingof2009Jan}, where the average density matrix is given by $\overline{\rho}=I_D/D$ on the $D$-dimensional symmetry sector from which the components are drawn. The remaining non-trivial terms in $W_\Psi(m)$ are closed loops of length four or pairs of loops of length two, and may be represented with a diagrammatic notation introduced in the Supplemental Material~\cite{Supplement} and summarized in Fig.~\ref{fig:table1}. In the diagrammatic formalism, each black node represents one of the four indices. A solid line denotes a {\it symmetric connection} which joins identical indices (i.e., a diagonal amplitude $\langle \phi_i|P|\phi_i\rangle$), while an arrow denotes a {\it directed connection} which joins distinct indices  (i.e., an off-diagonal amplitude $\langle \phi_i|P|\phi_j\rangle$). Lastly, a dashed line joins indices that are equal but which cannot be merged due to the presence of one or more directed connections. 

\indent All non-trivial diagrams may be evaluated via two contraction rules: one for symmetric connections, and one for directed connections. For the former, contraction is shown diagrammatically as "pinching" adjacent nodes connected by a solid line, and represents a simplification to {\it symmetric moments} $\mu_k(P)\equiv\mathbb{E}[\langle\phi|P|\phi\rangle^k]$. As such, the diagonal fourth moment $\sum_i p_i^4\langle\phi_i|P|\phi_i\rangle^4$ evaluates to $\|p\|_4^4\,\mu_4(P)$, while the pair term $\sum_{i\neq j}p_i^2p_j^2\langle\phi_i|P|\phi_i\rangle^2\langle\phi_j|P|\phi_j\rangle^2$evaluates to $(\|p\|_2^4-\|p\|_4^4)\mu_2^2(P)$. This allows us to write the purely diagonal component of the non-stabilizer witness as 
 \begin{align}
 \sum_{P\in \widetilde{\mathbb{P}}_N}\mathbb{E} &\left[\sigma^4_{\textrm{diag}}(P)\right]=3\big[\sum_{P\in{\mathbb{P}}_N}\mu_2^2(P)-n_{\textrm{fix}}\big]\|p\|_2^4\notag\\
    &\phantom{lll}+\bigg[\langle W_\Psi(1)\rangle+2n_{\textrm{fix}}-3\sum_{P\in{\mathbb{P}}_N}\mu_2^2(P)\bigg]\|p\|_4^4,
 \end{align}
\noindent where $\langle W_\Psi(1)\rangle$ is the average single-component witness over all (i.e., the un-filtered) Pauli strings.  In contrast, contraction of directed connections results in terms that go as $v=\textrm{Tr}(P\overline{\rho}P\overline{\rho})=D^{-1}$ (via Schur's lemma), which we define as the {\it directed second moment} and which takes the same value on every string that acts within the sector.

From these two contraction rules, the full $m$-component non-stabilizer witness may be evaluated. The {\it Parseval sum rule}~\cite{Nadimpalli2023Nov}, $\sum_{P\in\mathbb{P}_N}\mu_2(P)=\sum_{P\in\mathbb{P}_N}v=2^N$ results in two of the ten diagrams scaling as $2^{-N}$ for every ensemble, and hence these diagrams disappear in the thermodynamic limit. One other diagram, denoted $\beta$ in Fig.~\ref{fig:table1}, does not generally reduce to some product of $\mu_2(P)$ and $v$, however it can be shown to obey $\beta \le n_{\textrm{fix}}$. The remaining diagrams result in terms that scale as $v\mu_2(P)$ or $v^2$, with both summing to $n_{\textrm{fix}}$ over the filtered Pauli strings. Including the norms of the weights $p_i$ and taking $N\rightarrow \infty$, we arrive at Eqn.~\eqref{GeneralLaw}, with $a=\langle W_\Psi(1)\rangle$, $b=6\left[\sum_{P\in\widetilde{\mathbb{P}}_N}\mu_2^2(P)-2(n_{\textrm{fix}}-\beta)\right]$, and $c=4n_{\textrm{fix}}$. For an equal-weighed superposition, $\|p\|_2^4=m^{-2}$ and $\|p\|_4^4=m^{-3}$, and Eqn.~\eqref{GeneralLaw} becomes
\begin{align}
&\langle W_\Psi(m)\rangle\notag\\
&=4n_{\textrm{fix}}+\dfrac{b}{m^2}+\dfrac{\langle W_\Psi(1)\rangle-(b+4n_{\textrm{fix}})}{m^3}+\mathcal{O}(2^{-N}).\label{General_Witness}
\end{align}

\noindent Eqn.~\eqref{General_Witness} separates the witness into a floor and a diluted excess. Upon increasing the number of components $m\rightarrow \infty$, the witness descends to the {\it Haar floor} $c=4n_{\textrm{fix}}$ for every ensemble which satisfies the conditions of Schur's lemma. As such, the SRE saturates at the symmetry-restricted Haar ceiling $N-2-\log_2(n_{\textrm{fix}})$. All terms which are dependent on the norms of the weights, and which therefore are diluted upon increasing $m$, we define as the {\it excess}. The excess then determines how the SRE ascends to the Haar maximum, and results in a universal logarithmic scaling law $\langle M_2(m)\rangle -\langle M_2(1)\rangle\approx 3\log_2(m)$ for terms where $\{a-(b+c)\}/m^3$ dominates and $\approx 2\log_2(m)$ for terms where $b/m^2$ dominates.

\begin{figure}[t]
    \hspace{-2mm}\includegraphics[width=1\columnwidth]{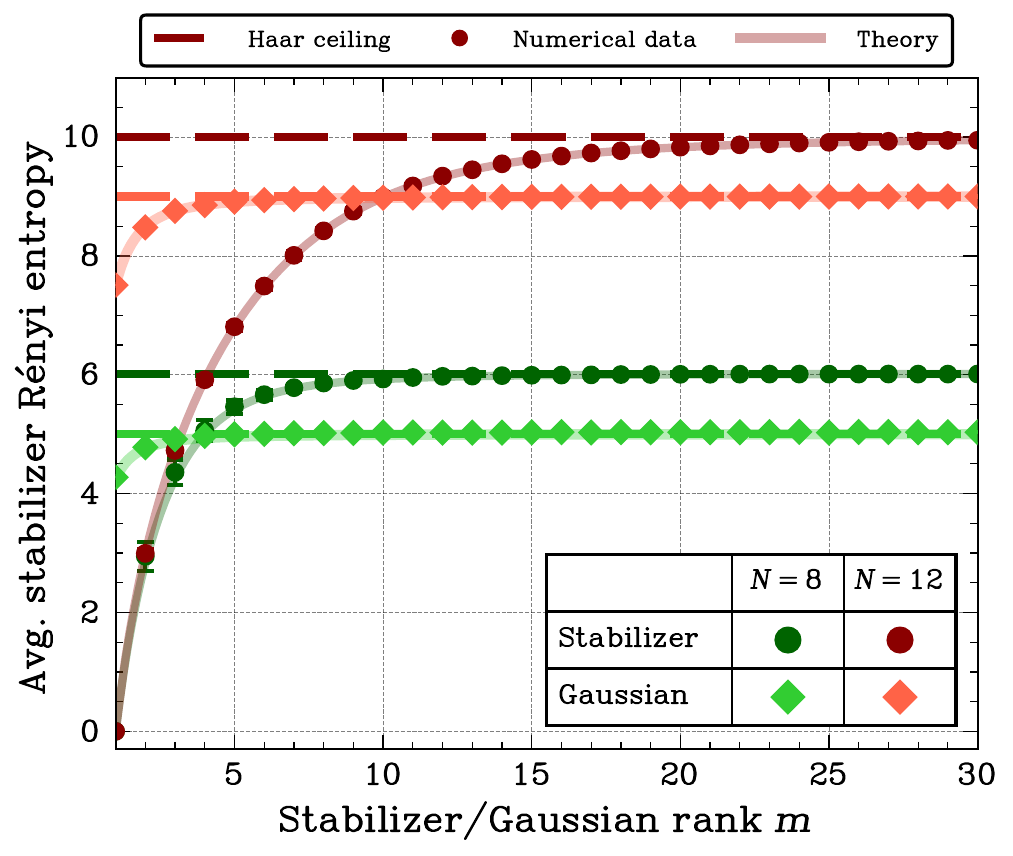}
    \caption{\small Average stabilizer R\'enyi entropy (SRE) for two classes of equal-weight random superpositions, stabilizer states (circles) and Gaussian states (diamonds), plotted versus the stabilizer rank and Gaussian rank, respectively. The SRE is averaged over $100$ samples for each value of $m$, with circles/diamonds representing numerical data, dashed lines the symmetry-restricted Haar ceiling, and solid lines the theoretical prediction. The SRE for random stabilizers is found with the fast Walsh-Hadamard transform~\cite{huang2026,Xiao2026}. Error bars are smaller than most data points.}
    \label{fig:Fig3}
\end{figure}

In the next two sections, $\langle W_\Psi(1)\rangle$, $b$, and $\sum_{\mathbb{P}_N}\mu_2^2(P)$ are explicitly evaluated for two cases: superpositions of typical stabilizer states and superpositions of typical fermionic Gaussian states. In both cases, the logarithmic scaling is demonstrated analytically and numerically.
\\\\
{\it SRE for superpositions of typical stabilizer states}-- For typical stabilizers, $m$ is identified as the stabilizer rank~\cite{Bravyi2016Jun,bravyi2019}, and as there is no symmetry constraint, the identity is the only fixed Pauli string (thus, $n_{\textrm{fix}}=1$). As a stabilizer state has $\langle\phi|P|\phi\rangle=\pm1$ on the $2^N$ strings of its stabilizer group and $0$ on every other string~\cite{AaronsonGottesman2004,passarelli2024piv}, $\langle W_\Psi(1)\rangle=2^N$. 
For the second moments, we write $\mu_2(P)=\textrm{Tr}[P^{\otimes 2}\mathbb{E}\{\rho^{\otimes 2}\}]$ and use the fact that random stabilizer states form a complex projective 2-design~\cite{Zhu2017,Zhu2016Sep}, from which the Schur--Weyl duality gives $\mathbb{E}\{\rho^{\otimes 2}\}=(1+\mathbb{F})/(D(D+1))$ where $\mathbb{F}$ is the swap operator and $D=2^N$~\cite{Gross2021Aug}. This results in the sum of $\mu_2^2(P)$ over all Pauli strings reducing to $2n_{\textrm{fix}}$ and $\beta\sim2^{-N}\rightarrow 0$, and hence Eqn.~\eqref{General_Witness} reduces to
\begin{align}
   \langle W_\Psi(m)\rangle^{\textrm{stab}}=4+\dfrac{2^N-4}{m^3}+\mathcal{O}(2^{-N}).\label{stab_witness}
\end{align}
In the above, the single-component excess $2^N-4$ is diluted as $m^{-3}$, so the SRE climbs as $3\log_2(m)$ from zero until it reaches the ceiling $N-2$ near $m\approx 2^{(N-2)/3}$. Fig.~\ref{fig:Fig3} shows the averaged SRE for $N=8$ and $12$, which follows the analytical prediction given above. By $N = 8$, fluctuations in the SRE are minimal and the average saturates to the maximal Haar value by $m= 4$. 

\vspace{2mm} 
{\it SRE for superpositions of typical Gaussian states}-- For random fermionic Gaussian states, $m$ is taken to be the Gaussian rank~\cite{Cudby2024,Hahn2025Oct,Wei2026Oct}. A Gaussian state has definite fermion parity, so if we superpose components with the same parity, both the parity string and the identity are fixed for this ensemble (thus, $n_{\textrm{fix}}=2$).  Unlike random stabilizer states, random Gaussian states do not form a unitary $2$-design~\cite{Bravyi2017Dec,Liu2021Dec,Paviglianiti2025Doped}, and thus the remaining two inputs follow from the graded moments of the Majorana-string amplitudes (see the Supplemental Material~\cite{Supplement} and Refs.~\cite{selberg1944bemerkninger, Forrester2010LogGases, debruyne2023linearstatisticscoulombgases, tiutiakina2023framepotentialbrowniansyk,collura2024,Turkeshi2026Sep}). In addition, the average single-component witness now has much richer structure than that of random stabilizer states:
\begin{align}
\langle W_\Psi(1)\rangle=\dfrac{(N+1)(N+2)}{2(2N+1)}\dfrac{4^N}{\binom{2N}{N}}\label{Gauss_inputs}.
\end{align}
Whereas $M_2(1)=0$ for random stabilizer states as a consequence of $\langle W_\Psi(1)\rangle=2^N$, a single Gaussian state has non-zero stabilizer magic~\cite{Supplement}. 

By recasting the Pauli strings in the Majorana basis, we similarly find that $\beta\rightarrow 0$ and $\sum_{P\in\mathbb{P}_N}\mu_2^2(P)=4^N/{\binom{2N}{N}}\sim\sqrt{\pi N}$ in the thermodynamic limit. This results in a  simplified form for the parity-restricted Gaussian witness in the limit of large $N$:

\begin{align}
    \langle W_\Psi(m)\rangle^{\textrm{Gauss}}&\simeq 8+\dfrac{6\sqrt{\pi N}-24}{m^2}\notag\\
    &+\dfrac{\tfrac{\sqrt{\pi}}{4}\big(N^{3/2}-\tfrac{171}{8}\sqrt{N}\big)+16}{m^3}+\mathcal{O}(N^{-1/2}).\label{Gauss_witness_asym}
\end{align}

\begin{figure}[t]
    \centering

    \begin{minipage}[b]{\columnwidth}
        \centering
        \makebox[\columnwidth][c]{{(a)}}\\[2pt]
        \includegraphics[width=1\columnwidth]{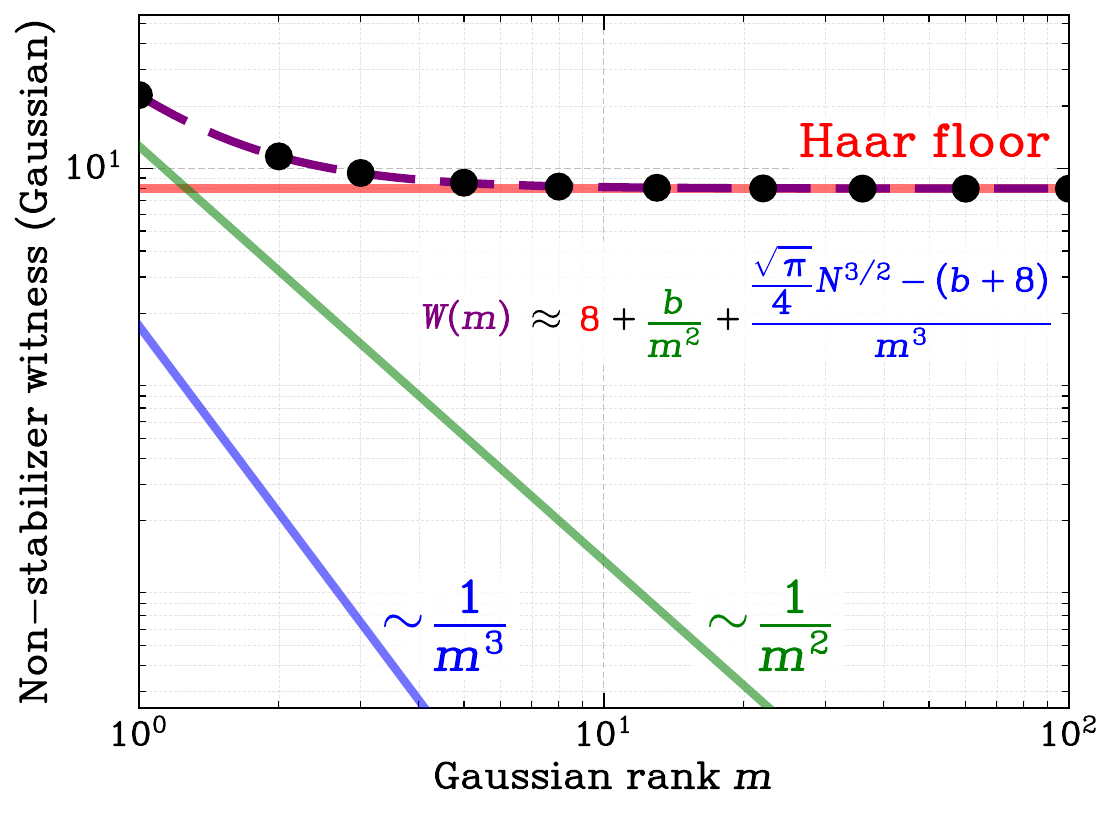}
    \end{minipage}
    
    \vspace{6pt}
    
    \begin{minipage}[b]{\columnwidth}
        \centering
        \makebox[\columnwidth][c]{{(b)}}\\[2pt]
        \includegraphics[width=1\columnwidth]{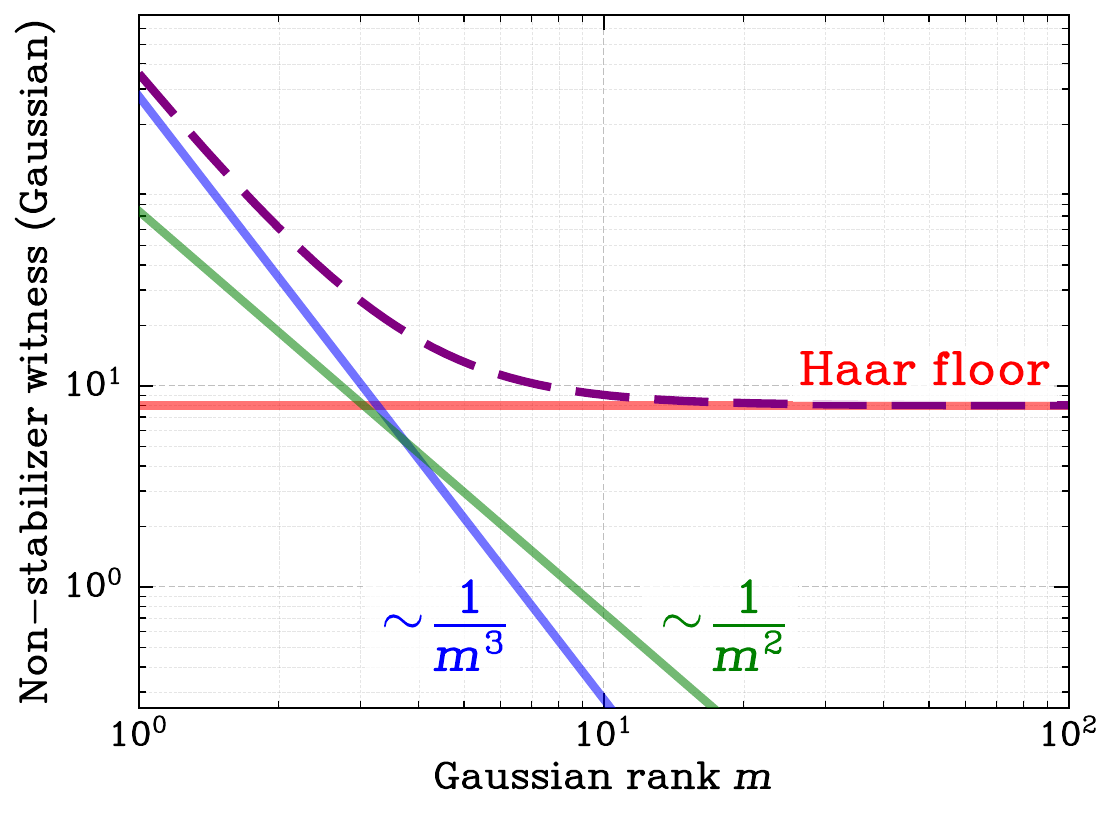}
    \end{minipage}

    \caption{\small (a) The non-stabilizer witness for equal-weight random Gaussian states versus Gaussian rank $m$ for $N=12$. Numerical data is represented by the black dots, while the full theoretical prediction is the dashed purple line. Whereas in the case of random stabilizer states the $1/m^3$ term dominates $M_2(m)$ for small $m$ (see Fig.~\ref{fig:FrontPage}), the growth of $M_2(m)$ for a small number of random Gaussian states is more strongly controlled by the $1/m^2$ dependence and the symmetry-restricted Haar floor of $4 n_{\textrm{fix}}=8$. (b) Same as (a), but only the theoretical prediction at $N=85$. Upon increasing $N$, the excess begins to scale as $1/m^3$.}    
    \label{fig:Last}
           \label{fig:Last2}
\end{figure}

\noindent Unlike the case of random stabilizer states, the $b$ term is non-zero in Eqn.~\eqref{Gauss_witness_asym}, while the single-component excess grows as $N^{3/2}$. In Fig.~\ref{fig:Fig3}, we plot the average SRE for random fermionic Gaussian states, in which we observe the sizeable R\'enyi entropy at $m=1$ and a logarithmic saturation to the Haar ceiling $N-3$. In Fig.~\ref{fig:Last}(a), we plot the non-stabilizer witness for random Gaussian states for $N=12$, which unlike random stabilizer states is dominated by the $b$-dependent term for small $m$. Nevertheless, upon increasing the system size, our theory predicts a dominant $a-(b+c)$ term and thus an emergent scaling $3\log_2(m)$ as $N\rightarrow \infty$, as for random stabilizers (see Fig.~\ref{fig:Last2}(b)).
\\\\
{\it Conclusions}-- We have analytically shown that large-$N$ superpositions of typical states, sampled uniformly from any ensemble which satisfies the conditions of Schur's lemma, will saturate the stabilizer Rényi entropy to the symmetry-restricted Haar-random maximum. The ensemble-averaged non-stabilizer witness for these superpositions universally follows the relation Eqn.~\eqref{GeneralLaw}, 
which in the thermodynamic limit can be split into a Haar-random floor and a diagonally-dominant excess. The former is universal for every ensemble, and coincides with the witness of a fully Haar-random state. The latter results in a logarithmic scaling law of the SRE in the thermodynamic limit, which shares similarities with the growth of the entanglement entropy of typical superpositions~\cite{TypicalEntanglement}. For both random stabilizer states and random Gaussian states, this scaling law takes the form $\langle \Delta M_2(m)\rangle\equiv \langle M_2(m)\rangle-\langle M_2(1)\rangle\approx 3\log_2(m)$. In the Supplement, we show that random computational-basis states instead ascend as $\langle \Delta M_2(m)\rangle\approx 2\log_2(m)$ for all $N$. However, for all cases under consideration, the increase in the SRE is universally controlled by terms diagonal in the Pauli spectrum, with the Haar ceiling set by terms purely off-diagonal.
\\\\
{\it Acknowledgements}-- We thank Ben Lehany-Fee, David Murtagh, Joost Slingerland, and Daniel Martín Velázquez for useful discussions. JTH would also like to acknowledge the hospitality of the Novo Nordisk Foundation Quantum Computing Programme (NQCP) at the Niels Bohr Institute in Copenhagen, Denmark. As of May 1st 2026, JTH has been funded by a Pathway Grant from Taighde \'Eireann--Research Ireland, and is employed by Maynooth University. Specifically, this publication has emanated from research conducted with the financial support of Taighde \'Eireann--Research Ireland, Grant Number 24/PATH-S/12630. For the purpose of Open Access, the authors have applied a CC BY public copyright licence to any Author Accepted Manuscript (AAM) version arising from this submission. The AAM will be available for free on the arXiv (a free distribution service and an open-access archive) upon completion of peer review. The original manuscript will also be available for free on the arXiv prior to completion of peer review.

\bibliography{main}

\begin{thebibliography}{7}%
\makeatletter
\providecommand \@ifxundefined [1]{%
 \@ifx{#1\undefined}
}%
\providecommand \@ifnum [1]{%
 \ifnum #1\expandafter \@firstoftwo
 \else \expandafter \@secondoftwo
 \fi
}%
\providecommand \@ifx [1]{%
 \ifx #1\expandafter \@firstoftwo
 \else \expandafter \@secondoftwo
 \fi
}%
\providecommand \natexlab [1]{#1}%
\providecommand \enquote  [1]{``#1''}%
\providecommand \bibnamefont  [1]{#1}%
\providecommand \bibfnamefont [1]{#1}%
\providecommand \citenamefont [1]{#1}%
\providecommand \href@noop [0]{\@secondoftwo}%
\providecommand \href [0]{\begingroup \@sanitize@url \@href}%
\providecommand \@href[1]{\@@startlink{#1}\@@href}%
\providecommand \@@href[1]{\endgroup#1\@@endlink}%
\providecommand \@sanitize@url [0]{\catcode `\\12\catcode `\$12\catcode
  `\&12\catcode `\#12\catcode `\^12\catcode `\_12\catcode `\%12\relax}%
\providecommand \@@startlink[1]{}%
\providecommand \@@endlink[0]{}%
\providecommand \url  [0]{\begingroup\@sanitize@url \@url }%
\providecommand \@url [1]{\endgroup\@href {#1}{\urlprefix }}%
\providecommand \urlprefix  [0]{URL }%
\providecommand \Eprint [0]{\href }%
\providecommand \doibase [0]{https://doi.org/}%
\providecommand \selectlanguage [0]{\@gobble}%
\providecommand \bibinfo  [0]{\@secondoftwo}%
\providecommand \bibfield  [0]{\@secondoftwo}%
\providecommand \translation [1]{[#1]}%
\providecommand \BibitemOpen [0]{}%
\providecommand \bibitemStop [0]{}%
\providecommand \bibitemNoStop [0]{.\EOS\space}%
\providecommand \EOS [0]{\spacefactor3000\relax}%
\providecommand \BibitemShut  [1]{\csname bibitem#1\endcsname}%
\let\auto@bib@innerbib\@empty
%</preamble>
\bibitem [{\citenamefont {Nadimpalli}\ \emph {et~al.}(2024)\citenamefont
  {Nadimpalli}, \citenamefont {Parham}, \citenamefont {Vasconcelos},\ and\
  \citenamefont {Yuen}}]{Nadimpalli2023Nov}%
  \BibitemOpen
  \bibfield  {author} {\bibinfo {author} {\bibfnamefont {S.}~\bibnamefont
  {Nadimpalli}}, \bibinfo {author} {\bibfnamefont {N.}~\bibnamefont {Parham}},
  \bibinfo {author} {\bibfnamefont {F.}~\bibnamefont {Vasconcelos}},\ and\
  \bibinfo {author} {\bibfnamefont {H.}~\bibnamefont {Yuen}},\ }\bibfield
  {title} {\bibinfo {title} {{On the Pauli Spectrum of QAC0}},\ }in\ \href
  {https://doi.org/10.1145/3618260.3649662} {\emph {\bibinfo {booktitle} {{ACM
  Conferences}}}}\ (\bibinfo  {publisher} {Association for Computing
  Machinery},\ \bibinfo {address} {New York, NY, USA},\ \bibinfo {year}
  {2024})\ pp.\ \bibinfo {pages} {1498--1506}\BibitemShut {NoStop}%
\bibitem [{\citenamefont {Bianchi}\ and\ \citenamefont
  {Don{\ifmmode\grave{a}\else\`{a}\fi}}(2019)}]{bianchi}%
  \BibitemOpen
  \bibfield  {author} {\bibinfo {author} {\bibfnamefont {E.}~\bibnamefont
  {Bianchi}}\ and\ \bibinfo {author} {\bibfnamefont {P.}~\bibnamefont
  {Don{\ifmmode\grave{a}\else\`{a}\fi}}},\ }\bibfield  {title} {\bibinfo
  {title} {{Typical entanglement entropy in the presence of a center: Page
  curve and its variance}},\ }\href
  {https://doi.org/10.1103/PhysRevD.100.105010} {\bibfield  {journal} {\bibinfo
   {journal} {Phys. Rev. D}\ }\textbf {\bibinfo {volume} {100}},\ \bibinfo
  {pages} {105010} (\bibinfo {year} {2019})}\BibitemShut {NoStop}%
\bibitem [{\citenamefont {Selberg}(1944)}]{selberg1944bemerkninger}%
  \BibitemOpen
  \bibfield  {author} {\bibinfo {author} {\bibfnamefont {A.}~\bibnamefont
  {Selberg}},\ }\bibfield  {title} {\bibinfo {title} {Bemerkninger om et
  multipelt integral},\ }\href
  {https://mathscinet.ams.org/mathscinet/article?mr=18287} {\bibfield
  {journal} {\bibinfo  {journal} {Norsk Matematisk Tidsskrift}\ }\textbf
  {\bibinfo {volume} {26}},\ \bibinfo {pages} {71} (\bibinfo {year}
  {1944})}\BibitemShut {NoStop}%
\bibitem [{\citenamefont {Forrester}(2010)}]{Forrester2010LogGases}%
  \BibitemOpen
  \bibfield  {author} {\bibinfo {author} {\bibfnamefont {P.~J.}\ \bibnamefont
  {Forrester}},\ }\href {https://doi.org/10.1515/9781400835416} {\emph
  {\bibinfo {title} {Log-Gases and Random Matrices}}},\ \bibinfo {series}
  {London Mathematical Society Monographs Series}, Vol.~\bibinfo {volume} {34}\
  (\bibinfo  {publisher} {Princeton University Press},\ \bibinfo {address}
  {Princeton, NJ},\ \bibinfo {year} {2010})\BibitemShut {NoStop}%
\bibitem [{\citenamefont {De~Bruyne}\ \emph {et~al.}(2024)\citenamefont
  {De~Bruyne}, \citenamefont {Le~Doussal}, \citenamefont {Majumdar},\ and\
  \citenamefont {Schehr}}]{debruyne2023linearstatisticscoulombgases}%
  \BibitemOpen
  \bibfield  {author} {\bibinfo {author} {\bibfnamefont {B.}~\bibnamefont
  {De~Bruyne}}, \bibinfo {author} {\bibfnamefont {P.}~\bibnamefont
  {Le~Doussal}}, \bibinfo {author} {\bibfnamefont {S.~N.}\ \bibnamefont
  {Majumdar}},\ and\ \bibinfo {author} {\bibfnamefont {G.}~\bibnamefont
  {Schehr}},\ }\bibfield  {title} {\bibinfo {title} {{Linear statistics for
  Coulomb gases: higher order cumulants}},\ }\href
  {https://doi.org/10.1088/1751-8121/ad329f} {\bibfield  {journal} {\bibinfo
  {journal} {J. Phys. A: Math. Theor.}\ }\textbf {\bibinfo {volume} {57}},\
  \bibinfo {pages} {155002} (\bibinfo {year} {2024})}\BibitemShut {NoStop}%
\bibitem [{\citenamefont {Tiutiakina}\ \emph {et~al.}(2024)\citenamefont
  {Tiutiakina}, \citenamefont {De~Luca},\ and\ \citenamefont
  {De~Nardis}}]{tiutiakina2023framepotentialbrowniansyk}%
  \BibitemOpen
  \bibfield  {author} {\bibinfo {author} {\bibfnamefont {A.}~\bibnamefont
  {Tiutiakina}}, \bibinfo {author} {\bibfnamefont {A.}~\bibnamefont
  {De~Luca}},\ and\ \bibinfo {author} {\bibfnamefont {J.}~\bibnamefont
  {De~Nardis}},\ }\bibfield  {title} {\bibinfo {title} {{Frame potential of
  Brownian SYK model of Majorana and Dirac fermions}},\ }\href
  {https://doi.org/10.1007/JHEP01(2024)115} {\bibfield  {journal} {\bibinfo
  {journal} {J. High Energy Phys.}\ }\textbf {\bibinfo {volume} {2024}}\bibinfo
   {number} { (1)},\ \bibinfo {pages} {115}}\BibitemShut {NoStop}%
\bibitem [{\citenamefont {Collura}\ \emph {et~al.}(2026)\citenamefont
  {Collura}, \citenamefont {De~Nardis}, \citenamefont {Alba},\ and\
  \citenamefont {Lami}}]{collura2024}%
  \BibitemOpen
\bibfield  {number} {  }\bibfield  {author} {\bibinfo {author} {\bibfnamefont
  {M.}~\bibnamefont {Collura}}, \bibinfo {author} {\bibfnamefont
  {J.}~\bibnamefont {De~Nardis}}, \bibinfo {author} {\bibfnamefont
  {V.}~\bibnamefont {Alba}},\ and\ \bibinfo {author} {\bibfnamefont
  {G.}~\bibnamefont {Lami}},\ }\bibfield  {title} {\bibinfo {title} {{The
  non-stabilizerness of fermionic Gaussian states}},\ }\href
  {https://doi.org/10.22331/q-2026-03-23-2036} {\bibfield  {journal} {\bibinfo
  {journal} {Quantum}\ }\textbf {\bibinfo {volume} {10}},\ \bibinfo {pages}
  {2036} (\bibinfo {year} {2026})},\ \Eprint
  {https://arxiv.org/abs/2412.05367v4} {2412.05367v4} \BibitemShut {NoStop}%
\end{thebibliography}%


\begin{thebibliography}{78}%
\makeatletter
\providecommand \@ifxundefined [1]{%
 \@ifx{#1\undefined}
}%
\providecommand \@ifnum [1]{%
 \ifnum #1\expandafter \@firstoftwo
 \else \expandafter \@secondoftwo
 \fi
}%
\providecommand \@ifx [1]{%
 \ifx #1\expandafter \@firstoftwo
 \else \expandafter \@secondoftwo
 \fi
}%
\providecommand \natexlab [1]{#1}%
\providecommand \enquote  [1]{``#1''}%
\providecommand \bibnamefont  [1]{#1}%
\providecommand \bibfnamefont [1]{#1}%
\providecommand \citenamefont [1]{#1}%
\providecommand \href@noop [0]{\@secondoftwo}%
\providecommand \href [0]{\begingroup \@sanitize@url \@href}%
\providecommand \@href[1]{\@@startlink{#1}\@@href}%
\providecommand \@@href[1]{\endgroup#1\@@endlink}%
\providecommand \@sanitize@url [0]{\catcode `\\12\catcode `\$12\catcode
  `\&12\catcode `\#12\catcode `\^12\catcode `\_12\catcode `\%12\relax}%
\providecommand \@@startlink[1]{}%
\providecommand \@@endlink[0]{}%
\providecommand \url  [0]{\begingroup\@sanitize@url \@url }%
\providecommand \@url [1]{\endgroup\@href {#1}{\urlprefix }}%
\providecommand \urlprefix  [0]{URL }%
\providecommand \Eprint [0]{\href }%
\providecommand \doibase [0]{https://doi.org/}%
\providecommand \selectlanguage [0]{\@gobble}%
\providecommand \bibinfo  [0]{\@secondoftwo}%
\providecommand \bibfield  [0]{\@secondoftwo}%
\providecommand \translation [1]{[#1]}%
\providecommand \BibitemOpen [0]{}%
\providecommand \bibitemStop [0]{}%
\providecommand \bibitemNoStop [0]{.\EOS\space}%
\providecommand \EOS [0]{\spacefactor3000\relax}%
\providecommand \BibitemShut  [1]{\csname bibitem#1\endcsname}%
\let\auto@bib@innerbib\@empty
%</preamble>
\bibitem [{\citenamefont {Nielsen}(2006)}]{Nielsen2006May}%
  \BibitemOpen
  \bibfield  {author} {\bibinfo {author} {\bibfnamefont {M.~A.}\ \bibnamefont
  {Nielsen}},\ }\bibfield  {title} {\bibinfo {title} {{A geometric approach to
  quantum circuit lower bounds}},\ }\href
  {https://dl.acm.org/doi/abs/10.5555/2011686.2011688} {\bibfield  {journal}
  {\bibinfo  {journal} {Quantum Inf. Comput.}\ }\textbf {\bibinfo {volume}
  {6}},\ \bibinfo {pages} {213} (\bibinfo {year} {2006})}\BibitemShut {NoStop}%
\bibitem [{\citenamefont {Plesch}\ and\ \citenamefont
  {Brukner}(2011)}]{Plesch2011Mar}%
  \BibitemOpen
  \bibfield  {author} {\bibinfo {author} {\bibfnamefont {M.}~\bibnamefont
  {Plesch}}\ and\ \bibinfo {author} {\bibfnamefont
  {{\ifmmode\check{C}\else\v{C}\fi}.}~\bibnamefont {Brukner}},\ }\bibfield
  {title} {\bibinfo {title} {{Quantum-state preparation with universal gate
  decompositions}},\ }\href {https://doi.org/10.1103/PhysRevA.83.032302}
  {\bibfield  {journal} {\bibinfo  {journal} {Phys. Rev. A}\ }\textbf {\bibinfo
  {volume} {83}},\ \bibinfo {pages} {032302} (\bibinfo {year}
  {2011})}\BibitemShut {NoStop}%
\bibitem [{\citenamefont {Aaronson}(2016)}]{Aaronson2016Jul}%
  \BibitemOpen
  \bibfield  {author} {\bibinfo {author} {\bibfnamefont {S.}~\bibnamefont
  {Aaronson}},\ }\bibfield  {title} {\bibinfo {title} {{The Complexity of
  Quantum States and Transformations: From Quantum Money to Black Holes}},\
  }\bibfield  {journal} {\bibinfo  {journal} {arXiv}\ }\href
  {https://doi.org/10.48550/arXiv.1607.05256} {10.48550/arXiv.1607.05256}
  (\bibinfo {year} {2016}),\ \Eprint {https://arxiv.org/abs/1607.05256}
  {1607.05256} \BibitemShut {NoStop}%
\bibitem [{\citenamefont {Hu}\ and\ \citenamefont {Kais}(2020)}]{Hu2020Sep}%
  \BibitemOpen
  \bibfield  {author} {\bibinfo {author} {\bibfnamefont {Z.}~\bibnamefont
  {Hu}}\ and\ \bibinfo {author} {\bibfnamefont {S.}~\bibnamefont {Kais}},\
  }\bibfield  {title} {\bibinfo {title} {{Characterization of Quantum States
  Based on Creation Complexity}},\ }\href
  {https://doi.org/10.1002/qute.202000043} {\bibfield  {journal} {\bibinfo
  {journal} {Adv. Quantum Technol.}\ }\textbf {\bibinfo {volume} {3}},\
  \bibinfo {pages} {2000043} (\bibinfo {year} {2020})}\BibitemShut {NoStop}%
\bibitem [{\citenamefont {Gottesman}(1998)}]{Gottesman1998Jul}%
  \BibitemOpen
  \bibfield  {author} {\bibinfo {author} {\bibfnamefont {D.}~\bibnamefont
  {Gottesman}},\ }\bibfield  {title} {\bibinfo {title} {{The Heisenberg
  Representation of Quantum Computers}},\ }\bibfield  {journal} {\bibinfo
  {journal} {arXiv}\ }\href {https://doi.org/10.48550/arXiv.quant-ph/9807006}
  {10.48550/arXiv.quant-ph/9807006} (\bibinfo {year} {1998}),\ \Eprint
  {https://arxiv.org/abs/quant-ph/9807006} {quant-ph/9807006} \BibitemShut
  {NoStop}%
\bibitem [{\citenamefont {Aaronson}\ and\ \citenamefont
  {Gottesman}(2004)}]{AaronsonGottesman2004}%
  \BibitemOpen
  \bibfield  {author} {\bibinfo {author} {\bibfnamefont {S.}~\bibnamefont
  {Aaronson}}\ and\ \bibinfo {author} {\bibfnamefont {D.}~\bibnamefont
  {Gottesman}},\ }\bibfield  {title} {\bibinfo {title} {{Improved simulation of
  stabilizer circuits}},\ }\href {https://doi.org/10.1103/PhysRevA.70.052328}
  {\bibfield  {journal} {\bibinfo  {journal} {Phys. Rev. A}\ }\textbf {\bibinfo
  {volume} {70}},\ \bibinfo {pages} {052328} (\bibinfo {year}
  {2004})}\BibitemShut {NoStop}%
\bibitem [{\citenamefont {Bravyi}\ \emph {et~al.}(2019)\citenamefont {Bravyi},
  \citenamefont {Browne}, \citenamefont {Calpin}, \citenamefont {Campbell},
  \citenamefont {Gosset},\ and\ \citenamefont {Howard}}]{bravyi2019}%
  \BibitemOpen
  \bibfield  {author} {\bibinfo {author} {\bibfnamefont {S.}~\bibnamefont
  {Bravyi}}, \bibinfo {author} {\bibfnamefont {D.}~\bibnamefont {Browne}},
  \bibinfo {author} {\bibfnamefont {P.}~\bibnamefont {Calpin}}, \bibinfo
  {author} {\bibfnamefont {E.}~\bibnamefont {Campbell}}, \bibinfo {author}
  {\bibfnamefont {D.}~\bibnamefont {Gosset}},\ and\ \bibinfo {author}
  {\bibfnamefont {M.}~\bibnamefont {Howard}},\ }\bibfield  {title} {\bibinfo
  {title} {{Simulation of quantum circuits by low-rank stabilizer
  decompositions}},\ }\href {https://doi.org/10.22331/q-2019-09-02-181}
  {\bibfield  {journal} {\bibinfo  {journal} {Quantum}\ }\textbf {\bibinfo
  {volume} {3}},\ \bibinfo {pages} {181} (\bibinfo {year} {2019})},\ \Eprint
  {https://arxiv.org/abs/1808.00128v2} {1808.00128v2} \BibitemShut {NoStop}%
\bibitem [{\citenamefont {Bravyi}\ and\ \citenamefont
  {Kitaev}(2005)}]{Bravyi2005Feb}%
  \BibitemOpen
  \bibfield  {author} {\bibinfo {author} {\bibfnamefont {S.}~\bibnamefont
  {Bravyi}}\ and\ \bibinfo {author} {\bibfnamefont {A.}~\bibnamefont
  {Kitaev}},\ }\bibfield  {title} {\bibinfo {title} {{Universal quantum
  computation with ideal Clifford gates and noisy ancillas}},\ }\href
  {https://doi.org/10.1103/PhysRevA.71.022316} {\bibfield  {journal} {\bibinfo
  {journal} {Phys. Rev. A}\ }\textbf {\bibinfo {volume} {71}},\ \bibinfo
  {pages} {022316} (\bibinfo {year} {2005})}\BibitemShut {NoStop}%
\bibitem [{\citenamefont {Howard}\ \emph {et~al.}(2014)\citenamefont {Howard},
  \citenamefont {Wallman}, \citenamefont {Veitch},\ and\ \citenamefont
  {Emerson}}]{Howard2014Jun}%
  \BibitemOpen
  \bibfield  {author} {\bibinfo {author} {\bibfnamefont {M.}~\bibnamefont
  {Howard}}, \bibinfo {author} {\bibfnamefont {J.}~\bibnamefont {Wallman}},
  \bibinfo {author} {\bibfnamefont {V.}~\bibnamefont {Veitch}},\ and\ \bibinfo
  {author} {\bibfnamefont {J.}~\bibnamefont {Emerson}},\ }\bibfield  {title}
  {\bibinfo {title} {{Contextuality supplies the {`}magic{'} for quantum
  computation}},\ }\href {https://doi.org/10.1038/nature13460} {\bibfield
  {journal} {\bibinfo  {journal} {Nature}\ }\textbf {\bibinfo {volume} {510}},\
  \bibinfo {pages} {351} (\bibinfo {year} {2014})}\BibitemShut {NoStop}%
\bibitem [{\citenamefont {Veitch}\ \emph {et~al.}(2014)\citenamefont {Veitch},
  \citenamefont {Mousavian}, \citenamefont {Gottesman},\ and\ \citenamefont
  {Emerson}}]{Veitch2014Jan}%
  \BibitemOpen
  \bibfield  {author} {\bibinfo {author} {\bibfnamefont {V.}~\bibnamefont
  {Veitch}}, \bibinfo {author} {\bibfnamefont {S.~A.~H.}\ \bibnamefont
  {Mousavian}}, \bibinfo {author} {\bibfnamefont {D.}~\bibnamefont
  {Gottesman}},\ and\ \bibinfo {author} {\bibfnamefont {J.}~\bibnamefont
  {Emerson}},\ }\bibfield  {title} {\bibinfo {title} {{The resource theory of
  stabilizer quantum computation}},\ }\href
  {https://doi.org/10.1088/1367-2630/16/1/013009} {\bibfield  {journal}
  {\bibinfo  {journal} {New J. Phys.}\ }\textbf {\bibinfo {volume} {16}},\
  \bibinfo {pages} {013009} (\bibinfo {year} {2014})}\BibitemShut {NoStop}%
\bibitem [{\citenamefont {Howard}\ and\ \citenamefont
  {Campbell}(2017)}]{Howard2017Mar}%
  \BibitemOpen
  \bibfield  {author} {\bibinfo {author} {\bibfnamefont {M.}~\bibnamefont
  {Howard}}\ and\ \bibinfo {author} {\bibfnamefont {E.}~\bibnamefont
  {Campbell}},\ }\bibfield  {title} {\bibinfo {title} {{Application of a
  Resource Theory for Magic States to Fault-Tolerant Quantum Computing}},\
  }\href {https://doi.org/10.1103/PhysRevLett.118.090501} {\bibfield  {journal}
  {\bibinfo  {journal} {Phys. Rev. Lett.}\ }\textbf {\bibinfo {volume} {118}},\
  \bibinfo {pages} {090501} (\bibinfo {year} {2017})}\BibitemShut {NoStop}%
\bibitem [{\citenamefont {Leone}\ \emph {et~al.}(2022)\citenamefont {Leone},
  \citenamefont {Oliviero},\ and\ \citenamefont {Hamma}}]{leone2022}%
  \BibitemOpen
  \bibfield  {author} {\bibinfo {author} {\bibfnamefont {L.}~\bibnamefont
  {Leone}}, \bibinfo {author} {\bibfnamefont {S.~F.~E.}\ \bibnamefont
  {Oliviero}},\ and\ \bibinfo {author} {\bibfnamefont {A.}~\bibnamefont
  {Hamma}},\ }\bibfield  {title} {\bibinfo {title} {{Stabilizer R\'enyi
  Entropy}},\ }\href {https://doi.org/10.1103/PhysRevLett.128.050402}
  {\bibfield  {journal} {\bibinfo  {journal} {Phys. Rev. Lett.}\ }\textbf
  {\bibinfo {volume} {128}},\ \bibinfo {pages} {050402} (\bibinfo {year}
  {2022})}\BibitemShut {NoStop}%
\bibitem [{\citenamefont {Projansky}\ \emph {et~al.}(2024)\citenamefont
  {Projansky}, \citenamefont {Heath},\ and\ \citenamefont
  {Whitfield}}]{Projansky2024Aug}%
  \BibitemOpen
  \bibfield  {author} {\bibinfo {author} {\bibfnamefont {A.~M.}\ \bibnamefont
  {Projansky}}, \bibinfo {author} {\bibfnamefont {J.~T.}\ \bibnamefont
  {Heath}},\ and\ \bibinfo {author} {\bibfnamefont {J.~D.}\ \bibnamefont
  {Whitfield}},\ }\bibfield  {title} {\bibinfo {title} {{Entanglement spectrum
  of matchgate circuits with universal and non-universal resources}},\ }\href
  {https://doi.org/10.22331/q-2024-08-07-1432} {\bibfield  {journal} {\bibinfo
  {journal} {Quantum}\ }\textbf {\bibinfo {volume} {8}},\ \bibinfo {pages}
  {1432} (\bibinfo {year} {2024})},\ \Eprint
  {https://arxiv.org/abs/2312.08447v3} {2312.08447v3} \BibitemShut {NoStop}%
\bibitem [{\citenamefont {White}\ and\ \citenamefont
  {White}(2024)}]{White2024Dec}%
  \BibitemOpen
  \bibfield  {author} {\bibinfo {author} {\bibfnamefont {C.~D.}\ \bibnamefont
  {White}}\ and\ \bibinfo {author} {\bibfnamefont {M.~J.}\ \bibnamefont
  {White}},\ }\bibfield  {title} {\bibinfo {title} {{Magic states of top
  quarks}},\ }\href {https://doi.org/10.1103/PhysRevD.110.116016} {\bibfield
  {journal} {\bibinfo  {journal} {Phys. Rev. D}\ }\textbf {\bibinfo {volume}
  {110}},\ \bibinfo {pages} {116016} (\bibinfo {year} {2024})}\BibitemShut
  {NoStop}%
\bibitem [{\citenamefont {White}\ \emph {et~al.}(2021)\citenamefont {White},
  \citenamefont {Cao},\ and\ \citenamefont {Swingle}}]{White2021Feb}%
  \BibitemOpen
  \bibfield  {author} {\bibinfo {author} {\bibfnamefont {C.~D.}\ \bibnamefont
  {White}}, \bibinfo {author} {\bibfnamefont {C.}~\bibnamefont {Cao}},\ and\
  \bibinfo {author} {\bibfnamefont {B.}~\bibnamefont {Swingle}},\ }\bibfield
  {title} {\bibinfo {title} {{Conformal field theories are magical}},\ }\href
  {https://doi.org/10.1103/PhysRevB.103.075145} {\bibfield  {journal} {\bibinfo
   {journal} {Phys. Rev. B}\ }\textbf {\bibinfo {volume} {103}},\ \bibinfo
  {pages} {075145} (\bibinfo {year} {2021})}\BibitemShut {NoStop}%
\bibitem [{\citenamefont {Liu}\ and\ \citenamefont
  {Winter}(2022)}]{Liu2022May}%
  \BibitemOpen
  \bibfield  {author} {\bibinfo {author} {\bibfnamefont {Z.-W.}\ \bibnamefont
  {Liu}}\ and\ \bibinfo {author} {\bibfnamefont {A.}~\bibnamefont {Winter}},\
  }\bibfield  {title} {\bibinfo {title} {{Many-Body Quantum Magic}},\ }\href
  {https://doi.org/10.1103/PRXQuantum.3.020333} {\bibfield  {journal} {\bibinfo
   {journal} {PRX Quantum}\ }\textbf {\bibinfo {volume} {3}},\ \bibinfo {pages}
  {020333} (\bibinfo {year} {2022})}\BibitemShut {NoStop}%
\bibitem [{\citenamefont {Oliviero}\ \emph {et~al.}(2022)\citenamefont
  {Oliviero}, \citenamefont {Leone},\ and\ \citenamefont
  {Hamma}}]{Oliviero2022Oct}%
  \BibitemOpen
  \bibfield  {author} {\bibinfo {author} {\bibfnamefont {S.~F.~E.}\
  \bibnamefont {Oliviero}}, \bibinfo {author} {\bibfnamefont {L.}~\bibnamefont
  {Leone}},\ and\ \bibinfo {author} {\bibfnamefont {A.}~\bibnamefont {Hamma}},\
  }\bibfield  {title} {\bibinfo {title} {{Magic-state resource theory for the
  ground state of the transverse-field Ising model}},\ }\href
  {https://doi.org/10.1103/PhysRevA.106.042426} {\bibfield  {journal} {\bibinfo
   {journal} {Phys. Rev. A}\ }\textbf {\bibinfo {volume} {106}},\ \bibinfo
  {pages} {042426} (\bibinfo {year} {2022})}\BibitemShut {NoStop}%
\bibitem [{\citenamefont {Haug}\ and\ \citenamefont {Kim}(2023)}]{Haug2023Jan}%
  \BibitemOpen
  \bibfield  {author} {\bibinfo {author} {\bibfnamefont {T.}~\bibnamefont
  {Haug}}\ and\ \bibinfo {author} {\bibfnamefont {M.~S.}\ \bibnamefont {Kim}},\
  }\bibfield  {title} {\bibinfo {title} {{Scalable Measures of Magic Resource
  for Quantum Computers}},\ }\href
  {https://doi.org/10.1103/PRXQuantum.4.010301} {\bibfield  {journal} {\bibinfo
   {journal} {PRX Quantum}\ }\textbf {\bibinfo {volume} {4}},\ \bibinfo {pages}
  {010301} (\bibinfo {year} {2023})}\BibitemShut {NoStop}%
\bibitem [{\citenamefont {Odavi{\ifmmode\acute{c}\else\'{c}\fi}}\ \emph
  {et~al.}(2023)\citenamefont {Odavi{\ifmmode\acute{c}\else\'{c}\fi}},
  \citenamefont {Haug}, \citenamefont {Torre}, \citenamefont {Hamma},
  \citenamefont {Franchini},\ and\ \citenamefont {Giampaolo}}]{Odavic2023Oct}%
  \BibitemOpen
  \bibfield  {author} {\bibinfo {author} {\bibfnamefont {J.}~\bibnamefont
  {Odavi{\ifmmode\acute{c}\else\'{c}\fi}}}, \bibinfo {author} {\bibfnamefont
  {T.}~\bibnamefont {Haug}}, \bibinfo {author} {\bibfnamefont {G.}~\bibnamefont
  {Torre}}, \bibinfo {author} {\bibfnamefont {A.}~\bibnamefont {Hamma}},
  \bibinfo {author} {\bibfnamefont {F.}~\bibnamefont {Franchini}},\ and\
  \bibinfo {author} {\bibfnamefont {S.~M.}\ \bibnamefont {Giampaolo}},\
  }\bibfield  {title} {\bibinfo {title} {{Complexity of frustration: A new
  source of non-local non-stabilizerness}},\ }\href
  {https://doi.org/10.21468/SciPostPhys.15.4.131} {\bibfield  {journal}
  {\bibinfo  {journal} {SciPost Phys.}\ }\textbf {\bibinfo {volume} {15}},\
  \bibinfo {pages} {131} (\bibinfo {year} {2023})}\BibitemShut {NoStop}%
\bibitem [{\citenamefont {Tarabunga}\ \emph {et~al.}(2023)\citenamefont
  {Tarabunga}, \citenamefont {Tirrito}, \citenamefont {Chanda},\ and\
  \citenamefont {Dalmonte}}]{Tarabunga2023Oct}%
  \BibitemOpen
  \bibfield  {author} {\bibinfo {author} {\bibfnamefont {P.~S.}\ \bibnamefont
  {Tarabunga}}, \bibinfo {author} {\bibfnamefont {E.}~\bibnamefont {Tirrito}},
  \bibinfo {author} {\bibfnamefont {T.}~\bibnamefont {Chanda}},\ and\ \bibinfo
  {author} {\bibfnamefont {M.}~\bibnamefont {Dalmonte}},\ }\bibfield  {title}
  {\bibinfo {title} {{Many-Body Magic Via Pauli-Markov Chains---From
  Criticality to Gauge Theories}},\ }\href
  {https://doi.org/10.1103/PRXQuantum.4.040317} {\bibfield  {journal} {\bibinfo
   {journal} {PRX Quantum}\ }\textbf {\bibinfo {volume} {4}},\ \bibinfo {pages}
  {040317} (\bibinfo {year} {2023})}\BibitemShut {NoStop}%
\bibitem [{\citenamefont {Collura}\ \emph {et~al.}(2026)\citenamefont
  {Collura}, \citenamefont {De~Nardis}, \citenamefont {Alba},\ and\
  \citenamefont {Lami}}]{collura2024}%
  \BibitemOpen
  \bibfield  {author} {\bibinfo {author} {\bibfnamefont {M.}~\bibnamefont
  {Collura}}, \bibinfo {author} {\bibfnamefont {J.}~\bibnamefont {De~Nardis}},
  \bibinfo {author} {\bibfnamefont {V.}~\bibnamefont {Alba}},\ and\ \bibinfo
  {author} {\bibfnamefont {G.}~\bibnamefont {Lami}},\ }\bibfield  {title}
  {\bibinfo {title} {{The non-stabilizerness of fermionic Gaussian states}},\
  }\href {https://doi.org/10.22331/q-2026-03-23-2036} {\bibfield  {journal}
  {\bibinfo  {journal} {Quantum}\ }\textbf {\bibinfo {volume} {10}},\ \bibinfo
  {pages} {2036} (\bibinfo {year} {2026})},\ \Eprint
  {https://arxiv.org/abs/2412.05367v4} {2412.05367v4} \BibitemShut {NoStop}%
\bibitem [{\citenamefont {Hoshino}\ \emph
  {et~al.}(2026{\natexlab{a}})\citenamefont {Hoshino}, \citenamefont
  {Oshikawa},\ and\ \citenamefont {Ashida}}]{Hoshino2026Feb}%
  \BibitemOpen
  \bibfield  {author} {\bibinfo {author} {\bibfnamefont {M.}~\bibnamefont
  {Hoshino}}, \bibinfo {author} {\bibfnamefont {M.}~\bibnamefont {Oshikawa}},\
  and\ \bibinfo {author} {\bibfnamefont {Y.}~\bibnamefont {Ashida}},\
  }\bibfield  {title} {\bibinfo {title} {{Stabilizer R\'enyi Entropy and
  Conformal Field Theory}},\ }\href {https://doi.org/10.1103/ylsz-dm3y}
  {\bibfield  {journal} {\bibinfo  {journal} {Phys. Rev. X}\ }\textbf {\bibinfo
  {volume} {16}},\ \bibinfo {pages} {011037} (\bibinfo {year}
  {2026}{\natexlab{a}})}\BibitemShut {NoStop}%
\bibitem [{\citenamefont {Ding}\ \emph {et~al.}(2025)\citenamefont {Ding},
  \citenamefont {Wang},\ and\ \citenamefont {Yan}}]{Ding2025Aug}%
  \BibitemOpen
  \bibfield  {author} {\bibinfo {author} {\bibfnamefont {Y.-M.}\ \bibnamefont
  {Ding}}, \bibinfo {author} {\bibfnamefont {Z.}~\bibnamefont {Wang}},\ and\
  \bibinfo {author} {\bibfnamefont {Z.}~\bibnamefont {Yan}},\ }\bibfield
  {title} {\bibinfo {title} {{Evaluating Many-Body Stabilizer R\'enyi Entropy
  by Sampling Reduced Pauli Strings: Singularities, Volume Law, and Nonlocal
  Magic}},\ }\href {https://doi.org/10.1103/pyzr-jmvw} {\bibfield  {journal}
  {\bibinfo  {journal} {PRX Quantum}\ }\textbf {\bibinfo {volume} {6}},\
  \bibinfo {pages} {030328} (\bibinfo {year} {2025})}\BibitemShut {NoStop}%
\bibitem [{\citenamefont {Russomanno}\ \emph {et~al.}(2025)\citenamefont
  {Russomanno}, \citenamefont {Passarelli}, \citenamefont {Rossini},\ and\
  \citenamefont {Lucignano}}]{Russomanno2025Aug}%
  \BibitemOpen
  \bibfield  {author} {\bibinfo {author} {\bibfnamefont {A.}~\bibnamefont
  {Russomanno}}, \bibinfo {author} {\bibfnamefont {G.}~\bibnamefont
  {Passarelli}}, \bibinfo {author} {\bibfnamefont {D.}~\bibnamefont
  {Rossini}},\ and\ \bibinfo {author} {\bibfnamefont {P.}~\bibnamefont
  {Lucignano}},\ }\bibfield  {title} {\bibinfo {title} {{Nonstabilizerness in
  the unitary and monitored quantum dynamics of XXZ-staggered and
  Sachdev-Ye-Kitaev models}},\ }\href {https://doi.org/10.1103/njgn-fksh}
  {\bibfield  {journal} {\bibinfo  {journal} {Phys. Rev. B}\ }\textbf {\bibinfo
  {volume} {112}},\ \bibinfo {pages} {064312} (\bibinfo {year}
  {2025})}\BibitemShut {NoStop}%
\bibitem [{\citenamefont {Lamma}\ \emph {et~al.}(2026)\citenamefont {Lamma},
  \citenamefont {Bauer}, \citenamefont {Dalmonte},\ and\ \citenamefont
  {Collura}}]{Lamma2026Aug}%
  \BibitemOpen
  \bibfield  {author} {\bibinfo {author} {\bibfnamefont {E.}~\bibnamefont
  {Lamma}}, \bibinfo {author} {\bibfnamefont {T.}~\bibnamefont {Bauer}},
  \bibinfo {author} {\bibfnamefont {M.}~\bibnamefont {Dalmonte}},\ and\
  \bibinfo {author} {\bibfnamefont {M.}~\bibnamefont {Collura}},\ }\bibfield
  {title} {\bibinfo {title} {{Magic of Kitaev spin liquids}},\ }\bibfield
  {journal} {\bibinfo  {journal} {arXiv}\ }\href
  {https://doi.org/10.48550/arXiv.2608.25929} {10.48550/arXiv.2608.25929}
  (\bibinfo {year} {2026}),\ \Eprint {https://arxiv.org/abs/2608.25929}
  {2608.25929} \BibitemShut {NoStop}%
\bibitem [{\citenamefont {Sabharwal}\ \emph {et~al.}(2026)\citenamefont
  {Sabharwal}, \citenamefont {Shannon},\ and\ \citenamefont
  {Skrzypczyk}}]{Sabharwal2026Sep}%
  \BibitemOpen
  \bibfield  {author} {\bibinfo {author} {\bibfnamefont {S.}~\bibnamefont
  {Sabharwal}}, \bibinfo {author} {\bibfnamefont {N.}~\bibnamefont {Shannon}},\
  and\ \bibinfo {author} {\bibfnamefont {P.}~\bibnamefont {Skrzypczyk}},\
  }\bibfield  {title} {\bibinfo {title} {{Such stuff as magic is made on:
  compact operator algebra, stabilizer polytope and the structure of reduced
  density matrices in a Kitaev spin liquid}},\ }\bibfield  {journal} {\bibinfo
  {journal} {arXiv}\ }\href {https://doi.org/10.48550/arXiv.2609.03485}
  {10.48550/arXiv.2609.03485} (\bibinfo {year} {2026}),\ \Eprint
  {https://arxiv.org/abs/2609.03485} {2609.03485} \BibitemShut {NoStop}%
\bibitem [{\citenamefont {Turkeshi}\ \emph {et~al.}(2025)\citenamefont
  {Turkeshi}, \citenamefont {Dymarsky},\ and\ \citenamefont
  {Sierant}}]{turkeshiPRB2025}%
  \BibitemOpen
  \bibfield  {author} {\bibinfo {author} {\bibfnamefont {X.}~\bibnamefont
  {Turkeshi}}, \bibinfo {author} {\bibfnamefont {A.}~\bibnamefont {Dymarsky}},\
  and\ \bibinfo {author} {\bibfnamefont {P.}~\bibnamefont {Sierant}},\
  }\bibfield  {title} {\bibinfo {title} {{Pauli spectrum and nonstabilizerness
  of typical quantum many-body states}},\ }\href
  {https://doi.org/10.1103/PhysRevB.111.054301} {\bibfield  {journal} {\bibinfo
   {journal} {Phys. Rev. B}\ }\textbf {\bibinfo {volume} {111}},\ \bibinfo
  {pages} {054301} (\bibinfo {year} {2025})}\BibitemShut {NoStop}%
\bibitem [{\citenamefont {Szombathy}\ \emph {et~al.}(2025)\citenamefont
  {Szombathy}, \citenamefont {Valli}, \citenamefont {Moca}, \citenamefont
  {Farkas},\ and\ \citenamefont
  {Zar{\ifmmode\acute{a}\else\'{a}\fi}nd}}]{Szombathy2025Oct}%
  \BibitemOpen
  \bibfield  {author} {\bibinfo {author} {\bibfnamefont {D.}~\bibnamefont
  {Szombathy}}, \bibinfo {author} {\bibfnamefont {A.}~\bibnamefont {Valli}},
  \bibinfo {author} {\bibfnamefont {C.~P.}\ \bibnamefont {Moca}}, \bibinfo
  {author} {\bibfnamefont {L.}~\bibnamefont {Farkas}},\ and\ \bibinfo {author}
  {\bibfnamefont {G.}~\bibnamefont {Zar{\ifmmode\acute{a}\else\'{a}\fi}nd}},\
  }\bibfield  {title} {\bibinfo {title} {{Asymptotically independent
  fluctuations of stabilizer R{\ifmmode\acute{e}\else\'{e}\fi}nyi entropy and
  entanglement in random unitary circuits}},\ }\href
  {https://doi.org/10.1103/jplh-zl35} {\bibfield  {journal} {\bibinfo
  {journal} {Phys. Rev. Res.}\ }\textbf {\bibinfo {volume} {7}},\ \bibinfo
  {pages} {043072} (\bibinfo {year} {2025})}\BibitemShut {NoStop}%
\bibitem [{\citenamefont {Page}(1993)}]{Page1993Aug}%
  \BibitemOpen
  \bibfield  {author} {\bibinfo {author} {\bibfnamefont {D.~N.}\ \bibnamefont
  {Page}},\ }\bibfield  {title} {\bibinfo {title} {{Average entropy of a
  subsystem}},\ }\href {https://doi.org/10.1103/PhysRevLett.71.1291} {\bibfield
   {journal} {\bibinfo  {journal} {Phys. Rev. Lett.}\ }\textbf {\bibinfo
  {volume} {71}},\ \bibinfo {pages} {1291} (\bibinfo {year}
  {1993})}\BibitemShut {NoStop}%
\bibitem [{\citenamefont {Iaconis}(2021)}]{Iaconis2021Feb}%
  \BibitemOpen
  \bibfield  {author} {\bibinfo {author} {\bibfnamefont {J.}~\bibnamefont
  {Iaconis}},\ }\bibfield  {title} {\bibinfo {title} {{Quantum State Complexity
  in Computationally Tractable Quantum Circuits}},\ }\href
  {https://doi.org/10.1103/PRXQuantum.2.010329} {\bibfield  {journal} {\bibinfo
   {journal} {PRX Quantum}\ }\textbf {\bibinfo {volume} {2}},\ \bibinfo {pages}
  {010329} (\bibinfo {year} {2021})}\BibitemShut {NoStop}%
\bibitem [{\citenamefont {Sierant}\ \emph {et~al.}(2026)\citenamefont
  {Sierant}, \citenamefont {Stornati},\ and\ \citenamefont
  {Turkeshi}}]{Sierant2026Jan}%
  \BibitemOpen
  \bibfield  {author} {\bibinfo {author} {\bibfnamefont {P.}~\bibnamefont
  {Sierant}}, \bibinfo {author} {\bibfnamefont {P.}~\bibnamefont {Stornati}},\
  and\ \bibinfo {author} {\bibfnamefont {X.}~\bibnamefont {Turkeshi}},\
  }\bibfield  {title} {\bibinfo {title} {{Fermionic Magic Resources of Quantum
  Many-Body Systems}},\ }\href {https://doi.org/10.1103/3yx4-1j27} {\bibfield
  {journal} {\bibinfo  {journal} {PRX Quantum}\ }\textbf {\bibinfo {volume}
  {7}},\ \bibinfo {pages} {010302} (\bibinfo {year} {2026})}\BibitemShut
  {NoStop}%
\bibitem [{\citenamefont {Hoshino}\ \emph
  {et~al.}(2026{\natexlab{b}})\citenamefont {Hoshino}, \citenamefont
  {Matsuda},\ and\ \citenamefont {Ashida}}]{Hoshino2026Oct}%
  \BibitemOpen
  \bibfield  {author} {\bibinfo {author} {\bibfnamefont {M.}~\bibnamefont
  {Hoshino}}, \bibinfo {author} {\bibfnamefont {R.}~\bibnamefont {Matsuda}},\
  and\ \bibinfo {author} {\bibfnamefont {Y.}~\bibnamefont {Ashida}},\
  }\bibfield  {title} {\bibinfo {title} {{Universal Bound and Phase Transition
  in Many-Body Fermionic Non-Gaussianity}},\ }\bibfield  {journal} {\bibinfo
  {journal} {arXiv}\ }\href {https://doi.org/10.48550/arXiv.2610.02075}
  {10.48550/arXiv.2610.02075} (\bibinfo {year} {2026}{\natexlab{b}}),\ \Eprint
  {https://arxiv.org/abs/2610.02075} {2610.02075} \BibitemShut {NoStop}%
\bibitem [{\citenamefont {Roberts}\ and\ \citenamefont
  {Yoshida}(2017)}]{Roberts2017Apr}%
  \BibitemOpen
  \bibfield  {author} {\bibinfo {author} {\bibfnamefont {D.~A.}\ \bibnamefont
  {Roberts}}\ and\ \bibinfo {author} {\bibfnamefont {B.}~\bibnamefont
  {Yoshida}},\ }\bibfield  {title} {\bibinfo {title} {{Chaos and complexity by
  design}},\ }\href {https://doi.org/10.1007/JHEP04(2017)121} {\bibfield
  {journal} {\bibinfo  {journal} {J. High Energy Phys.}\ }\textbf {\bibinfo
  {volume} {2017}}\bibinfo  {number} { (4)},\ \bibinfo {pages}
  {121}}\BibitemShut {NoStop}%
\bibitem [{\citenamefont {Bravyi}\ \emph {et~al.}(2016)\citenamefont {Bravyi},
  \citenamefont {Smith},\ and\ \citenamefont {Smolin}}]{Bravyi2016Jun}%
  \BibitemOpen
\bibfield  {number} {  }\bibfield  {author} {\bibinfo {author} {\bibfnamefont
  {S.}~\bibnamefont {Bravyi}}, \bibinfo {author} {\bibfnamefont
  {G.}~\bibnamefont {Smith}},\ and\ \bibinfo {author} {\bibfnamefont {J.~A.}\
  \bibnamefont {Smolin}},\ }\bibfield  {title} {\bibinfo {title} {{Trading
  Classical and Quantum Computational Resources}},\ }\href
  {https://doi.org/10.1103/PhysRevX.6.021043} {\bibfield  {journal} {\bibinfo
  {journal} {Phys. Rev. X}\ }\textbf {\bibinfo {volume} {6}},\ \bibinfo {pages}
  {021043} (\bibinfo {year} {2016})}\BibitemShut {NoStop}%
\bibitem [{\citenamefont {Wang}\ \emph {et~al.}(2020)\citenamefont {Wang},
  \citenamefont {Wilde},\ and\ \citenamefont {Su}}]{Wang2020Mar}%
  \BibitemOpen
  \bibfield  {author} {\bibinfo {author} {\bibfnamefont {X.}~\bibnamefont
  {Wang}}, \bibinfo {author} {\bibfnamefont {M.~M.}\ \bibnamefont {Wilde}},\
  and\ \bibinfo {author} {\bibfnamefont {Y.}~\bibnamefont {Su}},\ }\bibfield
  {title} {\bibinfo {title} {{Efficiently Computable Bounds for Magic State
  Distillation}},\ }\href {https://doi.org/10.1103/PhysRevLett.124.090505}
  {\bibfield  {journal} {\bibinfo  {journal} {Phys. Rev. Lett.}\ }\textbf
  {\bibinfo {volume} {124}},\ \bibinfo {pages} {090505} (\bibinfo {year}
  {2020})}\BibitemShut {NoStop}%
\bibitem [{\citenamefont {Haug}\ and\ \citenamefont
  {Piroli}(2023)}]{HaugPiroli2023Jan}%
  \BibitemOpen
  \bibfield  {author} {\bibinfo {author} {\bibfnamefont {T.}~\bibnamefont
  {Haug}}\ and\ \bibinfo {author} {\bibfnamefont {L.}~\bibnamefont {Piroli}},\
  }\bibfield  {title} {\bibinfo {title} {{Quantifying nonstabilizerness of
  matrix product states}},\ }\href
  {https://doi.org/10.1103/PhysRevB.107.035148} {\bibfield  {journal} {\bibinfo
   {journal} {Phys. Rev. B}\ }\textbf {\bibinfo {volume} {107}},\ \bibinfo
  {pages} {035148} (\bibinfo {year} {2023})}\BibitemShut {NoStop}%
\bibitem [{\citenamefont {Liu}\ and\ \citenamefont {Clark}(2025)}]{Liu2025Feb}%
  \BibitemOpen
  \bibfield  {author} {\bibinfo {author} {\bibfnamefont {Z.}~\bibnamefont
  {Liu}}\ and\ \bibinfo {author} {\bibfnamefont {B.~K.}\ \bibnamefont
  {Clark}},\ }\bibfield  {title} {\bibinfo {title} {{Nonequilibrium quantum
  Monte Carlo algorithm for stabilizer R{\ifmmode\acute{e}\else\'{e}\fi}nyi
  entropy in spin systems}},\ }\href
  {https://doi.org/10.1103/PhysRevB.111.085144} {\bibfield  {journal} {\bibinfo
   {journal} {Phys. Rev. B}\ }\textbf {\bibinfo {volume} {111}},\ \bibinfo
  {pages} {085144} (\bibinfo {year} {2025})}\BibitemShut {NoStop}%
\bibitem [{\citenamefont {Huang}\ \emph {et~al.}(2025)\citenamefont {Huang},
  \citenamefont {Li}, \citenamefont {Lee},\ and\ \citenamefont
  {Zhong}}]{huang2026}%
  \BibitemOpen
  \bibfield  {author} {\bibinfo {author} {\bibfnamefont {X.}~\bibnamefont
  {Huang}}, \bibinfo {author} {\bibfnamefont {H.-Z.}\ \bibnamefont {Li}},
  \bibinfo {author} {\bibfnamefont {C.~H.}\ \bibnamefont {Lee}},\ and\ \bibinfo
  {author} {\bibfnamefont {J.-X.}\ \bibnamefont {Zhong}},\ }\bibfield  {title}
  {\bibinfo {title} {{A fast and exact approach for stabilizer R\'enyi entropy
  via the XOR-FWHT algorithm}},\ }\bibfield  {journal} {\bibinfo  {journal}
  {arXiv}\ }\href {https://doi.org/10.48550/arXiv.2512.24685}
  {10.48550/arXiv.2512.24685} (\bibinfo {year} {2025}),\ \Eprint
  {https://arxiv.org/abs/2512.24685} {2512.24685} \BibitemShut {NoStop}%
\bibitem [{\citenamefont {Xiao}\ and\ \citenamefont {Ryu}(2026)}]{Xiao2026}%
  \BibitemOpen
  \bibfield  {author} {\bibinfo {author} {\bibfnamefont {Z.}~\bibnamefont
  {Xiao}}\ and\ \bibinfo {author} {\bibfnamefont {S.}~\bibnamefont {Ryu}},\
  }\bibfield  {title} {\bibinfo {title} {{Exponentially Accelerated Sampling of
  Pauli Strings for Nonstabilizerness}},\ }\href
  {https://doi.org/10.1103/v3f2-ldr1} {\bibfield  {journal} {\bibinfo
  {journal} {Phys. Rev. Lett.}\ }\textbf {\bibinfo {volume} {137}},\ \bibinfo
  {pages} {070402} (\bibinfo {year} {2026})}\BibitemShut {NoStop}%
\bibitem [{\citenamefont {Bittel}\ and\ \citenamefont
  {Leone}(2026)}]{Bittel2026Apr}%
  \BibitemOpen
  \bibfield  {author} {\bibinfo {author} {\bibfnamefont {L.}~\bibnamefont
  {Bittel}}\ and\ \bibinfo {author} {\bibfnamefont {L.}~\bibnamefont {Leone}},\
  }\bibfield  {title} {\bibinfo {title} {{Operational interpretation of the
  Stabilizer Entropy}},\ }\href {https://doi.org/10.22331/q-2026-04-15-2069}
  {\bibfield  {journal} {\bibinfo  {journal} {Quantum}\ }\textbf {\bibinfo
  {volume} {10}},\ \bibinfo {pages} {2069} (\bibinfo {year} {2026})},\ \Eprint
  {https://arxiv.org/abs/2507.22883v3} {2507.22883v3} \BibitemShut {NoStop}%
\bibitem [{\citenamefont {Goldstein}\ \emph {et~al.}(2006)\citenamefont
  {Goldstein}, \citenamefont {Lebowitz}, \citenamefont {Tumulka},\ and\
  \citenamefont
  {Zangh{\ifmmode\grave{\imath}\else\`{\i}\fi}}}]{Goldstein2006Feb}%
  \BibitemOpen
  \bibfield  {author} {\bibinfo {author} {\bibfnamefont {S.}~\bibnamefont
  {Goldstein}}, \bibinfo {author} {\bibfnamefont {J.~L.}\ \bibnamefont
  {Lebowitz}}, \bibinfo {author} {\bibfnamefont {R.}~\bibnamefont {Tumulka}},\
  and\ \bibinfo {author} {\bibfnamefont {N.}~\bibnamefont
  {Zangh{\ifmmode\grave{\imath}\else\`{\i}\fi}}},\ }\bibfield  {title}
  {\bibinfo {title} {{Canonical Typicality}},\ }\href
  {https://doi.org/10.1103/PhysRevLett.96.050403} {\bibfield  {journal}
  {\bibinfo  {journal} {Phys. Rev. Lett.}\ }\textbf {\bibinfo {volume} {96}},\
  \bibinfo {pages} {050403} (\bibinfo {year} {2006})}\BibitemShut {NoStop}%
\bibitem [{\citenamefont {Reimann}(2007)}]{Reimann2007Oct}%
  \BibitemOpen
  \bibfield  {author} {\bibinfo {author} {\bibfnamefont {P.}~\bibnamefont
  {Reimann}},\ }\bibfield  {title} {\bibinfo {title} {{Typicality for
  Generalized Microcanonical Ensembles}},\ }\href
  {https://doi.org/10.1103/PhysRevLett.99.160404} {\bibfield  {journal}
  {\bibinfo  {journal} {Phys. Rev. Lett.}\ }\textbf {\bibinfo {volume} {99}},\
  \bibinfo {pages} {160404} (\bibinfo {year} {2007})}\BibitemShut {NoStop}%
\bibitem [{\citenamefont {Wilde}(2017)}]{Wilde2017Feb}%
  \BibitemOpen
  \bibfield  {author} {\bibinfo {author} {\bibfnamefont {M.~M.}\ \bibnamefont
  {Wilde}},\ }\bibfield  {title} {\bibinfo {title} {{Quantum Typicality}},\
  }in\ \href {https://doi.org/10.1017/9781316809976.018} {\emph {\bibinfo
  {booktitle} {{Quantum Information Theory}}}}\ (\bibinfo  {publisher}
  {Cambridge University Press},\ \bibinfo {address} {Cambridge, England, UK},\
  \bibinfo {year} {2017})\ pp.\ \bibinfo {pages} {443--466}\BibitemShut
  {NoStop}%
\bibitem [{\citenamefont {Quinn}\ \emph {et~al.}(2026)\citenamefont {Quinn},
  \citenamefont {Heath},\ and\ \citenamefont {Kells}}]{TypicalEntanglement}%
  \BibitemOpen
  \bibfield  {author} {\bibinfo {author} {\bibfnamefont {D.}~\bibnamefont
  {Quinn}}, \bibinfo {author} {\bibfnamefont {J.~T.}\ \bibnamefont {Heath}},\
  and\ \bibinfo {author} {\bibfnamefont {G.}~\bibnamefont {Kells}},\ }\bibfield
   {title} {\bibinfo {title} {{Typical Entanglement of Superpositions}},\
  }\bibfield  {journal} {\bibinfo  {journal} {arXiv}\ }\href
  {https://doi.org/10.48550/arXiv.2607.06474} {10.48550/arXiv.2607.06474}
  (\bibinfo {year} {2026}),\ \Eprint {https://arxiv.org/abs/2607.06474}
  {2607.06474} \BibitemShut {NoStop}%
\bibitem [{Note1()}]{Note1}%
  \BibitemOpen
  \bibinfo {note} {We fix the weights by $\DOTSB \sum@ \slimits@ _ip_i=1$
  rather than normalizing each realization of $|\Psi _m\rangle $. As the
  components are not orthogonal, this introduces an average error of order
  $2^{-N}$ in the witness. \par}\BibitemShut {NoStop}%
\bibitem [{\citenamefont {Georgi}(2000)}]{georgi}%
  \BibitemOpen
  \bibfield  {author} {\bibinfo {author} {\bibfnamefont {H.}~\bibnamefont
  {Georgi}},\ }\href {https://doi.org/10.1201/9780429499210} {\emph {\bibinfo
  {title} {{Lie Algebras In Particle Physics : from Isospin To Unified
  Theories}}}}\ (\bibinfo  {publisher} {Taylor {\&} Francis},\ \bibinfo
  {address} {Andover, England, UK},\ \bibinfo {year} {2000})\BibitemShut
  {NoStop}%
\bibitem [{\citenamefont {Serre}(1977)}]{Serre}%
  \BibitemOpen
  \bibfield  {author} {\bibinfo {author} {\bibfnamefont {J.-P.}\ \bibnamefont
  {Serre}},\ }\href {https://link.springer.com/book/10.1007/978-1-4684-9458-7}
  {\emph {\bibinfo {title} {{Linear Representations of Finite Groups}}}}\
  (\bibinfo  {publisher} {Springer},\ \bibinfo {address} {New York, NY, USA},\
  \bibinfo {year} {1977})\BibitemShut {NoStop}%
\bibitem [{\citenamefont {Etingof}\ \emph {et~al.}(2009)\citenamefont
  {Etingof}, \citenamefont {Golberg}, \citenamefont {Hensel}, \citenamefont
  {Liu}, \citenamefont {Schwendner}, \citenamefont {Vaintrob},\ and\
  \citenamefont {Yudovina}}]{Etingof2009Jan}%
  \BibitemOpen
  \bibfield  {author} {\bibinfo {author} {\bibfnamefont {P.}~\bibnamefont
  {Etingof}}, \bibinfo {author} {\bibfnamefont {O.}~\bibnamefont {Golberg}},
  \bibinfo {author} {\bibfnamefont {S.}~\bibnamefont {Hensel}}, \bibinfo
  {author} {\bibfnamefont {T.}~\bibnamefont {Liu}}, \bibinfo {author}
  {\bibfnamefont {A.}~\bibnamefont {Schwendner}}, \bibinfo {author}
  {\bibfnamefont {D.}~\bibnamefont {Vaintrob}},\ and\ \bibinfo {author}
  {\bibfnamefont {E.}~\bibnamefont {Yudovina}},\ }\bibfield  {title} {\bibinfo
  {title} {{Introduction to representation theory}},\ }\bibfield  {journal}
  {\bibinfo  {journal} {arXiv}\ }\href
  {https://doi.org/10.48550/arXiv.0901.0827} {10.48550/arXiv.0901.0827}
  (\bibinfo {year} {2009}),\ \Eprint {https://arxiv.org/abs/0901.0827}
  {0901.0827} \BibitemShut {NoStop}%
\bibitem [{\citenamefont {Renes}\ \emph {et~al.}(2004)\citenamefont {Renes},
  \citenamefont {Blume-Kohout}, \citenamefont {Scott},\ and\ \citenamefont
  {Caves}}]{Renes2004}%
  \BibitemOpen
  \bibfield  {author} {\bibinfo {author} {\bibfnamefont {J.~M.}\ \bibnamefont
  {Renes}}, \bibinfo {author} {\bibfnamefont {R.}~\bibnamefont {Blume-Kohout}},
  \bibinfo {author} {\bibfnamefont {A.~J.}\ \bibnamefont {Scott}},\ and\
  \bibinfo {author} {\bibfnamefont {C.~M.}\ \bibnamefont {Caves}},\ }\bibfield
  {title} {\bibinfo {title} {Symmetric informationally complete quantum
  measurements},\ }\href {https://doi.org/10.1063/1.1737053} {\bibfield
  {journal} {\bibinfo  {journal} {Journal of Mathematical Physics}\ }\textbf
  {\bibinfo {volume} {45}},\ \bibinfo {pages} {2171} (\bibinfo {year}
  {2004})}\BibitemShut {NoStop}%
\bibitem [{\citenamefont {Klappenecker}\ and\ \citenamefont
  {R{\"o}tteler}(2005)}]{KlappeneckerRoetteler2005}%
  \BibitemOpen
  \bibfield  {author} {\bibinfo {author} {\bibfnamefont {A.}~\bibnamefont
  {Klappenecker}}\ and\ \bibinfo {author} {\bibfnamefont {M.}~\bibnamefont
  {R{\"o}tteler}},\ }\bibfield  {title} {\bibinfo {title} {Mutually unbiased
  bases are complex projective 2-designs},\ }in\ \href
  {https://ieeexplore.ieee.org/document/1523643} {\emph {\bibinfo {booktitle}
  {Proceedings of the 2005 IEEE International Symposium on Information
  Theory}}}\ (\bibinfo  {publisher} {IEEE},\ \bibinfo {year} {2005})\ pp.\
  \bibinfo {pages} {1740--1744}\BibitemShut {NoStop}%
\bibitem [{\citenamefont {Ambainis}\ and\ \citenamefont
  {Emerson}(2007)}]{AmbainisEmerson2007}%
  \BibitemOpen
  \bibfield  {author} {\bibinfo {author} {\bibfnamefont {A.}~\bibnamefont
  {Ambainis}}\ and\ \bibinfo {author} {\bibfnamefont {J.}~\bibnamefont
  {Emerson}},\ }\bibfield  {title} {\bibinfo {title} {Quantum {$t$}-designs:
  {$t$}-wise independence in the quantum world},\ }in\ \href
  {https://doi.org/10.1109/CCC.2007.26} {\emph {\bibinfo {booktitle} {2007 22nd
  Annual IEEE Conference on Computational Complexity}}}\ (\bibinfo  {publisher}
  {IEEE},\ \bibinfo {year} {2007})\ pp.\ \bibinfo {pages}
  {129--140}\BibitemShut {NoStop}%
\bibitem [{\citenamefont {Dankert}\ \emph {et~al.}(2009)\citenamefont
  {Dankert}, \citenamefont {Cleve}, \citenamefont {Emerson},\ and\
  \citenamefont {Livine}}]{Dankert2009Jul}%
  \BibitemOpen
  \bibfield  {author} {\bibinfo {author} {\bibfnamefont {C.}~\bibnamefont
  {Dankert}}, \bibinfo {author} {\bibfnamefont {R.}~\bibnamefont {Cleve}},
  \bibinfo {author} {\bibfnamefont {J.}~\bibnamefont {Emerson}},\ and\ \bibinfo
  {author} {\bibfnamefont {E.}~\bibnamefont {Livine}},\ }\bibfield  {title}
  {\bibinfo {title} {{Exact and approximate unitary 2-designs and their
  application to fidelity estimation}},\ }\href
  {https://doi.org/10.1103/PhysRevA.80.012304} {\bibfield  {journal} {\bibinfo
  {journal} {Phys. Rev. A}\ }\textbf {\bibinfo {volume} {80}},\ \bibinfo
  {pages} {012304} (\bibinfo {year} {2009})}\BibitemShut {NoStop}%
\bibitem [{\citenamefont {Smith}\ and\ \citenamefont
  {Leung}(2006)}]{Smith2006Dec}%
  \BibitemOpen
  \bibfield  {author} {\bibinfo {author} {\bibfnamefont {G.}~\bibnamefont
  {Smith}}\ and\ \bibinfo {author} {\bibfnamefont {D.}~\bibnamefont {Leung}},\
  }\bibfield  {title} {\bibinfo {title} {{Typical entanglement of stabilizer
  states}},\ }\href {https://doi.org/10.1103/PhysRevA.74.062314} {\bibfield
  {journal} {\bibinfo  {journal} {Phys. Rev. A}\ }\textbf {\bibinfo {volume}
  {74}},\ \bibinfo {pages} {062314} (\bibinfo {year} {2006})}\BibitemShut
  {NoStop}%
\bibitem [{\citenamefont {Koenig}\ and\ \citenamefont
  {Smolin}(2014)}]{Koenig2014Dec}%
  \BibitemOpen
  \bibfield  {author} {\bibinfo {author} {\bibfnamefont {R.}~\bibnamefont
  {Koenig}}\ and\ \bibinfo {author} {\bibfnamefont {J.~A.}\ \bibnamefont
  {Smolin}},\ }\bibfield  {title} {\bibinfo {title} {{How to efficiently select
  an arbitrary Clifford group element}},\ }\href
  {https://doi.org/10.1063/1.4903507} {\bibfield  {journal} {\bibinfo
  {journal} {J. Math. Phys.}\ }\textbf {\bibinfo {volume} {55}},\ \bibinfo
  {pages} {122202} (\bibinfo {year} {2014})}\BibitemShut {NoStop}%
\bibitem [{\citenamefont {Struchalin}\ \emph {et~al.}(2021)\citenamefont
  {Struchalin}, \citenamefont {Zagorovskii}, \citenamefont {Kovlakov},
  \citenamefont {Straupe},\ and\ \citenamefont {Kulik}}]{Struchalin2021Jan}%
  \BibitemOpen
  \bibfield  {author} {\bibinfo {author} {\bibfnamefont {G.~I.}\ \bibnamefont
  {Struchalin}}, \bibinfo {author} {\bibfnamefont {{\relax Ya}.~A.}\
  \bibnamefont {Zagorovskii}}, \bibinfo {author} {\bibfnamefont {E.~V.}\
  \bibnamefont {Kovlakov}}, \bibinfo {author} {\bibfnamefont {S.~S.}\
  \bibnamefont {Straupe}},\ and\ \bibinfo {author} {\bibfnamefont {S.~P.}\
  \bibnamefont {Kulik}},\ }\bibfield  {title} {\bibinfo {title} {{Experimental
  Estimation of Quantum State Properties from Classical Shadows}},\ }\href
  {https://doi.org/10.1103/PRXQuantum.2.010307} {\bibfield  {journal} {\bibinfo
   {journal} {PRX Quantum}\ }\textbf {\bibinfo {volume} {2}},\ \bibinfo {pages}
  {010307} (\bibinfo {year} {2021})}\BibitemShut {NoStop}%
\bibitem [{\citenamefont
  {Mag{\ifmmode\acute{a}\else\'{a}\fi}n}(2016)}]{Magan2016Jan}%
  \BibitemOpen
  \bibfield  {author} {\bibinfo {author} {\bibfnamefont {J.~M.}\ \bibnamefont
  {Mag{\ifmmode\acute{a}\else\'{a}\fi}n}},\ }\bibfield  {title} {\bibinfo
  {title} {{Random Free Fermions: An Analytical Example of Eigenstate
  Thermalization}},\ }\href {https://doi.org/10.1103/PhysRevLett.116.030401}
  {\bibfield  {journal} {\bibinfo  {journal} {Phys. Rev. Lett.}\ }\textbf
  {\bibinfo {volume} {116}},\ \bibinfo {pages} {030401} (\bibinfo {year}
  {2016})}\BibitemShut {NoStop}%
\bibitem [{\citenamefont {Boutin}\ and\ \citenamefont
  {Bauer}(2021)}]{BoutinPRR2021}%
  \BibitemOpen
  \bibfield  {author} {\bibinfo {author} {\bibfnamefont {S.}~\bibnamefont
  {Boutin}}\ and\ \bibinfo {author} {\bibfnamefont {B.}~\bibnamefont {Bauer}},\
  }\bibfield  {title} {\bibinfo {title} {{Quantum impurity models using
  superpositions of fermionic Gaussian states: Practical methods and
  applications}},\ }\href {https://doi.org/10.1103/PhysRevResearch.3.033188}
  {\bibfield  {journal} {\bibinfo  {journal} {Phys. Rev. Res.}\ }\textbf
  {\bibinfo {volume} {3}},\ \bibinfo {pages} {033188} (\bibinfo {year}
  {2021})}\BibitemShut {NoStop}%
\bibitem [{\citenamefont {Bhattacharjee}\ \emph {et~al.}(2021)\citenamefont
  {Bhattacharjee}, \citenamefont {Nandy},\ and\ \citenamefont
  {Pathak}}]{Bhattacharjee2021Dec}%
  \BibitemOpen
  \bibfield  {author} {\bibinfo {author} {\bibfnamefont {B.}~\bibnamefont
  {Bhattacharjee}}, \bibinfo {author} {\bibfnamefont {P.}~\bibnamefont
  {Nandy}},\ and\ \bibinfo {author} {\bibfnamefont {T.}~\bibnamefont
  {Pathak}},\ }\bibfield  {title} {\bibinfo {title} {{Eigenstate capacity and
  Page curve in fermionic Gaussian states}},\ }\href
  {https://doi.org/10.1103/PhysRevB.104.214306} {\bibfield  {journal} {\bibinfo
   {journal} {Phys. Rev. B}\ }\textbf {\bibinfo {volume} {104}},\ \bibinfo
  {pages} {214306} (\bibinfo {year} {2021})}\BibitemShut {NoStop}%
\bibitem [{\citenamefont {Bianchi}\ \emph {et~al.}(2021)\citenamefont
  {Bianchi}, \citenamefont {Hackl},\ and\ \citenamefont {Kieburg}}]{bhk}%
  \BibitemOpen
  \bibfield  {author} {\bibinfo {author} {\bibfnamefont {E.}~\bibnamefont
  {Bianchi}}, \bibinfo {author} {\bibfnamefont {L.}~\bibnamefont {Hackl}},\
  and\ \bibinfo {author} {\bibfnamefont {M.}~\bibnamefont {Kieburg}},\
  }\bibfield  {title} {\bibinfo {title} {{Page curve for fermionic Gaussian
  states}},\ }\href {https://doi.org/10.1103/PhysRevB.103.L241118} {\bibfield
  {journal} {\bibinfo  {journal} {Phys. Rev. B}\ }\textbf {\bibinfo {volume}
  {103}},\ \bibinfo {pages} {L241118} (\bibinfo {year} {2021})}\BibitemShut
  {NoStop}%
\bibitem [{\citenamefont {Yu}\ \emph {et~al.}(2023)\citenamefont {Yu},
  \citenamefont {Gong},\ and\ \citenamefont {Cirac}}]{Yu2023Jan}%
  \BibitemOpen
  \bibfield  {author} {\bibinfo {author} {\bibfnamefont {X.-H.}\ \bibnamefont
  {Yu}}, \bibinfo {author} {\bibfnamefont {Z.}~\bibnamefont {Gong}},\ and\
  \bibinfo {author} {\bibfnamefont {J.~I.}\ \bibnamefont {Cirac}},\ }\bibfield
  {title} {\bibinfo {title} {{Free-fermion Page curve: Canonical typicality and
  dynamical emergence}},\ }\href
  {https://doi.org/10.1103/PhysRevResearch.5.013044} {\bibfield  {journal}
  {\bibinfo  {journal} {Phys. Rev. Res.}\ }\textbf {\bibinfo {volume} {5}},\
  \bibinfo {pages} {013044} (\bibinfo {year} {2023})}\BibitemShut {NoStop}%
\bibitem [{\citenamefont {{\ifmmode\acute{S}\else\'{S}\fi}wi{\k{e}}tek}\ \emph
  {et~al.}(2026)\citenamefont {{\ifmmode\acute{S}\else\'{S}\fi}wi{\k{e}}tek},
  \citenamefont {Kliczkowski}, \citenamefont {Vidmar},\ and\ \citenamefont
  {Rigol}}]{swietek2026onebody}%
  \BibitemOpen
  \bibfield  {author} {\bibinfo {author} {\bibfnamefont {R.}~\bibnamefont
  {{\ifmmode\acute{S}\else\'{S}\fi}wi{\k{e}}tek}}, \bibinfo {author}
  {\bibfnamefont {M.}~\bibnamefont {Kliczkowski}}, \bibinfo {author}
  {\bibfnamefont {L.}~\bibnamefont {Vidmar}},\ and\ \bibinfo {author}
  {\bibfnamefont {M.}~\bibnamefont {Rigol}},\ }\bibfield  {title} {\bibinfo
  {title} {{One-Body Purity, Non-Gaussianity, and Entanglement in Interacting
  Integrable Models}},\ }\bibfield  {journal} {\bibinfo  {journal} {arXiv}\
  }\href {https://doi.org/10.48550/arXiv.2607.01326}
  {10.48550/arXiv.2607.01326} (\bibinfo {year} {2026}),\ \Eprint
  {https://arxiv.org/abs/2607.01326} {2607.01326} \BibitemShut {NoStop}%
\bibitem [{\citenamefont {Turkeshi}\ \emph {et~al.}(2026)\citenamefont
  {Turkeshi}, \citenamefont {Sierant},\ and\ \citenamefont
  {Tarabunga}}]{Turkeshi2026Sep}%
  \BibitemOpen
  \bibfield  {author} {\bibinfo {author} {\bibfnamefont {X.}~\bibnamefont
  {Turkeshi}}, \bibinfo {author} {\bibfnamefont {P.}~\bibnamefont {Sierant}},\
  and\ \bibinfo {author} {\bibfnamefont {P.~S.}\ \bibnamefont {Tarabunga}},\
  }\bibfield  {title} {\bibinfo {title} {{Pauli spectrum and nonstabilizerness
  of random fermionic Gaussian states}},\ }\bibfield  {journal} {\bibinfo
  {journal} {arXiv}\ }\href {https://doi.org/10.48550/arXiv.2609.15739}
  {10.48550/arXiv.2609.15739} (\bibinfo {year} {2026}),\ \Eprint
  {https://arxiv.org/abs/2609.15739} {2609.15739} \BibitemShut {NoStop}%
\bibitem [{Sup()}]{Supplement}%
  \BibitemOpen
  \href@noop {} {\ }\bibinfo {note} {J.T. Heath, A. Urilyon, D. Quinn, and G.
  Kells. "Supplemental Material: Typical Superpositions are
  Magical"}\BibitemShut {NoStop}%
\bibitem [{\citenamefont {Nadimpalli}\ \emph {et~al.}(2024)\citenamefont
  {Nadimpalli}, \citenamefont {Parham}, \citenamefont {Vasconcelos},\ and\
  \citenamefont {Yuen}}]{Nadimpalli2023Nov}%
  \BibitemOpen
  \bibfield  {author} {\bibinfo {author} {\bibfnamefont {S.}~\bibnamefont
  {Nadimpalli}}, \bibinfo {author} {\bibfnamefont {N.}~\bibnamefont {Parham}},
  \bibinfo {author} {\bibfnamefont {F.}~\bibnamefont {Vasconcelos}},\ and\
  \bibinfo {author} {\bibfnamefont {H.}~\bibnamefont {Yuen}},\ }\bibfield
  {title} {\bibinfo {title} {{On the Pauli Spectrum of QAC0}},\ }in\ \href
  {https://doi.org/10.1145/3618260.3649662} {\emph {\bibinfo {booktitle} {{ACM
  Conferences}}}}\ (\bibinfo  {publisher} {Association for Computing
  Machinery},\ \bibinfo {address} {New York, NY, USA},\ \bibinfo {year}
  {2024})\ pp.\ \bibinfo {pages} {1498--1506}\BibitemShut {NoStop}%
\bibitem [{\citenamefont {Passarelli}\ \emph {et~al.}(2024)\citenamefont
  {Passarelli}, \citenamefont {Fazio},\ and\ \citenamefont
  {Lucignano}}]{passarelli2024piv}%
  \BibitemOpen
  \bibfield  {author} {\bibinfo {author} {\bibfnamefont {G.}~\bibnamefont
  {Passarelli}}, \bibinfo {author} {\bibfnamefont {R.}~\bibnamefont {Fazio}},\
  and\ \bibinfo {author} {\bibfnamefont {P.}~\bibnamefont {Lucignano}},\
  }\bibfield  {title} {\bibinfo {title} {{Nonstabilizerness of permutationally
  invariant systems}},\ }\href {https://doi.org/10.1103/PhysRevA.110.022436}
  {\bibfield  {journal} {\bibinfo  {journal} {Phys. Rev. A}\ }\textbf {\bibinfo
  {volume} {110}},\ \bibinfo {pages} {022436} (\bibinfo {year}
  {2024})}\BibitemShut {NoStop}%
\bibitem [{\citenamefont {Zhu}(2017)}]{Zhu2017}%
  \BibitemOpen
  \bibfield  {author} {\bibinfo {author} {\bibfnamefont {H.}~\bibnamefont
  {Zhu}},\ }\bibfield  {title} {\bibinfo {title} {{Multiqubit Clifford groups
  are unitary 3-designs}},\ }\href {https://doi.org/10.1103/PhysRevA.96.062336}
  {\bibfield  {journal} {\bibinfo  {journal} {Phys. Rev. A}\ }\textbf {\bibinfo
  {volume} {96}},\ \bibinfo {pages} {062336} (\bibinfo {year}
  {2017})}\BibitemShut {NoStop}%
\bibitem [{\citenamefont {Zhu}\ \emph {et~al.}(2016)\citenamefont {Zhu},
  \citenamefont {Kueng}, \citenamefont {Grassl},\ and\ \citenamefont
  {Gross}}]{Zhu2016Sep}%
  \BibitemOpen
  \bibfield  {author} {\bibinfo {author} {\bibfnamefont {H.}~\bibnamefont
  {Zhu}}, \bibinfo {author} {\bibfnamefont {R.}~\bibnamefont {Kueng}}, \bibinfo
  {author} {\bibfnamefont {M.}~\bibnamefont {Grassl}},\ and\ \bibinfo {author}
  {\bibfnamefont {D.}~\bibnamefont {Gross}},\ }\bibfield  {title} {\bibinfo
  {title} {{The Clifford group fails gracefully to be a unitary 4-design}},\
  }\bibfield  {journal} {\bibinfo  {journal} {arXiv}\ }\href
  {https://doi.org/10.48550/arXiv.1609.08172} {10.48550/arXiv.1609.08172}
  (\bibinfo {year} {2016}),\ \Eprint {https://arxiv.org/abs/1609.08172}
  {1609.08172} \BibitemShut {NoStop}%
\bibitem [{\citenamefont {Gross}\ \emph {et~al.}(2021)\citenamefont {Gross},
  \citenamefont {Nezami},\ and\ \citenamefont {Walter}}]{Gross2021Aug}%
  \BibitemOpen
  \bibfield  {author} {\bibinfo {author} {\bibfnamefont {D.}~\bibnamefont
  {Gross}}, \bibinfo {author} {\bibfnamefont {S.}~\bibnamefont {Nezami}},\ and\
  \bibinfo {author} {\bibfnamefont {M.}~\bibnamefont {Walter}},\ }\bibfield
  {title} {\bibinfo {title} {{Schur{\textendash}Weyl Duality for the Clifford
  Group with Applications: Property Testing, a Robust Hudson Theorem, and de
  Finetti Representations}},\ }\href
  {https://doi.org/10.1007/s00220-021-04118-7} {\bibfield  {journal} {\bibinfo
  {journal} {Commun. Math. Phys.}\ }\textbf {\bibinfo {volume} {385}},\
  \bibinfo {pages} {1325} (\bibinfo {year} {2021})}\BibitemShut {NoStop}%
\bibitem [{\citenamefont {Cudby}\ and\ \citenamefont
  {Strelchuk}(2023)}]{Cudby2024}%
  \BibitemOpen
  \bibfield  {author} {\bibinfo {author} {\bibfnamefont {J.}~\bibnamefont
  {Cudby}}\ and\ \bibinfo {author} {\bibfnamefont {S.}~\bibnamefont
  {Strelchuk}},\ }\bibfield  {title} {\bibinfo {title} {{Gaussian decomposition
  of magic states for matchgate computations}},\ }\bibfield  {journal}
  {\bibinfo  {journal} {arXiv}\ }\href
  {https://doi.org/10.48550/arXiv.2307.12654} {10.48550/arXiv.2307.12654}
  (\bibinfo {year} {2023}),\ \Eprint {https://arxiv.org/abs/2307.12654}
  {2307.12654} \BibitemShut {NoStop}%
\bibitem [{\citenamefont {Hahn}\ \emph {et~al.}(2025)\citenamefont {Hahn},
  \citenamefont {Takagi}, \citenamefont {Ferrini},\ and\ \citenamefont
  {Yamasaki}}]{Hahn2025Oct}%
  \BibitemOpen
  \bibfield  {author} {\bibinfo {author} {\bibfnamefont {O.}~\bibnamefont
  {Hahn}}, \bibinfo {author} {\bibfnamefont {R.}~\bibnamefont {Takagi}},
  \bibinfo {author} {\bibfnamefont {G.}~\bibnamefont {Ferrini}},\ and\ \bibinfo
  {author} {\bibfnamefont {H.}~\bibnamefont {Yamasaki}},\ }\bibfield  {title}
  {\bibinfo {title} {{Classical simulation and quantum resource theory of
  non-Gaussian optics}},\ }\href {https://doi.org/10.22331/q-2025-10-13-1881}
  {\bibfield  {journal} {\bibinfo  {journal} {Quantum}\ }\textbf {\bibinfo
  {volume} {9}},\ \bibinfo {pages} {1881} (\bibinfo {year} {2025})},\ \Eprint
  {https://arxiv.org/abs/2404.07115v5} {2404.07115v5} \BibitemShut {NoStop}%
\bibitem [{\citenamefont {Wei}\ \emph {et~al.}(2026)\citenamefont {Wei},
  \citenamefont {Wu}, \citenamefont {Liu},\ and\ \citenamefont
  {Liu}}]{Wei2026Oct}%
  \BibitemOpen
  \bibfield  {author} {\bibinfo {author} {\bibfnamefont {F.}~\bibnamefont
  {Wei}}, \bibinfo {author} {\bibfnamefont {K.-W.}\ \bibnamefont {Wu}},
  \bibinfo {author} {\bibfnamefont {Z.}~\bibnamefont {Liu}},\ and\ \bibinfo
  {author} {\bibfnamefont {Z.-W.}\ \bibnamefont {Liu}},\ }\bibfield  {title}
  {\bibinfo {title} {{Robust exponential lower bounds for fermionic and bosonic
  Gaussian ranks}},\ }\bibfield  {journal} {\bibinfo  {journal} {arXiv}\ }\href
  {https://doi.org/10.48550/arXiv.2610.02172} {10.48550/arXiv.2610.02172}
  (\bibinfo {year} {2026}),\ \Eprint {https://arxiv.org/abs/2610.02172}
  {2610.02172} \BibitemShut {NoStop}%
\bibitem [{\citenamefont {Bravyi}\ and\ \citenamefont
  {Gosset}(2017)}]{Bravyi2017Dec}%
  \BibitemOpen
  \bibfield  {author} {\bibinfo {author} {\bibfnamefont {S.}~\bibnamefont
  {Bravyi}}\ and\ \bibinfo {author} {\bibfnamefont {D.}~\bibnamefont
  {Gosset}},\ }\bibfield  {title} {\bibinfo {title} {{Complexity of Quantum
  Impurity Problems}},\ }\href {https://doi.org/10.1007/s00220-017-2976-9}
  {\bibfield  {journal} {\bibinfo  {journal} {Commun. Math. Phys.}\ }\textbf
  {\bibinfo {volume} {356}},\ \bibinfo {pages} {451} (\bibinfo {year}
  {2017})}\BibitemShut {NoStop}%
\bibitem [{\citenamefont {Liu}(2021)}]{Liu2021Dec}%
  \BibitemOpen
  \bibfield  {author} {\bibinfo {author} {\bibfnamefont {Y.}~\bibnamefont
  {Liu}},\ }\bibfield  {title} {\bibinfo {title} {{Moments of Random Quantum
  Circuits and Applications in Random Circuit Sampling}},\ }\href
  {https://uwspace.uwaterloo.ca/items/32123824-cfea-4f52-a6d6-01f1a2e33bf3}
  {\bibfield  {journal} {\bibinfo  {journal} {University of Waterloo}\ }
  (\bibinfo {year} {2021})}\BibitemShut {NoStop}%
\bibitem [{\citenamefont {Paviglianiti}\ \emph {et~al.}(2026)\citenamefont
  {Paviglianiti}, \citenamefont {Lumia}, \citenamefont {Tirrito}, \citenamefont
  {Silva}, \citenamefont {Collura}, \citenamefont {Turkeshi},\ and\
  \citenamefont {Lami}}]{Paviglianiti2025Doped}%
  \BibitemOpen
  \bibfield  {author} {\bibinfo {author} {\bibfnamefont {A.}~\bibnamefont
  {Paviglianiti}}, \bibinfo {author} {\bibfnamefont {L.}~\bibnamefont {Lumia}},
  \bibinfo {author} {\bibfnamefont {E.}~\bibnamefont {Tirrito}}, \bibinfo
  {author} {\bibfnamefont {A.}~\bibnamefont {Silva}}, \bibinfo {author}
  {\bibfnamefont {M.}~\bibnamefont {Collura}}, \bibinfo {author} {\bibfnamefont
  {X.}~\bibnamefont {Turkeshi}},\ and\ \bibinfo {author} {\bibfnamefont
  {G.}~\bibnamefont {Lami}},\ }\bibfield  {title} {\bibinfo {title} {{Emergence
  of Generic Entanglement Structure in Doped Matchgate Circuits}},\ }\href
  {https://doi.org/10.1103/w97w-7zny} {\bibfield  {journal} {\bibinfo
  {journal} {Phys. Rev. Lett.}\ }\textbf {\bibinfo {volume} {136}},\ \bibinfo
  {pages} {020403} (\bibinfo {year} {2026})}\BibitemShut {NoStop}%
\bibitem [{\citenamefont {Selberg}(1944)}]{selberg1944bemerkninger}%
  \BibitemOpen
  \bibfield  {author} {\bibinfo {author} {\bibfnamefont {A.}~\bibnamefont
  {Selberg}},\ }\bibfield  {title} {\bibinfo {title} {Bemerkninger om et
  multipelt integral},\ }\href
  {https://mathscinet.ams.org/mathscinet/article?mr=18287} {\bibfield
  {journal} {\bibinfo  {journal} {Norsk Matematisk Tidsskrift}\ }\textbf
  {\bibinfo {volume} {26}},\ \bibinfo {pages} {71} (\bibinfo {year}
  {1944})}\BibitemShut {NoStop}%
\bibitem [{\citenamefont {Forrester}(2010)}]{Forrester2010LogGases}%
  \BibitemOpen
  \bibfield  {author} {\bibinfo {author} {\bibfnamefont {P.~J.}\ \bibnamefont
  {Forrester}},\ }\href {https://doi.org/10.1515/9781400835416} {\emph
  {\bibinfo {title} {Log-Gases and Random Matrices}}},\ \bibinfo {series}
  {London Mathematical Society Monographs Series}, Vol.~\bibinfo {volume} {34}\
  (\bibinfo  {publisher} {Princeton University Press},\ \bibinfo {address}
  {Princeton, NJ},\ \bibinfo {year} {2010})\BibitemShut {NoStop}%
\bibitem [{\citenamefont {De~Bruyne}\ \emph {et~al.}(2024)\citenamefont
  {De~Bruyne}, \citenamefont {Le~Doussal}, \citenamefont {Majumdar},\ and\
  \citenamefont {Schehr}}]{debruyne2023linearstatisticscoulombgases}%
  \BibitemOpen
  \bibfield  {author} {\bibinfo {author} {\bibfnamefont {B.}~\bibnamefont
  {De~Bruyne}}, \bibinfo {author} {\bibfnamefont {P.}~\bibnamefont
  {Le~Doussal}}, \bibinfo {author} {\bibfnamefont {S.~N.}\ \bibnamefont
  {Majumdar}},\ and\ \bibinfo {author} {\bibfnamefont {G.}~\bibnamefont
  {Schehr}},\ }\bibfield  {title} {\bibinfo {title} {{Linear statistics for
  Coulomb gases: higher order cumulants}},\ }\href
  {https://doi.org/10.1088/1751-8121/ad329f} {\bibfield  {journal} {\bibinfo
  {journal} {J. Phys. A: Math. Theor.}\ }\textbf {\bibinfo {volume} {57}},\
  \bibinfo {pages} {155002} (\bibinfo {year} {2024})}\BibitemShut {NoStop}%
\bibitem [{\citenamefont {Tiutiakina}\ \emph {et~al.}(2024)\citenamefont
  {Tiutiakina}, \citenamefont {De~Luca},\ and\ \citenamefont
  {De~Nardis}}]{tiutiakina2023framepotentialbrowniansyk}%
  \BibitemOpen
  \bibfield  {author} {\bibinfo {author} {\bibfnamefont {A.}~\bibnamefont
  {Tiutiakina}}, \bibinfo {author} {\bibfnamefont {A.}~\bibnamefont
  {De~Luca}},\ and\ \bibinfo {author} {\bibfnamefont {J.}~\bibnamefont
  {De~Nardis}},\ }\bibfield  {title} {\bibinfo {title} {{Frame potential of
  Brownian SYK model of Majorana and Dirac fermions}},\ }\href
  {https://doi.org/10.1007/JHEP01(2024)115} {\bibfield  {journal} {\bibinfo
  {journal} {J. High Energy Phys.}\ }\textbf {\bibinfo {volume} {2024}}\bibinfo
   {number} { (1)},\ \bibinfo {pages} {115}}\BibitemShut {NoStop}%
\end{thebibliography}%

\end{document}

\setcounter{secnumdepth}{3}

\title{Supplemental Material: \\ Typical Superpositions are Magical}

\author{Joshuah T. Heath}
\email{joshuah.heath@mu.ie}
\affiliation{Department of Physics, Science Building, Maynooth University, Maynooth, Co. Kildare W23 F2H6, Ireland}
\affiliation{Hamilton Institute, Eolas Building, Maynooth University, Maynooth, Co. Kildare W23 A3HY, Ireland}

\author{Andrew Urilyon}
\email{andrew.urilyon@mu.ie}
\affiliation{Department of Physics, Science Building, Maynooth University, Maynooth, Co. Kildare W23 F2H6, Ireland}

\author{Damien Quinn}
\email{damien.quinn.2026@mumail.ie}
\affiliation{Department of Physics, Science Building, Maynooth University, Maynooth, Co. Kildare W23 F2H6, Ireland}

\author{Graham Kells}
\email{Graham.Kells@mu.ie}
\affiliation{Department of Physics, Science Building, Maynooth University, Maynooth, Co. Kildare W23 F2H6, Ireland}
\affiliation{Hamilton Institute, Eolas Building, Maynooth University, Maynooth, Co. Kildare W23 A3HY, Ireland}
\affiliation{Dublin Institute for Advanced Studies, School of Theoretical Physics, 10 Burlington Road, Dublin, Co. Dublin D04 C932, Ireland}

\maketitle

\tableofcontents

\begin{table}[t]
\caption{\small Symbols used in the main text.}
\label{tab:glossary}
\small
\begin{tabular}{@{}p{0.18\columnwidth}p{0.75\columnwidth}@{}}
\toprule
$\mathbb{P}_N$ & $\bullet$ Pauli group on $N$ qubits ($4^N$ strings); the same with the fixed strings removed \\
$\widetilde{\mathbb{P}}_N$ & $\bullet$ Pauli group on $N$ qubits ($4^N$ strings) with the fixed strings removed \\
$n_{\textrm{fix}}$ & $\bullet$ Number of fixed strings; i.e., Pauli strings where $\langle\phi|P|\phi\rangle^2=1$ for every state of the ensemble \\
$D$ & $\bullet$ Dimension of the symmetry sector \\
$\overline{\rho}$ & $\bullet$ Average density matrix $\overline{\rho}=I_D/D$ (via Schur's lemma) \\
$\langle W_\Psi(m)\rangle $ & $\bullet$ Average non-stabilizer witness of the $m$-component superposition $|\Psi\rangle$, $M_2=N-\log_2 \langle W(m)\rangle$ \\
$\langle W_\Psi(1)\rangle$,\,\, a & $\bullet$ Average witness of a single component in the superposition\\
$p_i$ & $\bullet$ Weights $|\lambda_i|^2$ \\
$\|p\|_l^k$ & $\bullet$ The $k$-th power of their $l$-norm, e.g. $\|p\|_2^4=m^{-2}$, $\|p\|_4^4=m^{-3}$ at equal weights\\
$\sigma_{\textrm{diag}}(P)$, $\sigma_{\textrm{off}}(P)$ & $\bullet$ Diagonal and off-diagonal parts of $\langle\Psi|P|\Psi\rangle$\\
$\mu_k(P)$ & $\bullet$ Symmetric moments $\mathbb{E}[\langle\phi|P|\phi\rangle^k]$ of a single component \\
$\kappa_4(P)$ & $\bullet$ Fourth cumulant $\mu_4(P)-3\mu_2^2(P)$ \\
$v$ & $\bullet$ Directed second moment $\mathbb{E}|\langle\phi_i|P|\phi_j\rangle|^2=1/D$, which is flat across the Pauli spectrum \\
$\beta$ & $\bullet$ The one diagram not reducible to $\mu_2$, $\mu_4$ and/or $v$ \\
$b$ & $\bullet$ The term in the general $m$-fold superposition proportional to $\|p\|_2^4$, found to be $6\left\{\sum_{P\in\mathbb{P}_N}\mu_2^2(P)-2(n_{\textrm{fix}}-\beta)\right\}$
\\
$W_{\textrm{Haar}}$,\,c,\,$4n_{\textrm{fix}}$ & $\bullet$ The Haar floor of the non-stabilizer witness \\
Diagonal excess & $\bullet$ The single-component witness above the Haar floor for random stabilizer states, given by $\langle W_\Psi(1)\rangle-4n_{\textrm{fix}}$ \\
\bottomrule
\end{tabular}
\end{table}

\section{Stabilizer Rényi entropy (SRE) for a generic superposition of typical states}

\subsection{Diagrammatic expansion in terms of the diagonal and off-diagonal Pauli spectrum}

We consider a state $|\Psi\rangle$ which is an $m$-fold superposition of states $|\phi_i\rangle$:
\begin{align}
    |\Psi\rangle=\sum_j \lambda_j|\phi_j\rangle
\end{align}
The states $|\phi_j\rangle$ are typical Haar random states drawn from some distribution. For the purposes of this section, we will consider some general typical states and not concern ourselves with case-specific examples. Later on, we will consider case-specific examples (specifically, random stabilizer states and random Gaussian states) in order to write down closed form predictions for the $m$-fold SRE. We will consider general superpositions where $p_i=|\lambda_i|^2$, and then specialize to the case of equal superpositions where $\lambda_j=e^{i\theta_j}/\sqrt{m}$. For general notation used in this Supplement and the main test, see Table~\ref{tab:glossary}.

Our primary interest in this work is the quantum simulation complexity inherent to an $m$-fold superposition of states which, by themselves, carry little to no complexity. We quantify the complexity via the stablizer R\'enyi entropy (SRE) for the $m$-state superposition is given by
\begin{align}
    \langle M_2(m)\rangle =N-\log_2(\langle W_\Psi(m)\rangle )
\end{align}
where we define the average non-stabilizer witness as
\begin{align}
    \langle W_\Psi(m)\rangle =\mathbb{E}\bigg[\sum_{P\in\mathbb{P}_N}\langle \Psi|P|\Psi\rangle^4\bigg]=n_{\textrm{fix}}+\mathbb{E}\bigg[\sum_{P\in\widetilde{\mathbb{P}}_N}\langle \Psi|P|\Psi\rangle^4\bigg]
\end{align}
In the above, we have separated those Pauli strings which are "fixed" from the main sum; i.e., those $P$ such that $\langle \Psi|P|\Psi\rangle=\pm 1$ regardless of the value of the state $|\Psi\rangle$. The remaining Pauli strings $\widetilde{\mathbb{P}}_N$ are then part of the "filtered Pauli group" which does not include the fixed strings.

\begin{figure}[t]
    \centering
        \includegraphics[width=0.5\columnwidth]{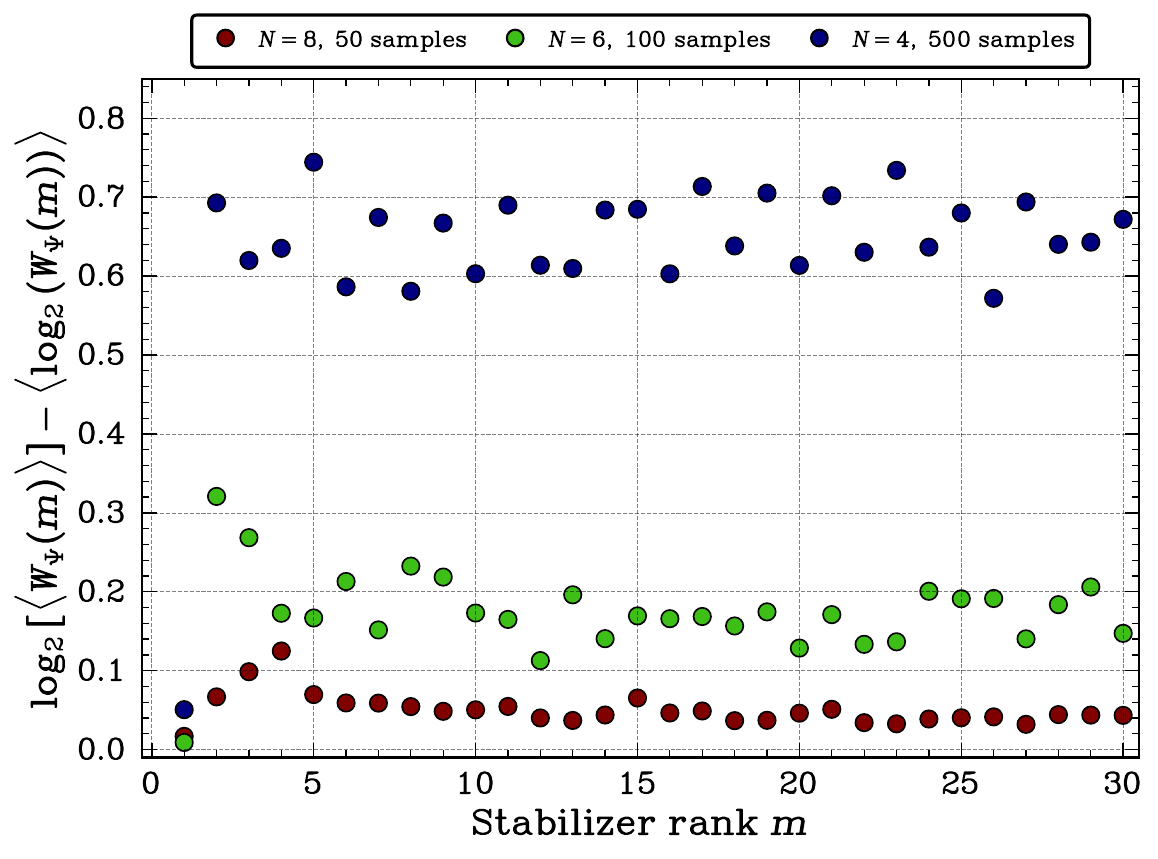}
    \caption{\small The Jensen gap for the non-stabilizer witness. For $N=4$, we see that the gap fluctuates even when taking $500$ samples, and is comparable to the maximum value of the SRE $N-2=2$. As $N$ is increased, we need less samples to reach convergence of the data, and the average gap over large values of $m$ also becomes negligible compared to the Haar ceiling $N-2$.}
    \label{fig:FigS1}
\end{figure}

Taking the fourth power of the average $\langle \Psi|P|\Psi\rangle^4$, we first proceed to split up terms which are diagonal or off-diagonal in the superposition index:

\begin{align}
    \sum_{P\in \widetilde{\mathbb{P}}_N}\langle \Psi|P|\Psi\rangle^4&=\sum_{P\in\widetilde{\mathbb{P}}_N}\left\{\left(\sum_i\langle \phi_i|\lambda_i^*\right)P\left(\sum_j\lambda_j|\phi_j\rangle\right)\right\}^4\notag\\
    &=\sum_{P\in\widetilde{\mathbb{P}}_N}\left(\sum_{i,\,j}\lambda_i\lambda_j^*\langle \phi_i|P|\phi_j\rangle\right)^4\notag\\
    &=\sum_{P\in\widetilde{\mathbb{P}}_N}\bigg\{\sum_{i}|\lambda_i|^2\langle \phi_i|P|\phi_i\rangle+\sum_{i\not=j}\lambda_i\lambda_j^*\langle \phi_i|P|\phi_j\rangle\bigg\}^4  \notag\\
    &\equiv\sum_{P\in\widetilde{\mathbb{P}}_N}\bigg\{\sum_{i}p_i\langle \phi_i|P|\phi_i\rangle+\sum_{i\not=j}p_{ij}\langle \phi_i|P|\phi_j\rangle\bigg\}^4,\qquad p_i\equiv |\lambda_i|^2,\quad p_{ij}\equiv \lambda_{i}\lambda_j^* \notag\\
    &=\sum_{P\in\widetilde{\mathbb{P}}_N}\left(\sigma_{\textrm{diag}}(P)+\sigma_{\textrm{off}}(P)\right)^4,\qquad\qquad \sigma_{\textrm{diag}}(P)\equiv \sum_i p_i \langle \phi_i|P|\phi_i\rangle,\quad \sigma_{\textrm{off}}\equiv \sum_{i\not=j}p_{ij}\langle \phi_i|P|\phi_j\rangle \notag \\
    &=\sum_{P\in\widetilde{\mathbb{P}}_N}\bigg[\sigma_{\textrm{diag}}^4(P)+4\sigma_{\textrm{diag}}^3(P)\sigma_{\textrm{off}}(P)+6\sigma_{\textrm{diag}}^2(P)\sigma_{\textrm{off}}^2(P)+4\sigma_{\textrm{diag}}\sigma_{\textrm{off}}^3(P)+\sigma_{\textrm{off}}^4(P)\bigg]\label{P4}
\end{align}
In the above, we have expressed the final expression in terms of $\sigma_{\textrm{diag}}(P)$ and $\sigma_{\textrm{off}}(P)$, which we take to represent the diagonal and off-diagonal components of the Pauli spectrum of the $m$-fold superposition, respectively. As we are considering large numbers of random states, we are concerned with the calculation of $\langle W(m)\rangle$, and as a consequence we take the expectation value of $\mathbb{E}[...]$ of each of the above. Note that there is a subtlety in that we are assuming that the average of the logarithm of the witness is equivalent to the logarithm of the average witness, or in other words that the Jensen gap $\log_2[\langle W_\Psi(m)\rangle]- \langle \log_2[ W_\Psi(m)]\rangle$. In Fig.~\ref{fig:FigS1}, we plot the Jensen gap for $N=4$, $6$, and $8$, and we see that while there is a finite Jensen gap it vanishes as $N$ grows.

Returning to Eqn.~\eqref{P4}, we see that that certain terms disappear by virtue of averaging. This can be seen by writing the full 

\begin{align}
    \mathbb{E}[\langle \Psi|P|\Psi\rangle^4]=\sum_{i_1,\,j_1}\sum_{i_2,\,j_2}\sum_{i_3,\,j_3}\sum_{i_4,\,j_4}\langle \prod_{s=1}^4 \lambda_{i_s}^*\lambda_{i_s}\rangle\prod_{\ell=1}^4X_{i_\ell j_\ell},\qquad X_{ij}\equiv \langle \phi_i|P|\phi_j\rangle
\end{align}
\noindent Note that in the above notation, terms in the non-stabilizer witness will now be of the form $X_{i}^2X_{ij}X_{ji}$, $X_{i}^4$, $X_{ij}X_{ji}X_{\ell i}X_{i\ell}$, etc, where we take $X_{ii}\equiv X_i$. In order to restrict the sums, the phases must be averaged over, such that for each index $j$ there must be the same number of occurrences of $\lambda_j$ must be equal to the number of occurrences of $\lambda_j^*$. If this rule is broken, then that term is trivially zero; i.e., terms with a dangling phase $e^{i(\theta_i-\theta_j)}$ such as $X_{i}^3X_{ij}$. The same is true for terms with a single dangling component $X_i$, such as $X_i X_{\ell}^3$, as these are independent random variables which are zero mean over the entire ensemble. From this rule, we immediately see that every term in $\sigma_{\textrm{diag}}^3(P)\sigma_{\textrm{off}}(P)$ is zero upon taking the expected value. 

Simplification of each non-zero diagram may follow by adopting a diagrammatic formalism. Each diagram initially consists of four nodes which represent the maximum four distinct indices inherent to each surviving term in the non-stabilizer witness. Each node has two protruding lines, which may characterize a symmetric (straight line) or directed (arrow) connection with the adjacent nodes. As such, a single node with two symmetric connections would represent a term $X_{ii}\equiv \langle \phi_i|P|\phi_i\rangle$, while a single node with one symmetric connection and one directed connection would represent $X_{ij}=\langle \phi_i|P|\phi_j\rangle$ with $i\not=j$. The only non-zero diagrams are those which are full four-cycles (connected) or two disconnected two-cycles (disconnected). Any symmetric connection that begins and ends at the same node would represent a floating average $X_{ii}$, and thus goes to zero upon averaging. Likewise, any node that has a single directed connection represents a floating phase, and thus will also average to zero. 

As nodes connected by a symmetric connection tell us that these indices are the same value, we may contract them diagrammatically so that the final irreducible diagram only contains as many nodes as there are distinct indices. This contraction results in closed symmetric self-cycles connected to the larger diagram. Two closed symmetric self-cycles on the same node are therefore a visual representation of the symmetric second moment, defined as 
\begin{align}
    \mu_2(P)\equiv \mathbb{E}\bigg[X_{i}^2 \bigg]= \mathbb{E}_i\bigg[\langle \phi_i|P|\phi_i\rangle^2\bigg]
\end{align}
It is worth noting that there might be two nodes which are equivalent but which cannot be contracted diagrammatically do to the presence of two directed cycles. In this instance, we will represent this connected by a dashed line. Nodes connected by a directed connection (arrow) will be represented by $v$, defined as

\begin{align}
    v\equiv  \mathbb{E}\bigg[X_{ij}X_{ji}\bigg]\equiv \mathbb{E}_i\mathbb{E}_j\bigg[\langle \phi_i|P|\phi_j\rangle \langle \phi_j|P|\phi_i\rangle\bigg]= \mathbb{E}_j\bigg[|\phi_j\rangle\langle \phi_j|\bigg]
\end{align}
Note that we have used in the above that $v$ is "flat" in the Pauli spectrum; i.e., that it is independent of $P$. We will show this explicitely in the next section. Contraction of the directed connections are not shown as a new diagram, however any two nodes connected by a closed directed loop are easily seen to reduce to a single $v$ term. Such contraction rules for symmetric and directed connections result in four fundamental diagrammatic representations, which will be used throughout the following derivations:

\begin{figure}[ht]
    \centering
    \hspace{10mm}\includegraphics[width=0.22\textwidth]{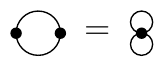}
% \end{figure}
% \begin{figure}[ht]
%     \centering
    \hspace{37mm}\includegraphics[width=0.23\textwidth]{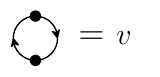}
\end{figure}
\begin{figure}[ht]
    \centering
    \hspace{7mm}\includegraphics[width=0.30\textwidth]{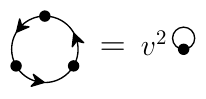}
        \hspace{26mm}\includegraphics[width=0.30\textwidth]{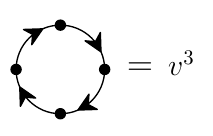}
\end{figure}\phantom{.}\\\\
\noindent From the above contractions, we can reduce any diagram as some single node times either $\mu_2(P)$ or some multiple of $v$.

Finally, note that we will use the shorthand $\gamma$ to represent a sum of some combination of the symmetric and directed second moments over the filtered Pauli strings. In total, five of these sums will appear in the following diagrams:

\begin{align}
    \gamma_{\mu\mu}\equiv \sum_{P\in \widetilde{\mathbb{P}}_N}\mu_2^2(P),
    \qquad \gamma_{vv}\equiv \sum_{P\in \widetilde{\mathbb{P}}} v^2,\qquad \gamma_{vvv}\equiv \sum_{P\in \widetilde{\mathbb{P}}} v^3,\qquad \gamma_{\mu v}\equiv \sum_{P\in \widetilde{\mathbb{P}}} v\mu_2(P),\qquad \gamma_{\mu vv}\equiv \sum_{P\in \widetilde{\mathbb{P}}} v^2\mu_2(P)
\end{align}

There are in total ten diagrams which do not vanish upon taking the expected value. Each diagram starts as either a fully connected four cycle or two disconnected two cycles, and may be put into an irreducible form via contraction of the symmetric connections. To facilitate ease of identification, for the purposes of this Supplement we will introduce a notation for each of the diagrams. Each diagram will be labelled via a partition notation with at most as a 4-tuple, with the number in the tuple denoting how many free indices there are. A superscript of a Roman numeral I, II, or III will denote of the diagram is fully connected, disconnected, or connected with one or more nodes which are equal but cannot be contracted further, respectively. As such, a diagram with four symmetric cycles will all have the same index, and thus be represented as $(4)^I$. A fully connected diagram with alternating symmetric and directed cycles corresponds to $(2,\,2)^I$, while a disconnected two-fold symmetric cycle would be $(2,\,2)^{II}$, and so on.

In addition to the diagrammatical notation itself, we must address two other technicalities before we can start calculating: namely, the multiplicity of each diagram, and the $m$ dependence. In the initial expansion of $(\sigma_{\textrm{diag}}(P)+\sigma_{\textrm{off}}(P))^4$, some repetition is included in the numerical factor $4$, $6$, etc., however we can be more careful by invoking the partition notation. Second, in regards to the $m$ dependence, when assuming equal weight superpositions $p_i\sim {1}/{m}$, we will multiply the "unrestricted" sum $\sum_{i,\,j,\,k,\,\ell}p_i^4=(1/m^4)m=1/m^3$ by a multiplicative factor depending upon the number of 
free indices. As such, two free parameters will correspond to a multiplicative factor $(m-1)$, three free parameters $(m-2)(m-1)$, and so on. The final expression is then expected to solely have dependence on constant terms plus terms that go as $1/m$, $1/m^2$, and $1/m^3$. For the case of general weight superpositions, we will use the notation $\|p\|_\ell^k=(\sum_ip_i^\ell)^{k/\ell}$.

All ten diagrams are calculated below. Where applicable, we discuss the subtleties in the $m$-dependence, contraction, or multiplicity. 

\begin{enumerate}
\item[1)] \underline{\bf Diagrams with no directed cycles.}

The fully diagonal Pauli spectrum is composed of two non-zero diagrams: a fully symmetric four-cycle and two disconnected symmetric two-cycles:  
\begin{align}
\mathbb{E}[\sigma_{\textrm{diag}}^4(P)]&=\sum_{P\in \widetilde{\mathbb{P}}_N}\mathbb{E}\bigg[\sum_{i,\,j} p_i^2p_j^2 X_i^2X_j^2\bigg]\notag\\
&= \sum_{P\in \widetilde{\mathbb{P}}_N}\mathbb{E}\bigg[\sum_ip_i^4 X_i^4\bigg]+3\sum_{P\in \widetilde{\mathbb{P}}_N} \mathbb{E}\bigg[\sum_{i\not=j} p_i^2 p_j^2 X_i^2 X_j^2\bigg]  \notag\\
&\equiv (4)^{{\textrm I}}+(2,\,2)^{\textrm{II}}
\end{align}

\begin{enumerate}

\item[1a)] \underline{Connected four-fold symmetric cycle.}\\
The diagram $(4)^{\textrm{I}}$ is special, as it is the only diagram that can be reduced to a single node with four self-cycles: 
\begin{figure}[ht]
    \centering
    \hspace{20mm}\includegraphics[width=0.5\textwidth]{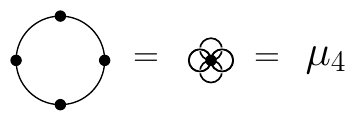}
\end{figure}\\
The calculation of this diagram can be reduced to the fourth moment $\mu_4(P)$, which upon summing over the Pauli strings becomes the average witness. However, note that we are summing over the filtered Pauli strings $\widetilde{{\mathbb{P}}}_N$, and if we define $\langle W_\Psi(1)\rangle$ as the average of the $m=1$ state witness over all Pauli strings then $\sum_{P\in\widetilde{\mathbb{P}}_N}\mu_4(P)=\langle W_\Psi(1)\rangle-n_{\textrm{fix}}$:
\begin{align}
(4)^I&=\sum_{P\in \widetilde{\mathbb{P}}_N}\mathbb{E}\bigg[\sum_i p_i^4 X_i^4\bigg]=\sum_{P\in \widetilde{\mathbb{P}}_N}\mathbb{E}\bigg[X_i^4\bigg]\sum_i p_i^4\equiv\sum_{P\in \widetilde{\mathbb{P}}_N}\mu_4(P)\|p\|_4^4\notag\\
&=\sum_{P\in \widetilde{\mathbb{P}}_N}\mu_4(P)\dfrac{1}{m^4}m=\dfrac{1}{m^3}\sum_{P\in \widetilde{\mathbb{P}}_N}\mu_4(P)\notag\\
&=\dfrac{\langle W_\Psi(1)\rangle -n_{\textrm{fix}}}{m^3}
\end{align}
The sum over the $p_i$ is pretty straight forward for this diagram, as there is only one free index.

\item[1b)] {\underline{Disconnected two-fold symmetric cycle.}}\\
For the case of the fully symmetric disconnected diagram, we are able to reduce each of the separate sub-diagrams to a single node with two symmetric self-cycles. Thus, from our previous discussion, we expect that the final result should go as $\mu_2^2(P)$:
\begin{figure}[h]
    \centering
    \hspace{20mm}\includegraphics[width=0.5\textwidth]{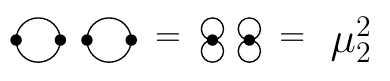}
\end{figure}\\
Note that this diagram has a non-trivial multiplicity. In total, assuming no restrictions, we have $4!=24$ possible terms with the same diagrammatic representation. For the case of $(2,\,2)^{II}$, $24$ double counts the same diagram twice (due to the square of the two Pauli string averages), and we additionally have a swap symmetry of the indices $i\leftrightarrow j$, and thus the multiplicity of this diagram is $4!/(2!2!2!)=3$. As there are two free indices, the summation over the weights $p_i^2 p_j^2$ becomes $m(m-1)$ times $1/m^4$:

\begin{align}
(2,\,2)^{II}&=3\sum_{P\in \widetilde{\mathbb{P}}_N}\mathbb{E}\bigg[\sum_{i,\,j}p_i^2 p_j^2 X_i^2 X_j^2\bigg]=3\sum_{P\in \widetilde{\mathbb{P}}_N}\sum_{i,\,j}p_i^2 p_j^2 \mathbb{E}\bigg[X_i^2 X_j^2\bigg]\notag\\
&=3\sum_{P\in \widetilde{\mathbb{P}}_N}\mu_2^2(P)\sum_{i,\,j}p_i^2 p_j^2\equiv 3\gamma_{\mu\mu}(\|p\|_2^4-\|p\|_4^4)\equiv 3\gamma_{\mu\mu}m(m-1)\dfrac{1}{m^4}\notag\\
&=3\gamma_{\mu\mu}\bigg(\dfrac{1}{m^2}-\dfrac{1}{m^3}\bigg)
\end{align}

Combining the above with $(4)^{\textrm{I}}$, we have the full contribution to the non-stabilizer witness from the purely diagonal Pauli spectrum:
\begin{align}
\sigma_{\textrm{diag}}^4(P)
=(4)^I+(2,\,2)^{II}
&=\bigg(\langle W_\Psi(1)\rangle-n_{\textrm{fix}}\bigg)\|p\|_4^4+3\gamma_{\mu\mu}\bigg(\|p\|_2^4-\|p\|_4^4\bigg)
\notag\\
&=\dfrac{\langle W_\Psi(1)\rangle-n_{\textrm{fix}}}{m^3}+3\gamma_{\mu\mu}\bigg(\dfrac{1}{m^2}-\dfrac{1}{m^3}\bigg)\notag\\
&=\boxed{\dfrac{3\gamma_{\mu \mu }}{m^2}+\dfrac{\langle W_\Psi(1)\rangle-n_{\textrm{fix}}-3\gamma_{\mu\mu}}{m^3}}
\end{align}
\end{enumerate}

\item[2)] {\underline {\bf Diagrams with two directed cycles.}}\\
As mentioned previously, all diagrams with one directed cycle average to zero, and thus we continue onto those with two directed cycles. This results in three separate diagrams:
\begin{align}
    \sigma_{\textrm{diag}}^2(P)\sigma_{\textrm{off}}^2(P)&=\sum_{P\in \widetilde{\mathbb{P}}_N}\mathbb{E}\bigg[\sum_{i\not=j,\,\ell,\,k}p_\ell p_k p_{i}p_j X_\ell X_k X_{ij}^2\bigg]\notag\\
    &=12\sum_{P\in \widetilde{\mathbb{P}}_N}\mathbb{E}\bigg[\sum_{i\not=j}p_i^2 p_j^2 X_iX_{ij}^2X_j\bigg]+12\sum_{P\in \widetilde{\mathbb{P}}_N}\mathbb{E}\bigg[\sum_{i\not=j}p_i^3 p_{j} X_i^2X_{ij}^2\bigg]+6\sum_{P\in \widetilde{\mathbb{P}}_N}\mathbb{E}\bigg[\sum_{i\not=j\not=\ell}p_\ell^2 p_{i} p_j X_\ell^2X_{ij}^2\bigg]\notag\\
    &\equiv (2,\,2)^{\textrm{I}}+(3,\,1)^{\textrm{I}}+(2,\,1,\,1)^{\textrm{II}}
\end{align}

\begin{enumerate}
\item[(2a)] {\underline{Connected two-fold directed cycle with double-pinned two-fold symmetric cycle.}}\\
This diagram, $(2,\,2)^{\textrm{I}}$, is one of two diagrams which cannot be simplified towards some product of the symmetric and directed second moments. This can be seen below, where in its irreducible form the contraction of the directed cycle to $v$ is inhibited by the presence of a pinned symmetric self-cycle on each node:\\
\begin{figure}[h]
    \centering
    \hspace{20mm}\includegraphics[width=0.52\textwidth]{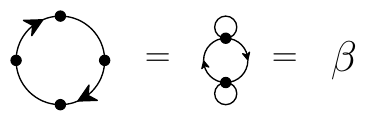}
\end{figure}\\
As a consequence of it's unique structure, we will have to postpone explicit evaluation until we consider the class-dependent calculations in the next section. As such, for now we'll denote this diagram simply as "$\beta$". The multiplicity of this diagram, we only have the index swap symmetry $i\leftrightarrow j$, and thus the multiplicity is $4!/2=12$. The $m$ dependence is equivalent to that of $(2,\,2)^{I}$, and thus the full sum is multiplied by $m(m-1)$ due to the two free indices:
\begin{align}
    (2,\,2)^{\textrm{I}}&=12\sum_{P\in \widetilde{\mathbb{P}}_N}\mathbb{E}\bigg[\sum_{i\not=j}p_i^2p_j^2 X_iX_{ij}^2X_j\bigg]\notag\\
    &=12\sum_{i\not=j}p_i^2 p_j^2\sum_{P\in \widetilde{\mathbb{P}}_N}\mathbb{E}\bigg[X_iX_{ij}X_j\bigg]\equiv 12\beta \sum_{i\not=j}p_i^2p_j^2\equiv 12\beta \bigg(\|p\|_2^4-\|p\|_4^4\bigg)=12\beta m(m-1)\dfrac{1}{m^4}\notag\\
    &=12\beta\bigg(\dfrac{1}{m^2}-\dfrac{1}{m^3}\bigg)
\end{align}

\item[(2b)] {\underline{Connected two-fold directed cycle with single-pinned two-fold symmetric cycle}}\\
Unlike the case of $(2,\,2)^{\textrm{I}}$, the diagram $(3,\,1)^{\textrm{I}}$ is not case dependent, as both the symmetric self-cycles are on the same node:\\
\begin{figure}[h]
    \centering
\hspace{20mm}\includegraphics[width=0.75\textwidth]{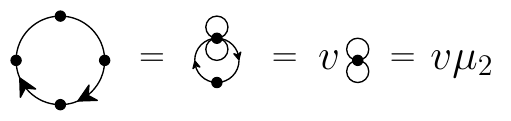}
\end{figure}\\
As we see below, this allows us to contract the directed cycles along the remaining free node. The multiplicity is the same as that of $\beta$ (that is, $4!/2=12$), however this is by virtue of the repeated term $X_i^2$ as opposed to the $i\leftrightarrow j$ symmetry. The $m$ dependence is identical to that of $\beta$ due to the two free indices:
\begin{align}
(3,\,1)^{\textrm{I}}&=12\sum_{P\in \widetilde{\mathbb{P}}_N}\mathbb{E}\bigg[\sum_{i\not=j}p_i^3 p_j X_i^2X_{ij}^2\bigg]\notag\\
&=12\sum_{i\not=j} p_i^3p_j \sum_{P\in\widetilde{\mathbb{P}}_N}\mathbb{E}\bigg[X_i^2 X_{ij}^2\bigg]\equiv12\sum_{P\in \widetilde{\mathbb{P}}_N}\mu_2(P)v\sum_{i\not=j}p_i^3p_j\equiv 12 \gamma_{\mu v}\sum_{i\not=j}p_i^3p_j\notag\\
&\equiv 12\gamma_{\mu v}\bigg(\|p\|_3^3-\|p\|_4^4\bigg)=12\gamma_{\mu v}\bigg(\dfrac{1}{m^2}-\dfrac{1}{m^3}\bigg)
\end{align}
As mentioned above, the single free node "un-pinned" by a symmetric cycle allows us to represent $\mathbb{E}[X_i^2 X_{ij}^2]$ in terms of $\mu_2(P)$ and $v$, as shown below:
\begin{align}
    \mathbb{E}[X_i^2X_{ij}^2]&=\mathbb{E}_i\mathbb{E}_j\bigg[\langle \phi_i|P|\phi_i\rangle^2\langle \phi_i|P|\phi_j\rangle\langle \phi_j |P|\phi_i\rangle\bigg]=v\mathbb{E}_i\bigg[\langle \phi_i|P|\phi_i\rangle^2\bigg]=v\mu_2(P)
\end{align}

\item[(2c)] \underline{Disconnected two-fold directed cycle and two-fold symmetric cycle.}\\
The disconnected diagram $(2,\,1,\,1)^{\textrm{II}}$ can be reduced to the product of a directed two-cycle and a symmetric self-cycle, suggesting that the final result will be proportional to $\mu_2(P)v$: \\
\begin{figure}[h]
    \centering
    \hspace{20mm}\includegraphics[width=0.6\textwidth]{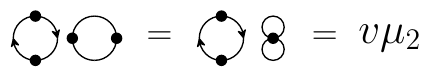}
\end{figure}\\
This diagram has an $i\leftrightarrow j$ symmetry and a double counting via $X_\ell^2$, and thus the multiplicity is $4!/(2!2)=6$. The sum over the weights $p_i$ should now be weighted by $m(m-1)(m-2)$ due to the three free indices:
\begin{align}
(2,\,1,\,1)^{\textrm{II}}&=6\sum_{P\in\widetilde{\mathbb{P}}_N}\mathbb{E}\bigg[\sum_{i\not=j\not=\ell}p_\ell^2p_{i} p_j X_\ell^2 X_{ij}^2\bigg]=6\sum_{i\not=j\not=\ell}p_\ell^2 p_{i} p_j\sum_{P\in\widetilde{\mathbb{P}}_N}\mathbb{E}\bigg[X_\ell^2X_{ij}^2\bigg]\notag\\
&=6\sum_{i\not=j\not=\ell}p_\ell^2p_ip_j\sum_{P\in\widetilde{\mathbb{P}}_N}\mu_2(P)v\equiv 6\gamma_{\mu v}\bigg(\|p\|_2^2-\|p\|_2^4-2\|p\|_3^3+2\|p\|_4^4\bigg)\notag\\
&=6\gamma_{\mu\nu}m(m-1)(m-2)\dfrac{1}{m^4}=6\gamma_{\mu \nu}\bigg(\dfrac{1}{m}-\dfrac{3}{m^2}+\dfrac{2}{m^3}\bigg)
\end{align}
\end{enumerate}
The sum of the three diagrams above now yields terms proportional to $1/m$, $1/m^2$, and $1/m^3$:
\begin{align}
\sigma_{\textrm{diag}}^2(P)\sigma_{\textrm{off}}^2(P)&=(2,\,2)^{\textrm{I}}+(3,\,1)^{\textrm{I}}+(2,\,1,\,1)^{\textrm{II}}\notag\\
&=12\beta\bigg(\|p\|_2^4-\|p\|_4^4\bigg)+12\gamma_{\mu v}\bigg(\|p\|_3^3-\|p\|_4^4\bigg)+6\gamma_{\mu v}\bigg(\|p\|_2^2-\|p\|_2^4-2\|p\|_3^3+2\|p\|_4^4\bigg)\notag\\
&=12\beta\bigg(\dfrac{1}{m^2}-\dfrac{1}{m^3}\bigg)+12\gamma_{\mu v}\bigg(\dfrac{1}{m^2}-\dfrac{1}{m^3}\bigg)+6\gamma_{\mu v}\bigg(\dfrac{1}{m}-\dfrac{3}{m^2}+\dfrac{2}{m^3}\bigg)\notag\\
&=\boxed{\dfrac{6\gamma_{\mu v}}{m}+\dfrac{12\beta-6\gamma_{\mu v}}{m^2}-\dfrac{12\beta}{m^3}}
\end{align}

\item[3)] {\underline{\bf Diagrams with three directed cycles}}
The majority of diagrams with three directed cycles average to zero, with only one surviving: 
\begin{align}
    \sigma_{\textrm{diag}}(P)\sigma_{\textrm{off}}^3(P)&=\sum_{P\in\widetilde{\mathbb{P}}_N}\mathbb{E}\bigg[\sum_{i\not= j\not=\ell,\,k}p_{i}p_{j}p_{\ell}p_k X_kX_{ij}X_{j\ell}X_{\ell i}\bigg]\notag\\
    &=24\sum_{P\in\widetilde{\mathbb{P}}_N}\mathbb{E}\bigg[\sum_{i\not=j\not=\ell}p_{i}^2p_{j}p_{\ell}X_iX_{ij}X_{j\ell}X_{\ell i}\bigg]\notag\\
    &\equiv  (2,\,1,\,1)^{\textrm{I}}
\end{align}

\begin{enumerate}
\item[(3a)] \underline{Connected three-fold directed cycle with single-pinned singular symmetric cycle} \\
This diagram reduces to a closed triple directed cycle with a single symmetric self-cycle attached to one node:\\
\begin{figure}[h]
    \centering
\hspace{50mm}\,\,\,\phantom{ddddddddd}\includegraphics[width=0.9\textwidth]{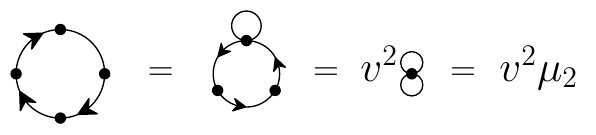}
\end{figure}\\
There is no redundency or symmetry of indices we can invoke, and thus the multiplicity of this diagram is $24$. Due to the three directed cycles, contraction results in $v^2$, with a single $\mu_2(P)$ due to the remaining symmetric cycle. As with $(2,\,1,\,1)^{\textrm{II}}$, there are three free indices, which leads to a multiplicity of $m(m-1)(m-2)$ for the sum over $p_i$:
\begin{align}
    \sigma_{\textrm{diag}}(P)\sigma_{\textrm{off}}^3(P)&=24\sum_{P\in\widetilde{\mathbb{P}}_N}\mathbb{E}\bigg[\sum_{i\not=j\not=\ell}p_{i}^2p_{j}p_{\ell}X_iX_{ij}X_{j\ell}X_{\ell i}\bigg]\notag\\
    &=24\mathbb{E}\bigg[\sum_{P\in\widetilde{\mathbb{P}}_N}X_iX_{ij}X_{j\ell}X_{\ell i}\bigg]\sum_{i\not=j\not=\ell}p_{i}^2p_{j}p_{\ell}\notag\\
    &\equiv 24\gamma_{\mu vv}\bigg(\|p\|_2^2-\|p\|_2^4-2\|p\|_3^3+2\|p\|_4^4\bigg)\notag\\
    &=24\gamma_{\mu vv}m(m-1)(m-2)\dfrac{1}{m^4}\notag\\
    &=24\gamma_{\mu vv}\bigg(\dfrac{1}{m}-\dfrac{3}{m^2}+\dfrac{2}{m^3}\bigg)
\end{align}
For completeness, the reduction to $\mu(P)v^2$ is shown below:
\begin{align}
    \mathbb{E}\bigg[X_iX_{ij}X_{j\ell}X_{\ell i}\bigg]&\equiv \mathbb{E}_i\mathbb{E}_j\mathbb{E}_\ell\bigg[\langle \phi_i|P|\phi_i\rangle\langle \phi_i|P|\phi_j\rangle \langle \phi_j|P|\phi_\ell\rangle\langle \phi_\ell|P|\phi_i\rangle\bigg]\notag\\
    &=v^2\mathbb{E}_i\bigg[\langle\phi_i|P|\phi_i\rangle\langle \phi_i|P|\phi_i\rangle\bigg]=v^2\mu_2(P)
\end{align}
The final expression for all terms with three directed cycles is therefore from one diagram:
\begin{align}
    \sigma_{\textrm{diag}}(P)\sigma_{\textrm{off}}^3(P)&=(2,\,1,\,1)^{\textrm{I}}\notag\\
    &=24\gamma_{\mu v v}\bigg(\|p\|_2^2-\|p\|_2^4-2\|p\|_3^3+2\|p\|_4^4\bigg)\notag\\
    &=\boxed{\dfrac{24\gamma_{\mu vv}}{m}-\dfrac{72\gamma_{\mu vv}}{m^2}+\dfrac{48\gamma_{\mu vv}}{m^3}}
\end{align}
\end{enumerate}

\item[4)] {\bf \underline{Diagrams with four directed cycles.}}\\
The our directed cycles are the final expressions we need to consider, and amount to four separate diagrams:
\begin{align}
    \sigma_{\textrm{off}}^4(P)&=\sum_{P\in\widetilde{\mathbb{P}}_N}\mathbb{E}\bigg[p_i p_j p_k p_\ell\sum_{i,\,j,\,k,\,\ell}X_{ij}X_{jk}X_{k\ell}X_{\ell i}\bigg]\notag\\
    &\notag\\
    &=6\sum_{P\in\widetilde{\mathbb{P}}_N}\mathbb{E}\bigg[\sum_{i,\not=j\not=k\not=\ell}p_i p_j p_k p_\ell X_{ij}X_{jk}X_{k\ell}X_{\ell i}\bigg]+3\sum_{P\in\widetilde{\mathbb{P}}_N}\mathbb{E}\bigg[\sum_{i\not=j\not=k\not=\ell}p_{i}p_jp_{k} p_\ell X_{ij}X_{ji}X_{\ell k}X_{k\ell}\bigg]\notag\\
    &\phantom{}+12\sum_{P\in\widetilde{\mathbb{P}}_N}\mathbb{E}\bigg[\sum_{i\not=j\not=k}p_{i}^2p_j  p_{k} X_{ij}X_{ji}X_{ik}X_{ki}\bigg]+3\sum_{P\in\widetilde{\mathbb{P}}_N}\mathbb{E}\bigg[\sum_{i\not=j}p_{i}^2 p_j^2 X_{ij}^2X_{ji}^2\bigg]
    &\notag\\
    &\notag\\
    &\equiv (1,\,1,\,1,\,1)^{\textrm{I}}+(1,\,1,\,1,\,1)^{\textrm{II}}+(2,\,1,\,1)^{\textrm{III}}+(2,\,2)^{\textrm{III}}
\end{align}

\begin{enumerate}
    \item[(4a)] \underline{Connected four-fold directed cycle.}\\
    The diagram $(1,\,1,\,1,\,1)^{\textrm{I}}$, as with all other diagrams with four directed cycles, cannot be further reduced diagrammatically from its starting form:
    \begin{figure}[h]
    \centering
    \hspace{20mm}\includegraphics[width=0.33\textwidth]{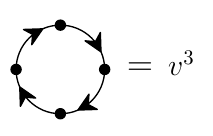}
\end{figure}\\
This diagram has a four-fold cyclic rotation symmetry, and thus it's multiplicity is $4!/4=6$. The contraction of the arrows results in the final expression being proportional to $v^3$, and it's four free indices results in a multiplicity of $m(m-1)(m-2)(m-3)$ multiplying the sum over $p_i$:
    \begin{align}
        (1,\,1,\,1,\,1)^{\textrm{I}}&=6\sum_{P\in\widetilde{\mathbb{P}}_N}\mathbb{E}\bigg[\sum_{i,\not=j\not=k\not=\ell}p_{i}p_{j}p_{k}p_{\ell}X_{ij}X_{jk}X_{k\ell}X_{\ell i}\bigg]\notag\\
        &=6\sum_{P\in\widetilde{\mathbb{P}}_N}\mathbb{E}\bigg[X_{ij}X_{jk}X_{k\ell}X_{\ell i}\bigg]\sum_{i\not=j\not=k\not=\ell}p_{i}p_{j}p_{k}p_{\ell}\notag\\
        &\equiv 6\gamma_{vvv}\bigg(1-6\|p\|_2^2+3\|p\|_2^4+8\|p\|_3^3-6\|p\|_4^4\bigg)\notag\\
        &=6\gamma_{vvv}\bigg(1-\dfrac{6}{m^3}+\dfrac{11}{m^2}-\dfrac{6}{m}\bigg)
    \end{align}
%\newpage 
\item[(4b)] \underline{Disconnected two-fold directed cycle.}\\
The diagram $(1,\,1,\,1,\,1)^{\textrm{II}}$, along with $(4)^{\textrm{I}}$, are the most diagrams in series, as will be explained further in the next section. The form of the diagram suggests a final form that will go as $v^2$:\newpage 
    \begin{figure}[h]
    \centering
    \hspace{20mm}\includegraphics[width=0.37\textwidth]{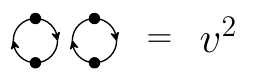}
\end{figure}
The indices between the cycles and the indices $i\leftrightarrow j$ and $\ell\leftrightarrow k$ can be exchanged, and thus the multiplicity is $4!/(2!2!2)=3$. The $m$-dependence is equivalent to $(1,\,1,\,1,\,1)^{\textrm{II}}$:
\begin{align}
    (1,\,1,\,1,\,1)^{\textrm{II}}&=3\sum_{P\in \widetilde{\mathbb{P}}_N}\mathbb{E}\bigg[\sum_{i\not=j\not=k\not=\ell}p_{i} p_j p_{k} p_\ell X_{ij}X_{ji}X_{\ell k}X_{k\ell}\bigg]\notag\\
    &=3\sum_{P\in\widetilde{\mathbb{P}}_N}\mathbb{E}\bigg[X_{ij}X_{ji}X_{\ell k}X_{k\ell }\bigg]\sum_{i\not=j\not =k\not=\ell}p_{i} p_j p_{k} p_\ell \notag\\
    &\equiv 3\gamma_{vv}\bigg(1-6\|p\|_2^2+3\|p\|_2^4+8\|p\|_3^3-6\|p\|_4^4\bigg)\notag\\
    &=3\gamma_{vv}m(m-1)(m-2)(m-3)\dfrac{1}{m^4}\notag\\
    &=3\gamma_{vv}\bigg(1-\dfrac{6}{m^3}+\dfrac{11}{m^2}-\dfrac{6}{m}\bigg)
\end{align}
Note the constant term $3\gamma_{vv}$, which will play a prominent role later on. 
\item[(4c)] \underline{Connected two-fold directed cycle with singular pinned node.}\\
This diagram is almost equivalent to $(1,\,1,\,1,\,1)^{\textrm{II}}$, except that now two of the nodes are "pinned" to be equal, as shown by the dashed black line:\\
    \begin{figure}[h]
    \centering
    \hspace{20mm}\includegraphics[width=0.18\textwidth]{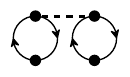}
\end{figure}\\
The diagram just contains the symmetry $j\leftrightarrow k$, and thus the multiplicity is $4!/2=12$. Reduction results in $v^2$, as in $(1,\,1,\,1,\,1)^{\textrm{II}}$, and the $m$-dependence is equivalent to $(2,\,1,\,1)^{\textrm{II}}$ by virtue of the three free indices:
\begin{align}
    (2,\,1,\,1)^{\textrm{III}}&=12\sum_{P\in\widetilde{\mathbb{P}}_N}\mathbb{E}\bigg[\sum_{i\not=j\not=k}p_{i}^2 p_{j} p_k X_{ij}X_{ji}X_{ik}X_{ki}\bigg]\notag\\
    &=12\sum_{P\in\widetilde{\mathbb{P}}_N}\mathbb{E}\bigg[X_{ij}X_{ji}X_{ik}X_{ki}\bigg]\sum_{i\not=j\not=k}p_{i}^2 p_{j} p_k  \notag\\
    &\equiv 12\gamma_{vv}\bigg(\|p\|_2^2-\|p\|_2^4-2\|p\|_3^3+2\|p\|_4^4\bigg)\notag\\
    &=12\gamma_{vv}m(m-1)(m-2)\dfrac{1}{m^4}\notag\\
    &=12\gamma_{vv}\bigg(\dfrac{1}{m}-\dfrac{3}{m^2}+\dfrac{2}{m^3}\bigg)
\end{align}
\item[(4d)] \underline{Connected two-fold directed cycle with double pinned node.}\\
The final diagram, $(2,\,2)^{\textrm{III}}$, is nearly identical in structure of $(2,\,1,\,1)^{\textrm{III}}$, except now the other two nodes are forced to be equal as well:\newpage
    \begin{figure}[h]
    \centering
    \hspace{20mm}\includegraphics[width=0.18\textwidth]{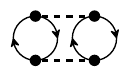}
\end{figure}
As with $\beta$ define above, this diagram cannot be reduced to some product of $v$ and $\mu_2(P)$ due to the pinning of the nodes blocking any further contraction. Explicit derivation is reserved for a later section, and thus we'll denote this term $\alpha$ for the time being. The argument of the multiplicity is equivalent to $(2,\,2)^{\textrm{II}}$, resulting in $3$. The $m$ dependence is also equivalent to $(2,\,2)^{\textrm{II}}$ due to the two free indices:
\begin{align}
    (2,\,2)^{\textrm{III}}&=3\sum_{P\in\widetilde{\mathbb{P}}_N}\mathbb{E}\bigg[\sum_{i\not=j}p_{i}^2 p_j^2 X_{ij}^2 X_{ji}^2\bigg]\notag\\
    &=3\sum_{P\in\widetilde{\mathbb{P}}_N}\mathbb{E}\bigg[X_{ij}^2X_{ji}^2\bigg]\sum_{i\not=j} p_{i}^2 p_j^2 \notag\\
    &\equiv 3 \alpha \bigg(\|p\|_2^4-\|p\|_4^4\bigg) \notag\\
    &=3\alpha m(m-1)\dfrac{1}{m^4}\notag\\
    &=3\alpha \bigg(\dfrac{1}{m^2}-\dfrac{1}{m^3}\bigg)
\end{align}

The summation of all four diagrams together is given below:
\begin{align}
    \sigma_{\textrm{off}}^4(P)&=(1,\,1,\,1,\,1)^{\textrm{I}}+(1,\,1,\,1,\,1)^{\textrm{II}}+(2, \,1,\,1)^{\textrm{III}}+(2, \,2)^{\textrm{III}}\notag\\
    &\notag\\
    &=6\gamma_{vvv}\bigg(1-6\|p\|_2^2+3\|p\|_2^4+8\|p\|_3^3-6\|p\|_4^4\bigg)+3\gamma_{vv}\bigg(1-6\|p\|_2^2+3\|p\|_2^4+8\|p\|_3^3-6\|p\|_4^4\bigg)\notag\\
    &\phantom{=}+12\gamma_{vv}\bigg(\|p\|_2^2-\|p\|_2^4-2\|p\|_3^3+2\|p\|_4^4\bigg)+3\alpha\bigg(\|p\|_2^4-\|p\|_4^4\bigg)
    \notag\\
    &\notag\\
    &=6\gamma_{vvv}\bigg(1-\dfrac{6}{m^3}+\dfrac{11}{m^2}-\dfrac{6}{m}\bigg)+3\gamma_{vv}\bigg(1-\dfrac{6}{m^3}+\dfrac{11}{m^2}-\dfrac{6}{m}\bigg)\notag\\
    &\phantom{=}+12\gamma_{vv}\bigg(\dfrac{1}{m}-\dfrac{3}{m^2}+\dfrac{2}{m^3}\bigg)+3\alpha \bigg(\dfrac{1}{m^2}-\dfrac{1}{m^3}\bigg)
    \notag\\
    &\notag\\
    &=\boxed{3\gamma_{vv}+6\gamma_{vvv}-\dfrac{6\gamma_{vv}+36\gamma_{vvv}}{m}+\dfrac{3\alpha-3\gamma_{vv}+66\gamma_{vvv}}{m^2}+\dfrac{-3\alpha+6\gamma_{vv}-36\gamma_{vvv}}{m^3}}
\end{align}
\end{enumerate}

\end{enumerate}

With all diagrams now derived, the full non-stabilizer witness may be easily evaluated:
\begin{align}
\langle W_\Psi(m)\rangle&=n_{\textrm{fix}}+\sum_{P\in\widetilde{\mathbb{P}}_N}\langle \Psi|P|\Psi\rangle^4\notag\\
&\notag\\
&=n_{\textrm{fix}}+(4)^{\textrm{I}}+(2,\,2)^{\textrm{II}}+(2,\,2)^{\textrm{I}}+(3,\,1)^{\textrm{I}}+(2,\,1,\,1)^{\textrm{II}}+(2,\,1,\,1)^{\textrm{I}}\notag\\
&\phantom{=n_{\textrm{fix}}l}+(1,\,1,\,1,\,1)^{\textrm{I}}+(1,\,1,\,1,\,1)^{\textrm{II}}+(2,\,1,\,1)^{\textrm{III}}+(2,\,2)^{\textrm{III}}\notag\\
&\notag\\
&=n_{\textrm{fix}}+\bigg(\langle W_\Psi(1)\rangle-n_{\textrm{fix}}-3\gamma_{\mu\mu}\bigg)\|p\|_4^4+3\gamma_{\mu\mu}\|p\|_2^4 \notag\\
&+6\gamma_{\mu v}\|p\|_2^2+\bigg(12\beta-6\gamma_{\mu v}\bigg)\|p\|_2^4-12\beta\|p\|_4^4\notag\\
&+24\gamma_{\mu vv}\|p\|_2^2-24\gamma_{\mu vv}\|p\|_2^4-48\gamma_{\mu vv}\|p\|_3^3+48\gamma_{\mu vv}\|p\|_4^4\notag\\
&+3\gamma_{vv}+6\gamma_{vvv}-\|p\|_2^2\bigg(6\gamma_{vv}+36\gamma_{vvv}\bigg)+\|p\|_4^4\bigg(-3\alpha+6\gamma_{vv}-36\gamma_{vvv}\bigg)+48\gamma_{vvv}\|p\|_3^3\notag\\
&+\|p\|_2^4\bigg(3\alpha-3\gamma_{vv}+18\gamma_{vvv}\bigg)\notag\\
&\notag\\
&=n_{\textrm{fix}}+3\gamma_{vv}+6\gamma_{vvv}+\|p\|_2^2\bigg(6\gamma_{\mu v}+24\gamma_{\mu vv}-6\gamma_{vv}-36\gamma_{vvv}\bigg)\notag\\
&\phantom{=n_{\textrm{fix}}+3\gamma_{vv}+6\gamma_{vvv}l}+\|p\|_3^3\bigg(-48\gamma_{\mu vv}+48\gamma_{vvv}\bigg)\notag\\
&\phantom{=n_{\textrm{fix}}+3\gamma_{vv}+6\gamma_{vvv}l}+\|p\|_2^4\bigg(3\gamma_{\mu\mu}+12\beta-6\gamma_{\mu v}-24\gamma_{\mu vv}+3\alpha-3\gamma_{vv}+18\gamma_{vvv}\bigg)\notag\\
&\phantom{=n_{\textrm{fix}}+3\gamma_{vv}+6\gamma_{vvv}ll}+
\|p\|_4^4\bigg(\langle W_\Psi(1)\rangle-n_{\textrm{fix}}-3\gamma_{\mu \mu}-12\beta +48\gamma_{\mu vv}-3\alpha+6\gamma_{vv}-36\gamma_{vvv}\bigg)
\end{align}
Assuming superpositions of equal weight, we then find that the average witness becomes 
\begin{align}
\langle W_\Psi(m)\rangle&=n_{\textrm{fix}}+\dfrac{3\gamma_{\mu \mu }}{m^2}+\dfrac{\langle W_\Psi(1)\rangle-n_{\textrm{fix}}-3\gamma_{\mu\mu}}{m^3}+\dfrac{6\gamma_{\mu v}}{m}+\dfrac{12\beta-6\gamma_{\mu v}}{m^2}-\dfrac{12\beta}{m^3}+\dfrac{24\gamma_{\mu vv}}{m}-\dfrac{72\gamma_{\mu vv}}{m^2}\notag\\
&\phantom{=n_{\textrm{fix}}l}+\dfrac{48\gamma_{\mu vv}}{m^3}+3\gamma_{vv}+6\gamma_{vvv}-\dfrac{6\gamma_{vv}+36\gamma_{vvv}}{m}+\dfrac{3\alpha-3\gamma_{vv}+66\gamma_{vvv}}{m^2}+\dfrac{-3\alpha+6\gamma_{vv}-36\gamma_{vvv}}{m^3}\notag\\
&\notag\\
&=\boxed{
\begin{aligned}
&n_{\textrm{fix}}+3\gamma_{vv}+6\gamma_{vvv}+\dfrac{1}{m}\bigg(6\gamma_{\mu v}+24\gamma_{\mu vv}-6\gamma_{vv}-36\gamma_{vvv}\bigg)\\
&\phantom{n_{\textrm{fix}}+3\gamma_{vv}+6\gamma_{vvv}}+\dfrac{1}{m^2}\bigg(3\gamma_{\mu\mu}+12\beta-6\gamma_{\mu v}-72\gamma_{\mu vv}+3\alpha -3\gamma_{vv}+66\gamma_{vvv}\bigg)\\
&\phantom{n_{\textrm{fix}}+3\gamma_{vv}+6\gamma_{vvv}}+\dfrac{1}{m^3}\bigg(\langle W_\Psi(1)\rangle-n_{\textrm{fix}}-3\gamma_{\mu \mu}-12\beta +48\gamma_{\mu vv}-3\alpha +6\gamma_{vv}-36\gamma_{vvv}\bigg)
\end{aligned}
}\label{Wm}
\end{align}

\begin{figure}[h]
    \includegraphics[width=\columnwidth]{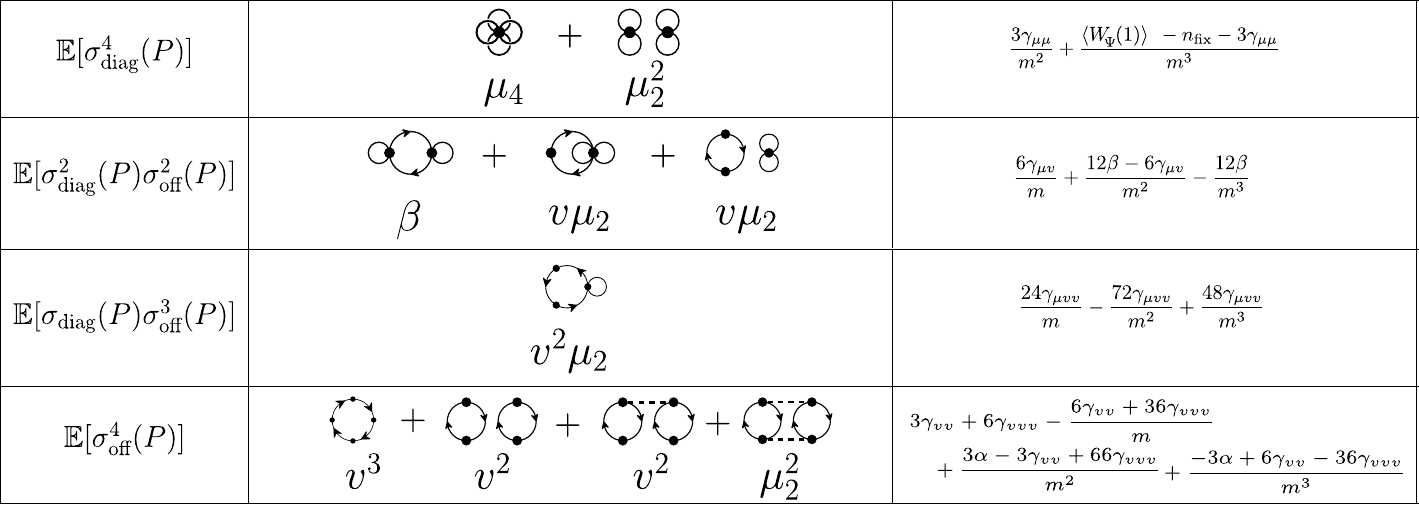}
    \caption{\small All the above calculation are collected into this table for easy reference, identifying each contributing diagram and the explicit contribution at leading $m$ in the rightmost column.}
    \label{fig:sm_calcTable}
\end{figure}

\subsection[Vanishing and non-vanishing diagrams]{Vanishing and non-vanishing diagrams: Schur's lemma, the Parseval sum rule, and ensemble-specific terms}

From Eqn.~\eqref{Wm}, we see that the form of $\langle W_\Psi(m)\rangle$ is dependent on second different combination of the symmetric and directed second moment; i.e., $\gamma_{vv}$, $\gamma_{vvv}$, $\gamma_{\mu v}$, $\gamma_{\mu vv}$, and $\gamma_{\mu \mu}$. As we only concerned about the thermodynamic limit, we may exploit two important mathematical tools to simplify this expression as $N\rightarrow \infty$; namely, the flat spectrum of $v$ and the Parseval sum rule.

The first statement that $v$ is flat in the Pauli spectrum is easily 

\begin{align}
    v&=\mathbb{E}\bigg[X_{ij}X_{ji}\bigg]=\mathbb{E}\bigg[\langle \phi_i|P|\phi_j\rangle\langle \phi_j|P|\phi_i\rangle\bigg]=\mathbb{E}_i\mathbb{E}_j\bigg[\langle \phi_i|P|\phi_j\rangle\langle \phi_j|P|\phi_i\rangle\bigg]=\mathbb{E}_i\bigg[\langle \phi_i|P\mathbb{E}_j\bigg[|\phi_j\rangle\langle \phi_j|\bigg]P|\phi_i\rangle\bigg]\notag\\
    &=\mathbb{E}_i\bigg[\langle \phi_i|P\bar{\rho}_jP|\phi_i\rangle\bigg]=\mathbb{E}_i\bigg[\textrm{Tr}\left(P\bar{\rho}_jP|\phi_i\rangle \langle \phi_i|\right)\bigg]=\textrm{Tr}\left(P\bar{\rho}_jP\bar{\rho}_i\right)=\textrm{Tr}\left(P\dfrac{I_D}{D}P\dfrac{I_D}{D}\right)=\dfrac{1}{D^2}\textrm{Tr}(P^2)=\boxed{\dfrac{1}{D}}
\end{align}
As such, $v$ is always $D^{-1}=n_{\textrm{fix}}/2^N$, regardless of the particular Pauli string.

The second mathematical tool we'll utilize is the Parseval sum rule, given by

\begin{align}
    \sum_{P\in \mathbb{P}_N}\mu_2(P)=    \sum_{P\in \mathbb{P}_N}v=2^N
\end{align}
We'll start with the sum over $\mu_2(P)$. Throughout this derivation, we will rely heavily upon the discussion of Pauli decomposition given by Nadimpalli, Parham, Vasconcelos, and Yuen~\cite{Nadimpalli2023Nov}.

We first note that, for any operator $A\in \mathcal{M}_N$ in a complex inner-product space $\mathcal{M}_N$, there is a unique representation of $A$ in terms of it's Pauli coefficients, which we'll denote as $B(P)$. We therefore may write
\begin{align}
    A=\sum_{P\in\mathbb{P}_N}B(P)P
\end{align}
The Pauli coefficient $B(P)$ can be solved for by taking the inner product $\langle P',\,\cdot\rangle$ of both sides with an arbitrary Pauli string $P'$, where $\langle X^\dagger,\, Y\rangle=\textrm{Tr}(X^\dagger Y)$ for some $X$ and $Y$:
\begin{align}
    \langle P',A\rangle&=\langle P',\sum_{P\in\mathbb{P}_N}B(P)P\rangle=B(P)\sum_{P\in\mathbb{P}_N}\langle P',P\rangle=B(P)2^N\delta_{P,\,P'}
\end{align}
\noindent where in the last line we have used the fact that, from the orthogonality of the Pauli group, 
\begin{align}
    \langle P,\,P'\rangle&=\textrm{Tr}(PP')=\prod_{i=1}^N\textrm{Tr}(P_iP_i')=\prod_{i=1}^N2\delta_{P_i,\,P_i'}=2^N\delta_{P,\,P'}\label{PP}
\end{align}
\noindent As such, we find that
\begin{align}
    B(P)&=2^{-N}\langle P,\,A\rangle\equiv 2^{-N}\textrm{Tr}(PA)\label{BP}
\end{align}
We now likewise use Eqn.~\eqref{PP} to derive a general formula for the Frobenius norm of $A$, given by $\|A\|_F^2\equiv \langle A,\,A\rangle$, in terms of the Pauli coefficients $B(P)$:
\begin{align}
    \langle A,\,A\rangle&=\textrm{Tr}(A^\dagger A)=\langle \sum_{P\in \mathbb{P}_N} B^*(P)P,\,\sum_{P'\in \mathbb{P}_N}B(P')P'\rangle=\sum_{P,\,P'\in\mathbb{P}_N}B^*(P)B(P')\langle P,\,P'\rangle=2^N\sum_P\|B(P)\|^2_2\notag\\
    &=2^{-N}\sum_P\|\textrm{Tr}(PA)\|^2_2
\end{align}
where we have utilized Eqn.~\ref{BP} in the last line. As discussed in Nadimpalli et al., we identify this result as a manifestation of Parseval's identity~\cite{Nadimpalli2023Nov} $\sum_n|\langle x,\,e_n\rangle|^2=\|x\|^2$ where $x\in H$ is an element in the Hilbert space $H$ with some inner product $\langle \cdot\, ,\,\cdot\rangle$ and $e_n$ is an orthonormal basis of $H$.

As a consequence of the above formula, we come to the most general form of the Parseval identity for some complex operator $A$, as introduced in Eqn.\cite{Nadimpalli2023Nov}:
\begin{align}
    \sum_{P\in \mathbb{P}_N}\|\textrm{Tr}(PA)\|^2_2=2^N\|A\|_F^2
\end{align}
We may now proceed with the sum over the symmetric second moment by recalling that $\mu_2(P)=\mathbb{E}_j[X_j^2(P)]$ where $X_j\equiv \langle \phi_j|P|\phi_j\rangle$ for some index $j$:
\begin{align}
    \sum_{P\in \mathbb{P}_N}\mu_2(P)&\equiv \sum_{P\in\mathbb{P}_N}\mathbb{E}_j[X_j^2(P)]=\mathbb{E}_j\bigg[\sum_{P\in\mathbb{P}_N}\langle \phi_j|P|\phi_j\rangle^2\bigg]=\mathbb{E}_j\bigg[\sum_{P\in\mathbb{P}_N}\|\textrm{Tr}(P\rho_j)\|_2^2\bigg]=2^N\mathbb{E}_j\|\rho_j\|_F^2=2^N\mathbb{E}_j [\textrm{Tr}(\rho_j^2)]^2\notag\\
    &=2^N
\end{align}
For $\sum_{P\in \mathbb{P}_N}v$, we note that we obtain the same result as the above if we recall that $v=\mathbb{E}_i\mathbb{E}_i[\langle \phi_i|P|\phi_j\rangle \langle \phi_j|P|\phi_i\rangle]\equiv \mathbb{E}_i\mathbb{E}_j[X_{ij}^2]$ for $i\not=j$, where $X_{ij}\equiv \langle \phi_i|P|\phi_j\rangle$:
\begin{align}
\sum_{P\in\mathbb{P}_N}v&\equiv \sum_{P\in \mathbb{P}_N}\mathbb{E}_i\mathbb{E}_j[\langle \phi_i|P|\phi_j\rangle\langle \phi_j|P|\phi_i\rangle]= \mathbb{E}_i\mathbb{E}_j\bigg[\sum_{P\in\mathbb{P}_N}\langle \phi_i|P|\phi_j\rangle\langle \phi_j|P|\phi_i\rangle\bigg]\notag\\
&=\mathbb{E}_i\mathbb{E}_j\bigg[\sum_{P\in\mathbb{P}_N}\|\textrm{Tr}(PM_{ij})\|_2^2\bigg],\quad M_{ij}\equiv |\phi_i\rangle \langle \phi_j|\notag\\
&=2^N\mathbb{E}_i\mathbb{E}_j[\|M_{ij}\|_F^2]=2^N\mathbb{E}_i\mathbb{E}_j [\textrm{Tr}(M_{ij}^\dagger M_{ij})]=2^N\mathbb{E}_j [\textrm{Tr}(\rho_j)]\notag\\
&=2^N
\end{align}
Thus, we have shown that $\sum_{P\in \mathbb{P}_N}\mu_2(P)=\sum_{P\in \mathbb{P}_N}v=2^N$. 

To proceed to the diagrammatic calculation we recall that sum over $P$ is restricted to the filtered Paulis $\widetilde{\mathbb P}_N$, where we have explicitely excluded the fixed Pauli strings. We can easily find the modification to the full sum below:
\begin{align}
    \sum_{P\in\widetilde{\mathbb{P}}_N}\mu_2(P)&= \sum_{P\in{\mathbb{P}}_N}\mu_2(P)- \sum_{P\in{\mathbb{P}}_N^{\textrm{fixed}}}\mu_2(P)=2^N-\sum_{P\in\mathbb{P}_N^{\textrm{fixed}}}\mathbb{E}_j\|\langle \phi_j|\phi_j\rangle\|_2^2 =2^N-n_{\textrm{fix}}\notag\\
    &\approx 2^N
\end{align}

\begin{align}
    \sum_{P\in\widetilde{\mathbb{P}}_N}v&=  \sum_{P\in{{\mathbb P}}_N}v-\sum_{P\in{\mathbb {P}}_N^{\textrm{fixed}}}v=2^N-\sum_{P\in{\mathbb {P}}_N^{\textrm{fixed}}}\mathbb{E}_i\mathbb{E}_j[\|\langle \phi_i|\phi_j\rangle\|^2_2]=2^N-\sum_{P\in{\mathbb {P}}_N^{\textrm{fixed}}}\mathbb{E}_i[\langle \phi_i|\bar{\rho}_j|\phi_i\rangle]=2^N-\dfrac{n_{\textrm{fix}}^2}{2^N}\notag\\
    &\approx 2^N
\end{align}
where in the second to last line we invoked $\mathbb{E}_j[\rho_j]=I_D/D$ where in general $D=2^N/n_{\textrm{fix}}$. As a consequence, in the thermodynamic limit and the cases we consider in this paper (i.e., typical stabilizers with $n_{\textrm{fix}}=1$ and typical Gaussian states with $n_{\textrm{fix}}=2$), restricting the sum to the filtered Pauli strings does not alter the $2^N$ result in the thermodynamic limit.  

Finally, we can combine the Parseval sum rule and the flat spectrum of the directed second moment to derive an expression for the number of filtered Pauli strings, $N_{\textrm{filter}}$:

\begin{align}
    \sum_{P\in \widetilde{\mathbb{P}}_N}v=\dfrac{1}{D}N_{\textrm{filter}}=2^N-\dfrac{n_{\textrm{fix}}^2}{2^N}\rightarrow N_{\textrm{filter}}=2^ND-\dfrac{n_{\textrm{fix}}^2}{2^N}D=\dfrac{4^N}{n_{\textrm{fix}}}-n_{\textrm{fix}}
\end{align}

With the flatness of $v$ and the Parseval sum rule derived, in addition to $N_{\textrm{filter}}$, we can exploit these two rule to tackle the sums over the directed and symmetric second moments. Throughout, we will use the fact that the total number of Pauli strings be summed over is $4^N-n_{\textrm{fix}}$:

\begin{align}
    \gamma_{vv}&\equiv \sum_{P\in\widetilde{\mathbb{P}}_N}v^2=\dfrac{1}{D^2}\left(\dfrac{4^N}{n_{\textrm{fix}}}-n_{\textrm{fix}}\right)=\dfrac{n_{\textrm{fix}}^2}{4^N}\left(\dfrac{4^N}{n_{\textrm{fix}}}-n_{\textrm{fix}}\right)=n_{\textrm{fix}}-\dfrac{n_{\textrm{fix}}^3}{4^N}\approx n_{\textrm{fix}}
\end{align}

\begin{align}
    \gamma_{vvv}&=\sum_{P\in\widetilde{\mathbb{P}}_N}v^3=\dfrac{1}{D^3}\bigg(\dfrac{4^N}{n_{\textrm{fix}}}-n_{\textrm{fix}}\bigg)=\dfrac{n_{\textrm{fix}}^3}{8^N}\left(\dfrac{4^N}{n_{\textrm{fix}}}-n_{\textrm{fix}}\right)=\dfrac{n_{\textrm{fix}}^2}{2^N}-\dfrac{n_{\textrm{fix}}^4}{8^N}\approx 0
\end{align}

\begin{align}
    \gamma_{\mu v}&=\sum_{P\in \widetilde{\mathbb{P}}_N}v\mu_2(P)=\dfrac{1}{D}\sum_{P\in \widetilde{\mathbb{P}}_N}\mu_2(P)=\dfrac{n_{\textrm{fix}}}{2^N}\left(2^N-n_{\textrm{fix}}\right)=n_{\textrm{fix}}-\dfrac{n_{\textrm{fix}}^2}{2^N}\approx n_{\textrm{fix}}
\end{align}

\begin{align}
    \gamma_{\mu vv}&=\sum_{P\in\widetilde{\mathbb{P}}_N}v^2 \mu_2(P)=\dfrac{1}{D^2}\sum_{P\in\widetilde{\mathbb{P}}_N} \mu_2(P)
    =\dfrac{n_{\textrm{fix}}^2}{4^N}\bigg(2^N-n_{\textrm{fix}}\bigg)
    =\dfrac{n_{\textrm{fix}}^2}{2^N}-\dfrac{n_{\textrm{fix}}^3}{4^N}
    \approx 0
\end{align}

\noindent As such, we see that $\gamma_{vv}\approx \gamma_{\mu v}\approx n_{\textrm{fix}}$ and $\gamma_{vvv}\approx \gamma_{\mu vv}\approx 0$. 

The final term, $\gamma_{\mu \mu}$, cannot be simplified further until we consider a case-specific study. This leaves the more non-trivial diagrams to evaluate: $\alpha$ and $\beta$. While we cannot utilize our diagrammatic rules to simplify these expressions, we can still evaluate these expressions with the Pauli decomposition previously introduced~\cite{Nadimpalli2023Nov}. We'll start with $\alpha$, which we'll rewrite in the following way:
\begin{align}
    \alpha &=\mathbb{E}\bigg[\sum_{P\in\widetilde{\mathbb P}_N}X_{ij}X_{ji}X_{ji}X_{ij}\bigg]=\mathbb{E}\bigg[\sum_{P\in\widetilde{\mathbb P}_N}\bigg(X_{ij}X_{ji}\bigg)^2\bigg]=\mathbb{E}\bigg[\sum_{P\in\widetilde{\mathbb P}_N}\bigg(\textrm{Tr}(P\rho_j P\rho_i)\bigg)^2\bigg]\notag\\
    &=\mathbb{E}\bigg[\sum_{P\in {\mathbb P}_N}\bigg(\textrm{Tr}(P\rho_j P\rho_i)\bigg)^2\bigg]-n_{\textrm{fix}}\mathbb{E}\bigg[\sum_{P\in {\mathbb P}^{\textrm{fixed}}_N}\bigg(\textrm{Tr}(P\rho_j P\rho_i)\bigg)^2\bigg]\notag\\
    &=\mathbb{E}\bigg[\sum_{P\in {\mathbb P}_N}\bigg(\textrm{Tr}(P\rho_j P\rho_i)\bigg)^2\bigg]-n_{\textrm{fix}}\mathbb{E}\bigg[\sum_{P\in {\mathbb P}^{\textrm{fixed}}_N}\bigg(\textrm{Tr}(\rho_j \rho_i)\bigg)^2\bigg]
\end{align}
In the above, note that we have extracted the fixed Pauli strings in the last line. The term in the fixed expression can be simplified to $\textrm{Tr}(\rho_j\rho_i)=\textrm{Tr}(|\phi_j\rangle\langle \phi_j|\phi_i\rangle\langle \phi_i|)$, which is zero as for this diagram we have $i\not =j$. To evaluate $\alpha$, we are then concerned with evaluating $\textrm{Tr}(P\rho_j P\rho_i)$. For this task, we will rewrite the density matrix in terms of the Pauli spectrum:

\begin{align}
    \rho_i=\sum_{R\in\mathbb{P}_N}B(R)R=\dfrac{1}{D}\sum_{R\in\mathbb{P}_N}\textrm{Tr}(R\rho_i)R\equiv \dfrac{1}{D}\sum_{R\in\mathbb{P}_N}X_i(R)R
\end{align}
where $R$ is taken to be another Pauli string and $X_i(R)\equiv \langle \phi_i|R|\phi_i\rangle$ as we defined before, except now we make the dependence on the specific Pauli string explicit. With the above, we can now represent the trace as follows:
\begin{align}
    \textrm{Tr}(P\rho_jP\rho_i)&=\textrm{Tr}(\rho_jP\rho_iP)=\textrm{Tr}\left(\rho_j P\dfrac{1}{D}\sum_{R\in\mathbb{P}_N}X_i(R)RP\right)=\dfrac{1}{D}\sum_{R\in\mathbb{P}_N}X_i(R)\textrm{Tr}(\rho_jPRP)
\end{align}
To simplify further, we will invoke the Pauli identity $PRP=\epsilon_{PR}R$, where
\begin{align}
    \sum_{P\in\mathbb{P}_N}\epsilon_{PR}\epsilon_{PR'}=D^2\delta_{R,\,R'},\qquad \epsilon_{PR}=\prod_{j=1}^N \epsilon_{P_jR_j},\qquad \epsilon_{P_jR_j}=\begin{cases}
        &-1,\qquad P_j\not=R_j,\\
        &+1,\qquad P_j=R_j
    \end{cases}
\end{align}
This allows us to simplify the trace above as follows:
\begin{align}
    \textrm{Tr}(P\rho_jP\rho_i)&=\dfrac{1}{D}\sum_{R\in\mathbb{P}_N}X_i(R)\textrm{Tr}(\rho_jPRP)=\dfrac{1}{D}\sum_{R\in\mathbb{P}_N}X_i(R)\textrm{Tr}\bigg(\rho_j\epsilon_{PR}R\bigg)=\dfrac{1}{D}\sum_{R\in\mathbb{P}_N}\epsilon_{PR}X_i(R)\textrm{Tr}\bigg(\rho_jR\bigg)\notag\\
    &=\dfrac{1}{D}\sum_{R\in\mathbb{P}_N}\epsilon_{PR}X_i(R)\textrm{Tr}\bigg(\dfrac{1}{D}\sum_{R'\in\mathbb{P}_N}X_j(R')R'R\bigg)=\dfrac{1}{D^2}\sum_{R\in\mathbb{P}_N}\epsilon_{PR}X_i(R)X_j(R')\textrm{Tr}(R'R)\notag\\
    &=\dfrac{1}{D}\sum_{R\in\mathbb{P}_N}\epsilon_{PR} X_i(R)X_j(R)
\end{align}
\noindent where we have used the identity $\textrm{Tr}(R'R)=N\delta_{R,\,R'}$. To find $\alpha$ from the above, we will square this expression, yielding 

\begin{align}
    \sum_{P\in\mathbb{P}_N}\bigg(\textrm{Tr}(P\rho_j P\rho_i)\bigg)^2&=\dfrac{1}{D^2}\sum_{P,\,R,\,R'\in\mathbb{P}_N}\epsilon_{PR}\epsilon_{PR'}X_i(R)X_j(R)X_i(R')X_j(R')\notag\\
    &=\sum_{R\in \mathbb{P}_N}X_i^2(R)X_j^2(R)
\end{align}
Taking the expectation value of the above then immediately yields an expression for $\alpha$ as effectively $\gamma_{\mu\mu}+n_{\textrm{fix}}$:
\begin{align}
    \alpha &=\mathbb{E}\bigg[\sum_{R\in \mathbb{P}_N}X_i^2(R)X_j^2(R)\bigg]=\mathbb{E}\bigg[\sum_{R\in \widetilde{\mathbb{P}}_N}X_i^2(R)X_j^2(R)\bigg]+n_{\textrm{fix}}\mathbb{E}\bigg[\sum_{R\in {\mathbb{P}}_N^{\textrm{fixed}}}X_i^2(R)X_j^2(R)\bigg]=\gamma_{\mu\mu}+n_{\textrm{fix}}
\end{align}

The above holds independent of the specific ensemble under consideration. For $\beta$, a nearly identical derivation results in the following simplification:

\begin{align}
    \beta=\mathbb{E}\bigg[\sum_{P\in \widetilde{\mathbb{P}}_N}X_i X_j X_{ij} X_{ji}\bigg]=\mathbb{E}\bigg[\sum_{P\in\mathbb{P}_N}\textrm{Tr}\left(\rho_i P \rho_i P \rho_j P \rho _j P\right)\bigg]=\mathbb{E}\bigg[\dfrac{1}{D}\sum_{R,\,R'\in\mathbb{P}_N}\epsilon_{RR'}X_i(R)X_i(R')X_j(R)X_j(R')\bigg]\label{beta}
\end{align}
Unlike $\alpha$, we have an additional factor of $\epsilon_{RR'}$ which cannot easily be removed from the sum. As a result, we will instead simplify the absolute value of $\beta$ in the sum via Cauchy-Schwartz inequality:
\begin{align}
    \bigg|\mathbb{E}[X_i X_j X_{ij}X_{ji}]\bigg| & \le \mathbb{E}[|X_iX_j||X_{ij}|^2]\le \dfrac{1}{2}\mathbb{E}[X_i^2|X_{ij}|^2]+\dfrac{1}{2}\mathbb{E}[X_i|X_{ij}^2|]=\mathbb{E}[X_i^2 |X_{ij}|^2]\notag\\
    &=\mathbb{E}_i[X_i^2]\mathbb{E}_j[|X_{ij}|^2]\notag\\
    &=\mu_2(P)v
\end{align}
As a consequence, if we sum over $P$, we can now invoke the Parseval sum rule:
\begin{align}
|\beta|= \bigg|\sum_{P\in\widetilde{\mathbb{P}}_N}\mathbb{E}[X_iX_jX_{ij}X_{ji}]\bigg|\le \sum_{P\in\widetilde{\mathbb{P}}_N}\bigg|\mathbb{E}[X_iX_jX_{ij}X_{ji}]\bigg|\le v\sum_{P\in \widetilde{\mathbb{P}}_N}\mu_2(P)\approx 
n_{\textrm{fix}}\end{align}
As a consequence, we see that all random ensembles must obey $\beta\le n_{\textrm{fix}}$ in the thermodynamic limit. Further progress can be made for Eqn.~\eqref{beta} whenever an ensemble is invariant under replacing a state $|\phi_i\rangle = S|\phi'_i\rangle$, where $S$ is some Pauli string. Since the ensemble is invariant under this identification it follows that
\begin{align}
\mathbb{E} \bigg[\langle \phi_i | R| \phi_i  \rangle \langle \phi_i | R'| \phi_i  \rangle \bigg] = \epsilon_{SR}\epsilon_{SR'} \mathbb{E} \bigg[\langle \phi'_i | R |\phi'_i  \rangle\langle \phi'_i | R' |\phi'_i  \rangle \bigg] =  \epsilon_{S,RR'} \mathbb{E} \bigg[\langle \phi'_i | R |\phi'_i  \rangle\langle \phi'_i | R' |\phi'_i  \rangle \bigg] .
\end{align}
This identity holds for any string $S$, thus implying $ R R' $ is an element of the fixed Pauli strings. It follows then that there are  $n_{\textrm{fix}}$ corresponding partners. Each corresponding pair of terms satisfy $X(R')=\pm \lambda X(R)$, where $\lambda$ is the eigenvalue of the fixed string $g$. The $\lambda$ terms are squared in the sum, resulting in an overall multiplicative factor of unity, and thus we are left with
\begin{align}
    \beta&=\mathbb{E}\bigg[\dfrac{1}{D}\sum_{R,\,R'\in\widetilde{\mathbb{P}}_N}\epsilon_{RR'}X_i(R) X_i(R')X_j(R)X_j(R') \bigg]\notag\\
    &=\mathbb{E}\bigg[\dfrac{n_{\textrm{fix}}}{D}\sum_{R\in\widetilde{\mathbb{P}}_N} X_i^2(R)X_j^2(R)\bigg]\notag\\
    &=\dfrac{n_{\textrm{fix}}}{D}\sum_{R\in\widetilde{\mathbb{P}}_N}\mu_2^2(R)\approx\dfrac{n_{\textrm{fix}}}{D}\sum_{R\in {\mathbb{P}}_N}\mu_2^2(R)
\end{align}
The above gives a closed form for $\beta$ that remains applicable and verified for random stabilizer states and random Gaussian states (shown later in the Supplemental Material). Furthermore, we can show that the prediction $\beta=n_{\textrm{fix}}\sum_{P\in\mathbb{P}_N}\mu_2^2(P)/D$ is consistent with our previous upper bound $\beta\le n_{\textrm{fix}}$. This can be seen by first noting that
\begin{align}
    \sum_{P\in\mathbb{P}_N}\mu_2^2(P)\le \textrm{max}_P[ \mu_2(P)]\sum_{P}\mu_2(P)=\textrm{max}_P[ \mu_2(P)]2^N
\end{align}
From Cauchy-Schwartz, we know that for states $|u\rangle$ and $|v\rangle$, then $|\langle w|v\rangle |^2\le \langle u|u\rangle \langle w|w\rangle$. Taking $|u\rangle=|\phi\rangle$ and $\langle w|=\langle \phi|P$, then
\begin{align}
    \mu_2(P)=\langle \phi|P|\phi\rangle^2\le \|\phi \|^2 \|P \phi\|^2=1
\end{align}
As such, we see that $\sum_{P\in \mathbb{P}_N}\mu_2^2(P)\le 2^N$, and thus $\beta$ must be less than $n_{\textrm{fix}}$, as introduced previously.  

For general weights of the superpositions, we find that the Parseval sum rule and the expressions for $\alpha$ and $\beta$ reduces the average witness greatly. Below we simplify the general form:
\begin{align}
    \langle W_\Psi(m)\rangle&=n_{\textrm{fix}}+3\gamma_{vv}+6\gamma_{vvv}+\|p\|_2^2\bigg(6\gamma_{\mu v}+24\gamma_{\mu vv}-6\gamma_{vv}-36\gamma_{vvv}\bigg)\notag\\
&\phantom{=n_{\textrm{fix}}+3\gamma_{vv}+6\gamma_{vvv}l}+\|p\|_3^3\bigg(-48\gamma_{\mu vv}+48\gamma_{vvv}\bigg)\notag\\
&\phantom{=n_{\textrm{fix}}+3\gamma_{vv}+6\gamma_{vvv}l}+\|p\|_2^4\bigg(3\gamma_{\mu\mu}+12\beta-6\gamma_{\mu v}-24\gamma_{\mu vv}+3\gamma_{\mu\mu}+3n_{\textrm{fix}}-3\gamma_{vv}+18\gamma_{vvv}\bigg)\notag\\
&\phantom{=n_{\textrm{fix}}+3\gamma_{vv}+6\gamma_{vvv}ll}+
\|p\|_4^4\bigg(\langle W_\Psi(1)\rangle-n_{\textrm{fix}}-3\gamma_{\mu \mu}-12\beta +48\gamma_{\mu vv}-3\gamma_{\mu\mu}-3n_{\textrm{fix}}+6\gamma_{vv}-36\gamma_{vvv}\bigg)\notag\\
&\notag\\
&\approx n_{\textrm{fix}}+3n_{\textrm{fix}}+\|p\|_2^2\bigg(6n_{\textrm{fix}}-6n_{\textrm{fix}}\bigg)\notag\\
&\phantom{=n_{\textrm{fix}}+3n_{\textrm{fix}}p}+\|p\|_2^4\bigg(3\gamma_{\mu\mu}+12\beta-6n_{\textrm{fix}}+3\gamma_{\mu\mu}+3n_{\textrm{fix}} -3n_{\textrm{fix}}\bigg)\notag\\
&\phantom{=n_{\textrm{fix}}+3n_{\textrm{fix}}p}+\|p\|_4^4\bigg(\langle W_\Psi(1)\rangle-n_{\textrm{fix}}-3\gamma_{\mu\mu}-12\beta -3\gamma_{\mu\mu}-3n_{\textrm{fix}} +6n_{\textrm{fix}}\bigg)\notag\\
&\notag\\
&=4n_{\textrm{fix}}+\|p\|_2^4\bigg(6\gamma_{\mu \mu} -6n_{\textrm{fix}}+12\beta\bigg)+\|p\|_4^4\bigg(\langle W_\Psi(1)\rangle-6\gamma_{\mu\mu}+2n_{\textrm{fix}}-12\beta\bigg)\notag\\
&\notag\\
&=4n_{\textrm{fix}}+\|p\|_2^4\bigg(6\sum_{P\in\mathbb{P}_N}\mu_2^2(P) -12n_{\textrm{fix}}+12\beta\bigg)+\|p\|_4^4\bigg(\langle W_\Psi(1)\rangle-6\sum_{P\in\mathbb{P}_N}\mu_2^2(P)+8n_{\textrm{fix}}-12\beta\bigg)\notag\\
&\notag\\
&\equiv 4n_{\textrm{fix}}+b\|p\|_2^4+\bigg(\langle W_\Psi(1)\rangle-(b+4_{\textrm{fix}})\bigg)\|p\|_4^4\notag\\
&\notag\\
&\equiv c +b\|p\|_2^4+\left\{a-(b+c)\right\}\|p\|_4^4
\end{align}
where we have taken the convention in the main text where $c=4n_{\textrm{fix}}$ is the $p$-independent term that determines the precise value of the Haar floor of the non-stabilizer witness, $a=\langle W(1)\rangle$ is the average witness for one superposition, and $b$ is defined as follows:
\begin{align}
    b&\equiv 
    6\bigg\{\sum_{P\in\mathbb{P}_N}\mu_2^2(P)-2(n_{\textrm{fix}}-\beta)\bigg\}
\end{align}
where we have re-wrote the sum to be over the full Pauli group in the third line, and which is the same form of $b$ we quote in the text. The average witness for equal weight superpositions follows immediately by recalling $\|p\|_2^4=1/m^2$ and $\|p\|_4^4=1/m^3$. 

\section{Non-stabilizer witness for three cases of typical superpositions}

\subsection{Case 1: Random stabilizer states}

As introduced in the previous section, the full calculation of the non-stabilizer witness and, thus, the stabilizer R\'enyi entropy requires five quantities which are specific to the ensemble under consideration: $n_{\textrm{fix}}$, $\langle W(1)\rangle_\Psi$, $\gamma_{\mu\mu}$, and the diagrams $\alpha$ and $\beta$. 

We'll start with random stabilizers. As mentioned in the main text, the average $\langle W_\Psi(1)\rangle$ taken over all Pauli strings reduces to $2^N$ as the individual averages only take on the values $0$ and $\pm 1$. In addition, as the only Pauli string which results in a constant average is the full identity, and thus $n_{\textrm{fix}}=1$ for random stabilizers.

The calculation of $\gamma_{\mu\mu}$ and the case-dependent diagrams $\alpha$ and $\beta$ are more complicated, and require more care. For all three, we will exploit the fact that the random stabilizers form a complex projective 2-design, and thus from the Schur-Weyl duality we may write
\begin{align}
    \mathbb{E}\left\{\rho^{\otimes}\right\}=\dfrac{{\mathbb{I}}^{\otimes 2}+\mathbb{F}}{D(D+1)}
\end{align}
where $\mathbb{F}$ is the swap operator. Our goal is to now represent $\gamma_{\mu\mu}$, $\alpha$, and $\beta$ as some function of $\mathbb{E}\left\{\rho^{\otimes 2}\right\}$. Starting with $\gamma_{\mu \mu}$, we find a clean closed form in the thermodynamic limit:

\begin{align}
    \mu_2^2(P)&=\mathbb{E}\bigg[\sum_{{P\in\widetilde{\mathbb{P}}_N}}\langle \phi_i|P|\phi_i\rangle^2\langle \phi_j|P|\phi_j\rangle^2\bigg]=\mathbb{E}\bigg[\sum_{{P\in\mathbb{P}_N}} [\textrm{Tr}(P\rho_i)]^2[\textrm{Tr}(P\rho_j)]^2\bigg]=\mathbb{E}\bigg[\sum_{{P\in\widetilde{\mathbb{P}}_N}}\textrm{Tr}\bigg\{P^{\otimes 2}\mathbb{E}(\rho_i^{\otimes 2})\bigg\}^2\bigg]\notag\\
    &=\mathbb{E}\bigg[\sum_{{P\in\widetilde{\mathbb{P}}_N}}\textrm{Tr}\bigg\{P^{\otimes 2}\left(\dfrac{I^{\otimes 2}+\mathbb{F}}{D(D+1)}\right)\bigg\}^2\bigg]\approx \left(\dfrac{D}{D(D+1)}\right)^2 2^ND=\dfrac{2^N D}{(D+1)^2}=n_{\textrm{fix}}\left(\dfrac{D}{D+1}\right)^2\approx \boxed{n_{\textrm{fix}}}
\end{align}
In the derivation above, we have utilized the identities 
\begin{align}
    \textrm{Tr}[(A\otimes B)(C\otimes D)]=\textrm{Tr}(AC)\textrm{Tr}(BD)
\end{align}
\begin{align}
    \textrm{Tr}[(A\otimes B)\mathbb{F}]=\textrm{Tr}(AB)
\end{align}
Thus,
\begin{align}
    \langle \phi_i|P|\phi_i\rangle^2=[\textrm{Tr}(P\rho_i)]^2=\textrm{Tr}(P^{\otimes 2}\rho_i^{\otimes 2})
\end{align}
\begin{align}
    \textrm{Tr}(P^{\otimes 2}{\mathbb{I}}^{\otimes 2})=\textrm{Tr}(P)^2=0
\end{align}
\begin{align}
    \textrm{Tr}(P^{\otimes 2}\mathbb{F})=\textrm{Tr}(P^2)=D
\end{align}
A similar calculation may be performed for $\beta$:

\begin{align}
    \beta&\equiv \mathbb{E}\bigg[\sum_{P\in\widetilde{\mathbb{P}}_N}\langle \phi_i|P|\phi_i\rangle \langle \phi_i|P|\phi_j\rangle \langle \phi_j|P|\phi_i\rangle\langle \phi_j|P|\phi_j\rangle\bigg]=\mathbb{E}\bigg[\sum_{P\in\widetilde{\mathbb{P}}_N} \textrm{Tr}(\rho_iP) \langle \phi_i|P\rho_jP|\phi_i\rangle \textrm{Tr}(\rho_j P)\bigg]\notag\\
        &=\mathbb{E}\bigg[\sum_{P\in\widetilde{\mathbb{P}}_N} \textrm{Tr}(\rho_iP) \langle \phi_i|P\rho_j \textrm{Tr}(\rho_j P)P|\phi_i\rangle \bigg]=\mathbb{E}_i\bigg[\sum_{P\in\widetilde{\mathbb{P}}_N} \textrm{Tr}(\rho_iP) \langle \phi_i|P\textrm{Tr}_2\left\{(\mathbb{I}\otimes P)\mathbb{E}_j(\rho_j^{\otimes 2})\right\}P|\phi_i\rangle \bigg]\notag\\
        &=\mathbb{E}_i\bigg[\sum_{P\in\widetilde{\mathbb{P}}_N} \textrm{Tr}(\rho_iP) \langle \phi_i|P\textrm{Tr}_2\left\{\dfrac{(\mathbb{I}\otimes P)(\mathbb{I}^{\otimes 2}+\mathbb{F})}{D(D+1)}\right\}P|\phi_i\rangle \bigg]=\mathbb{E}_i\bigg[\sum_{P\in\widetilde{\mathbb{P}}_N} \textrm{Tr}(\rho_iP) \langle \phi_i|P\left\{\dfrac{P}{D(D+1)}\right\}P|\phi_i\rangle \bigg]\notag\\
        &=\sum_{P\in\widetilde{\mathbb{P}}_N}\dfrac{1}{D(D+1)}\mu_2(P)\approx \sum_{P\in\widetilde{\mathbb{P}}_N}v^2\mu_2(P)\equiv \gamma_{\mu vv}\approx \boxed{0}
\end{align}
In the derivation of the above, we have utilized the identities 

\begin{align}
    \textrm{Tr}_2[({\mathbb{I}}\otimes P)(\mathbb{I}\otimes \mathbb{I})]=\textrm{Tr}_2({\mathbb{I}\otimes \mathbb{P}})=\mathbb{I}\textrm{Tr}(P)=0,\qquad \textrm{(For filtered Pauli strings only)}
\end{align}
\begin{align}
    \textrm{Tr}_2\left\{(\mathbb{I}\otimes P)\mathbb{F}\right\}=P
\end{align}

The same principle applies to the calculation of $\alpha$, except now this diagram evaluates to a finite value, unlike $\beta$:
\begin{align}
    \alpha&\equiv\mathbb{E}\bigg[\sum_{P\in\widetilde{\mathbb{P}}_N}\langle \phi_i|P|\phi_j\rangle\langle \phi_j|P|\phi_i\rangle\langle \phi_j|P|\phi_i\rangle \langle \phi_i|P|\phi_j\rangle\bigg]=\mathbb{E}\bigg[\sum_{P\in\widetilde{\mathbb{P}}_N}\langle \phi_i|P\rho_jP|\phi_i\rangle^2\bigg]=\mathbb{E}\bigg[\sum_{P\in\widetilde{\mathbb{P}}_N}\left\{\textrm{Tr}(P\rho_jP\rho_i)\right\}^2\bigg]\notag\\
    &=\mathbb{E}\bigg[\sum_{P\in\widetilde{\mathbb{P}}_N}\left\{\textrm{Tr}(P^{\otimes 2}\rho_j^{\otimes 2}P^{\otimes 2}\rho_i^{\otimes 2})\right\}\bigg]=\sum_{P\in\widetilde{\mathbb{P}}_N}\textrm{Tr}\bigg\{\dfrac{P^{\otimes 2}({\mathbb{I}^{\otimes 2}}+\mathbb{F})P^{\otimes 2}({\mathbb{I}^{\otimes 2}+\mathbb{F}})}{D(D+1)D(D+1)}\bigg\}\notag\\
    &=\sum_{P\in\widetilde{\mathbb{P}}_N}\dfrac{1}{D^2(D+1)^2}\textrm{Tr}\bigg\{P^{\otimes 2}P^{\otimes 2}+P^{\otimes 2}P^{\otimes 2}\mathbb{F}+P^{\otimes 2}\mathbb{F}P^{\otimes 2}+P^{\otimes 2}\mathbb{F}P^{\otimes 2}\mathbb{F}\bigg\}\notag\\
    &=\sum_{P\in\widetilde{\mathbb{P}}_N}\dfrac{1}{D^2(D+1)^2}\left\{\textrm{Tr}(\mathbb{I})+\textrm{Tr}(\mathbb{F})+\textrm{Tr}(\mathbb{F})+\textrm{Tr}(\mathbb{I})\right\}=\sum_{P\in\widetilde{\mathbb{P}}_N}\dfrac{1}{D^2(D+1)^2}\left\{D^2+D+D+D^2\right\}\notag\\
    &=\sum_{P\in\widetilde{\mathbb{P}}_N}\dfrac{2D(D+1)}{D^2(D+1)^2}\approx 2\sum_{P\in\widetilde{\mathbb{P}}_N} v^2\equiv 2\gamma_{vv}\approx \boxed{2n_{\textrm{fix}}}
\end{align}
where in the above we have utilized the fact that
\begin{align}
    \textrm{Tr}(P^{\otimes 2}P^{\otimes 2})=\textrm{Tr}(P^{\otimes 2}\mathbb{F}P^{\otimes 2}\mathbb{F})=\textrm{Tr}(P^2)\textrm{Tr}(P^2)=D^2
\end{align}
\begin{align}
\textrm{Tr}(P^{\otimes 2}\mathbb{F}P^{\otimes 2})= \textrm{Tr}(P^{\otimes 2}P^{\otimes 2}\mathbb{F})=\textrm{Tr}(\mathbb{I}^{\otimes 2}\mathbb{F})=D
\end{align}
With these terms now fully derived, we note that the previously defined $b$ reduces to
\begin{align}
    b\equiv 3\alpha+12\beta-9n_{\textrm{fix}}+3\gamma_{\mu\mu}=6n_{\textrm{fix}}-9n_{\textrm{fix}}+3n_{\textrm{fix}}=0
\end{align}
All together, the non-stabilizer witness for the $m$-fold superposition of random stabilizer states becomes
\begin{align}
    \langle W_\Psi(m)\rangle&=4+\|p\|_4^4(2^N-4)+\mathcal{O}(2^{-N})\notag\\
    &=4+\dfrac{2^N-4}{m^3}+\mathcal{O}(2^{-N}),
\end{align}
which is the form given in the main text and where in the last line we specialized to equal weight superpositions. 

\begin{figure}[t]
    \centering

    \begin{subfigure}[b]{\columnwidth}
        \centering
        \makebox[\columnwidth][c]{{(a)}}\\[2pt]
        \includegraphics[width=0.675\columnwidth]{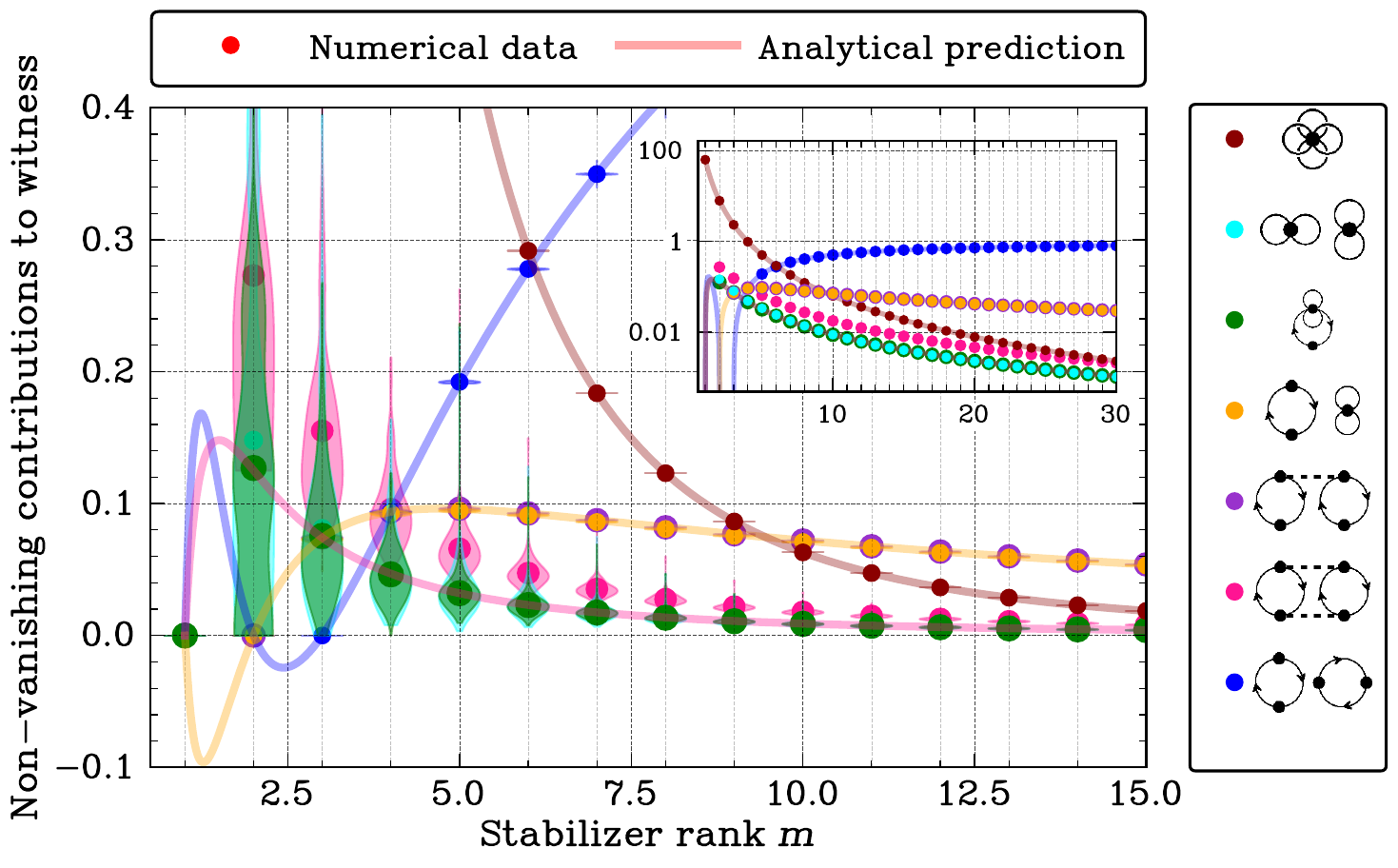}
        \label{fig:Fig3a}
    \end{subfigure}
    
    \vspace{6pt}
    
    \begin{subfigure}[b]{1\columnwidth}
        \centering
        \makebox[\columnwidth][c]{{(b)}}\\[2pt]
        \includegraphics[width=0.675\columnwidth]{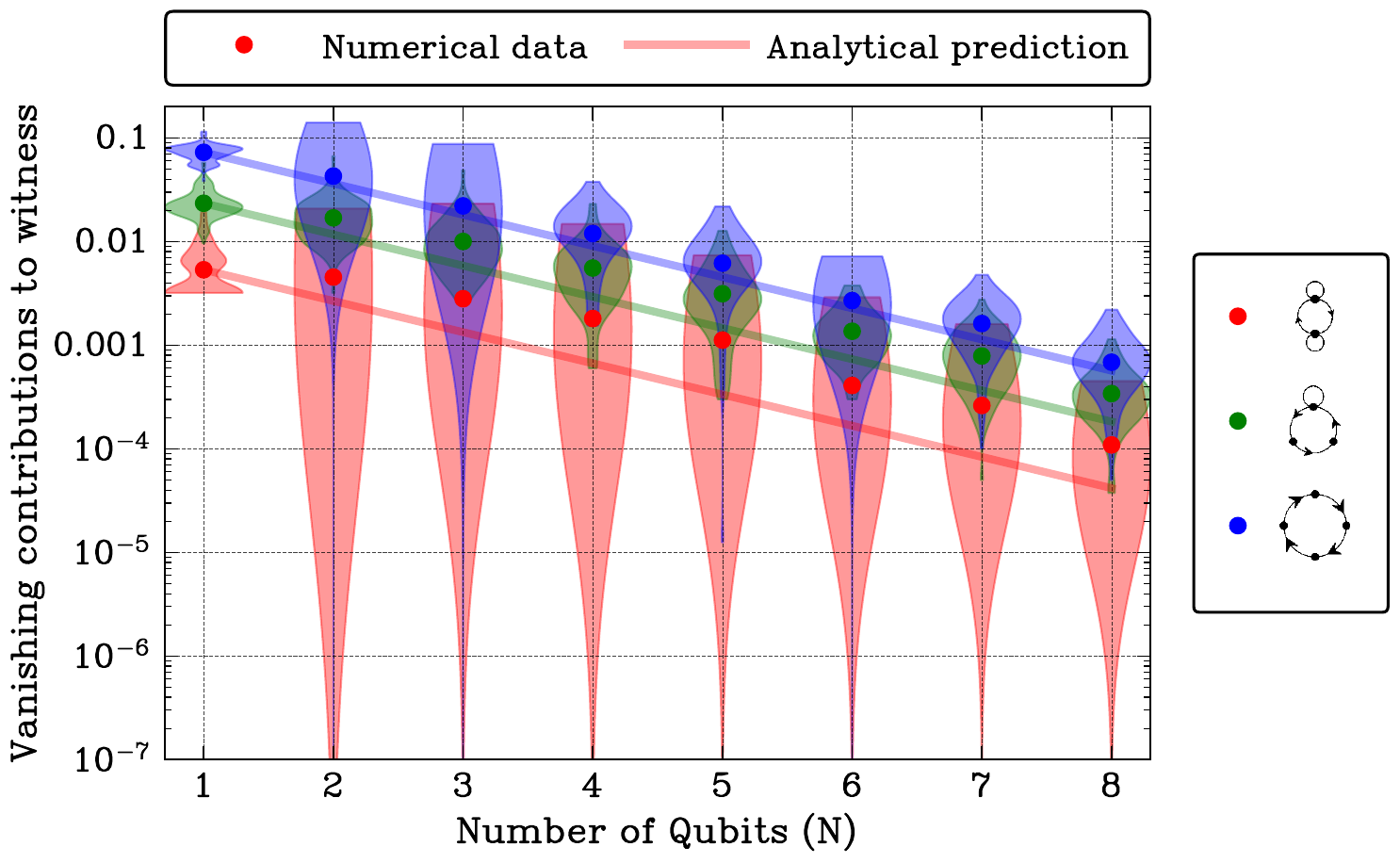}
        \label{fig:Fig3b}
    \end{subfigure}

    \caption{\small The non-vanishing contributions to the non-stabilizer witness versus the stabilizer rank $m$ for $N=6$ (a) and the vanishing contributions to the stabilizer witness versus number of qubits $N$ for $m=5$ (b) for random stabilizer states. In this context, a "vanishing" diagram refers to a predicted $2^{-N}$ dependence. In both figures, we average over $100$ trials for each $m$ and we utilize the diagrammatic shorthand introduced in Fig.~2 of the main text. In (a), we see that two diagrams dominate as we tune $m$: the four-fold symmetric cycle for small $m$ (maroon) and the disconnected two-fold directed cycle for large $m$ (dark blue). In (b), we see the remaining three contributions indeed fall roughly as $2^{-N}$, as given by the solid lines.}
    
    \label{fig:Fig3}
\end{figure}

In Fig.~3a and 3b, we plot the contribution to the non-stabilizer witness from each of the diagrams introduced in Fig.~2 of the main text. In Fig.~3a, we plot the diagrammatic terms that are expected to contribute in the thermodynamic limit versus $m$ for fixed $N=6$. For all $m$, the main contribution to the witness is dominated by two diagrams: the four-fold symmetric cycle and the disconnected two-fold directed cycle. The former results in the logarithmic growth for small $m$, while the saturation to the Haar value of $4$ is controlled by the latter. All other diagrams are negligible to these two contributions. In Fig.~3b, we plot the remaining three diagrams as a function of $N$ for $m=5$. For all three, we see that these diagrams vanish as $\sim D^{-1}=2^{-N}$ as $N\rightarrow \infty$, as predicted from the Parseval sum rule.

\subsection{Case 2: Random Gaussian states}

The case dependent diagrams: $\alpha$, $\gamma_{\mu\mu}$,  $\beta$, and $\langle W(1) \rangle$ for random Gaussian states are evaluated below in the order listed. Before proceeding to the explicit calculation the relevant identities are collected, many being repeated from above for convenient reference. As before the primary object of investigation is the density matrix given by
\begin{equation}
        \rho_i = |\phi_i \rangle \langle \phi_i | , \qquad \mathbb{E}_j[\rho_j]=I_D/D, \qquad D=2^N/n_{\textrm{fix}}
\end{equation}The set of Pauli operators $\mathbb{P}_N$ are a particularly convenient basis for the density matrix
\begin{align}
    \rho_i  = \sum_{P\in\mathbb{P}_N}B_i(P)P, \qquad B_i(P) &= \frac 1 {2^N} \textrm{Tr}(\rho_i P), \qquad \textrm{Tr}( P' P) = 2^N \delta_{PP'}.
\end{align}
These identities are further supplemented by the Pauli operator identity with 
\begin{align}
    P R P = \epsilon_{PR} R, &\qquad \sum_{P \in  \mathbb{P}} \epsilon_{PR} \epsilon_{PR'} = 4^N \delta_{RR'},\\
    \epsilon_{PR} = \prod_{j=1}^N \epsilon_{P_j R_j} &\qquad \epsilon_{P_j R_j} = \begin{cases}
        -1 , \quad  P, R \in \{X,Y,Z\}, P_j \neq R_j \\
        +1, \quad {\rm else}.
    \end{cases}.
\end{align}
These identities are the key initial step, which are used to simplify the diagrams into expressions determined by moments of the random state distribution. To compute these moments, the Pauli strings are recast into a Majorana representation. In this representation the eigenvalues of the minors of the correlation matrix are exactly known, and are just a mapping of moments for Pauli strings of fixed string length. After some massaging, an exact moment formula is obtained for these fixed length Pauli strings. With this route in mind, these steps are carried out explicitly below.

A random fermionic Gaussian state, $\rho$, is characterized by a real antisymmetric $2N \times 2N$ correlation matrix $\Gamma$ with $\Gamma^2 = - \mathbb{I}_{2N}$. The minors of these matrices, $\Gamma_A \subseteq \Gamma$, are $|A| \times |A|$ matrices with $|A| \leq 2N$. Consider the action on such a state by a Pauli operator, with the latter recast as products of an orthogonal basis of Majorana operators. With the caveat that a Pauli strings is uniquely identified by this Majorana form up to an overall phase
\begin{align}
M_A = i^{m(m-1)/2} \gamma_{a_1} \dots \gamma_{a_m}    .
\end{align}
It follows that $\textrm{Tr}(\rho M_A) = \textrm{Pf}(\Gamma_A)$, which vanishes for odd subsets based on parity considerations. Now the type of Pauli operator considered can be grouped by string length, thus fixing a corresponding minor $\Gamma_{A_K} $ such that $|A_k| = 2k$ resulting in the sum over Pauli strings to be recast and written in terms of the eigenvalues as
\begin{equation}
    \sum_{R \in \mathbb{P}_N} \bigg(\mathbb{E} \bigg[b(R)^2 \bigg] \bigg) = \sum_{A \subseteq 2N } \bigg(\mathbb{E}\bigg[  \textrm{Pf}(\Gamma_A)^2 \bigg] \bigg) = \sum_{k=0}^N \sum_{|A_k| = 2k } \bigg(\mathbb{E}\bigg[  \textrm{Pf}(\Gamma_{A_k})^2 \bigg] \bigg).
\end{equation}

Now the eigenvalues, $\lambda_i \in [0,1]$ of a minor $\Gamma_A$ are known to be given by the joint distribution for a Haar $2k$-frame in $\mathbb{R}^{2N}$, with $2k \leq 2N$, originally found in~\cite{bianchi} as
\begin{align}
    \mathbb{M}_A(\lambda_1, \dots ,\lambda_k) = \frac{1}{Z_{N,k}} \prod_{i<j}(\lambda_i^2 - \lambda_j^2)^2 \prod_{j=1}^k(1-\lambda_j^2)^{N-2k}
\end{align}
With $Z_{N,k}$ defined so that the total probability after integrating over all $\lambda_i$ of the distribution is unity. The moments of this distribution are, upon identifying $\lambda_i^2 = t_i$ and consequently $ d\lambda_i =dt_i / \sqrt{t_i}$, then the $s$-moment is given by
\begin{equation}
    \mathbb{E}\bigg[  \textrm{Pf}(\Gamma_{A_k})^{2s} \bigg] =\mathbb{E}\bigg[ \prod_{i=1}^k t^s_i \bigg] = \frac{1}{Z_{N,k}} \int_{[0,1]^k} \bigg[\prod_{i=1}^k d t_i \bigg]\prod_{i< j} (t_i - t_j)^2 \bigg(\prod_{i}^k t_i^{s -\frac 1 2} (1 - t_i)^{N-2k} \bigg)
\end{equation}
with $Z_{N,k}$ being defined so the $s=0$ moment is $1$, which just ensures a normalized distribution. Thus, the $s$-moment is directly expressed in terms of Selberg integrals~\cite{selberg1944bemerkninger, Forrester2010LogGases, debruyne2023linearstatisticscoulombgases, tiutiakina2023framepotentialbrowniansyk,collura2024} defined as
\begin{equation}
    S_k(a,b) = \int_{[0,1]^k} \bigg[\prod_{i=1}^k d t_i \bigg]\prod_{i< j} (t_i - t_j)^2 \bigg(\prod_{i}^k t_i^{a-1} (1 - t_i)^{b-1} \bigg).
\end{equation}
identical to the $s$-eigenvalue moment when $a = s+ 1/2$ and $b = N -2k+1$. This leads to a Selberg moment formula 
\begin{equation}
\label{eq:sm_selbergid}
    \mathbb{E}\bigg[ \prod_{i=1}^k t^s_i \bigg] = \frac{S(s+\frac 1 2, N-2k+1)}{S(\frac 1 2, N - 2k+1)} = \prod_{j=0}^{k-1} \frac{ \Gamma( \frac 1 2 +s +j) \Gamma(  N - k +j+\frac 1 2)}{\Gamma( \frac 1 2 + j)\Gamma( N  - k + j +\frac 1 2+ s)}.
\end{equation}
Now the main calculations can be carried out for the case dependent diagrams. To make full use of the orthogonality conditions of the Pauli basis a key first step will frequently involve reintroducing the fixed Pauli elements. 

The first quantity considered is the $\alpha$, whose form is taken from a slightly massaged form using universal steps already carried out in the stabilizer case. The definition is recalled below as
\begin{align}
        \alpha  = 
\mathbb{E} \bigg[ \sum_{P \in \widetilde{\mathbb{P}}_N} \bigg\{ \textrm{Tr}( P \rho_j P \rho_i ) \bigg\}^2  \bigg]  = \mathbb{E} \bigg[ \sum_{P \in \mathbb{P}_N} \bigg\{ \textrm{Tr}( P \rho_j P \rho_i ) \bigg\}^2  \bigg] - n_{\textrm{fix}} \mathbb{E}\bigg[ \bigg( \textrm{Tr}(\rho_i \rho_j)\bigg)^2 \bigg].
\end{align}
The first term is considered in some detail below, where the expression is rewritten in a Pauli string basis and identified with a corresponding $s$-moment formula in terms of the minors 
\begin{align}
\mathbb{E} \bigg[ \sum_{P \in \mathbb{P}_N} \bigg\{ \textrm{Tr}( P \rho_j P \rho_i ) \bigg\}^2  \bigg] &= \mathbb{E} \bigg[ \sum_{P \in \mathbb{P}_N} \prod_{a =1}^2\bigg\{ \sum_{R_a, R'_a} B_i(R_a) B_j(R'_a) \epsilon_{PR_a}\textrm{Tr}( R_a R'_a ) \bigg\}  \bigg] \notag \\
&= \mathbb{E} \bigg[ \sum_{R_1 \in \mathbb{P}_N} b_i^2(R_1) b_j^2(R_1) \bigg] = \sum_{R_1 \in \mathbb{P}_N} \bigg( \mathbb{E}\bigg[ b^2(R_1)\bigg]\bigg)^2 = \sum_{k=0}^N \sum_{|A_k| = 2k } \bigg(\mathbb{E}\bigg[ \prod^k_{i=1} t_i \bigg]\bigg)^2.
\end{align}
Now Eqn.~\eqref{eq:sm_selbergid} can be applied fixing $s = 1$ and noting that the sum over all possible strings at length $|A_k| = 2k$ only contributes a multiplicity. So the two pieces are
\begin{align}
    \mathbb{E}\bigg[ \prod^k_{i=1} t_i \bigg] = \binom{ N }{ k } \bigg/ \binom{ 2N}{2k} \qquad , \qquad 
    \sum_{|A| = 2k} 1 = \binom{2N}{2k}.
\end{align}
From these identities it quickly follows that the expectation value is
\begin{align}
%    \alpha +\mathbb{E}\bigg[ \textrm{Tr}(\rho_i^4) \bigg]=
\sum_{R \in \mathbb{P}_N} \bigg(\mathbb{E} \bigg[b(R)^2 \bigg] \bigg)^2 = \sum_{k=0}^N \frac{ \binom{N}{k}^2}{\binom{2N}{2k}} = \sum_{k=0}^N \frac{ \binom{2k}{k} \binom{ 2N - 2k}{N-k}}{\binom{2N }{N}} = \frac{4^N}{\binom{2N}{N}} \sim \sqrt{\pi N} + \frac{ \sqrt{\pi}}{8 \sqrt{N}} + O(N^{-3/2})
\end{align}
having used a combinatorial identity to rewrite the last line. The second term calculated using similar methods as 
\begin{align}
    \mathbb{E}\bigg[\bigg( \textrm{Tr}(\rho_i \rho_j) \bigg)^2\bigg] &= \frac{1}{4^N} \mathbb{E}\bigg[ \sum_{R,R' \in \mathbb{P}_N} b_i(R) b_j(R) b_i(R') b_j(R') \bigg] = \sum_{k=0}^N \sum_{ |A'|=2k }\sum_{ |A|=2k } \frac{1}{4^N} \bigg(\mathbb{E}\bigg[ \textrm{Pf}(A) \textrm{Pf}(A') \bigg] \bigg)^2 \notag \\
        &= \sum_{k=0}^N \sum_{ |A|=2k } \frac{2}{4^N} \bigg(\mathbb{E}\bigg[(\textrm{Pf}(A))^2] \bigg)^2 = \frac{2}{\binom{2N }{N}} \sim \frac{\sqrt{\pi N}}{2^N} + O( N^{-1/2} 2^{-N})
\end{align}
To reach the second line the orthogonality of the basis (up to a phase) has been used, which requires either $A = A'$ or $A' = A^c$ ($A'$ is the complement of $A$) for the contribution to be non-vanishing. Consequently, evaluation of one of the sums contributes a factor of $2$. And the result for the $s=1$ moment has been inserted. Thus finding $\alpha$
\begin{equation}
\boxed{    \alpha = \frac{ 4^N - 2 n_{\textrm{fix}}}{\binom{2N}{N}} \simeq \sqrt{\pi N} + \frac{ \sqrt{\pi}}{8 \sqrt{N}} + O\bigg(\textrm{max}(N^{-3/2},2^{-N} \sqrt{N} )\bigg)
}
\end{equation}

Now second case dependent term $\gamma_{\mu\mu}$ is considered. Steps that were already demonstrated above are omitted for brevity. The calculation here is much simpler given the previous work
\begin{align}
    %\boxed{
    \gamma_{\mu\mu} &= \mathbb{E}\bigg[ \sum_{P \in \widetilde{\mathbb{P}} } \textrm{Tr}( \rho_i P \rho_i P)\textrm{Tr}( \rho_j P \rho_j P)\bigg] =  \sum_{R \in \mathbb{P} } \bigg( \mathbb{E}\bigg[ b(R)^2 \bigg] \bigg)^2 - n_{\textrm{fix}} \mathbb{E}\bigg[  \textrm{Tr}( \rho_i^2 )\textrm{Tr}( \rho_j  ^2 )\bigg] \nonumber
\\       &= \sum_{R \in \mathbb{P} } \bigg( \mathbb{E}\bigg[ b(R)^2 \bigg] \bigg)^2- n_{\textrm{fix}} .
\end{align}
From the first to second line having applied both $\rho_i^2 = \rho_i$ and $\mathbb{E}_i[ \textrm{Tr}(\rho_i)] = 1$. The first term has already been determined during the calculation of $\alpha$ so
\begin{equation}
    \begin{boxed}{
         \gamma_{\mu\mu }= \frac{4^N }{\binom{2N}{N}} -n_{\textrm{fix}} \simeq \sqrt{\pi N} - n_{\textrm{fix}}+ O\bigg(N^{-1/2}\bigg) }
    \end{boxed}
\end{equation}
Recall now the definition of $\beta$ 
\begin{align}
    \beta &= \mathbb{E}\bigg[ \sum_{P \in \widetilde{\mathbb{P}} } \textrm{Tr}( \rho_i P \rho_j P \rho_j P \rho_i P)\bigg] = \mathbb{E}\bigg[ \sum_{P \in \mathbb{P} } \textrm{Tr}( \rho_i P \rho_j P \rho_j P \rho_i P)\bigg] - n_{\textrm{fix}} \mathbb{E}\bigg[  \textrm{Tr}( \rho_i \rho_j  \rho_j  \rho_i )\bigg]
    \end{align}
Whose first term is again simplified by rewriting the expression in the Pauli basis
\begin{align}
\mathbb{E}\bigg[ \sum_{P \in \mathbb{P} } \textrm{Tr}( \rho_i P \rho_j P \rho_j P \rho_i P)\bigg] &= \mathbb{E}\bigg[ \sum_{(P, R_1,R_1',R_2,R_2') \in \mathbb{P}}  \bigg(\prod^2_{a=1} b_i(R_a) b_j(R'_a) \bigg) \frac{\epsilon_{PR_1'} \epsilon_{PR_2} }{4^{2N}}\textrm{Tr}(R_1 R_1' R_2' R_2)\bigg], \notag \\
&= \frac{1}{4^N} \sum_{(R_1,R'_1 R_2') \in \mathbb{P}_N} \epsilon_{R_1' R_2'} \textrm{Tr}(R_1 R_2') \mathbb{E}\bigg[b_i(R_1) b_i(R_1') b_j(R_1') b_j(R_2')\bigg] \bigg], \notag \\
&= \frac{1}{2^N} \sum_{(R_1,R'_1 ) \in \mathbb{P}_N} \epsilon_{R_1' R_1}  \mathbb{E}\bigg[b_i(R_1) b_i(R'_1) b_j(R_1') b_j(R_1)\bigg], \notag \\ 
&= \frac 1 {2^N} \sum_{A_1,A_1'  \subseteq 2N} \epsilon_{A_1 A_1'} \bigg( \mathbb{E}_i\bigg[\textrm{Pf}(A_1) \textrm{Pf}(A'_1) \bigg] \bigg)^2 \notag \\
&= \frac 1 {2^N} \sum_{A_1 \subseteq 2N}  \bigg(\epsilon_{A_1 A_1} \mathbb{E}\bigg[\textrm{Pf}^2(A_1) \bigg]^2 +\epsilon_{A_1 A^c_1} \mathbb{E}\bigg[\textrm{Pf}(A_1) \textrm{Pf}(A^c_1) \bigg]^2 \bigg) \notag \\
%&= \frac{2}{2^N} \sum_{R_1  \in \mathbb{P}_N}  \bigg(\mathbb{E}\bigg[b(R_1)^2\bigg] \bigg)^2 
&=\frac 2 {2^N} \sum_{A_1 \subseteq 2N}  \bigg(\mathbb{E}\bigg[\textrm{Pf}^2(A_1) \bigg] \bigg)^2 = \frac{2^{N+1}}{\binom{2N}{N}}.
\end{align}
with $\epsilon_{AB} = (-1)^{|A| |B| - | A \cap B|}$ being defined through the Pauli string to Majorana mapping, generally  so $\epsilon_{A_1,A_1} = 1$ and $\epsilon_{A_1,A^c_1} = 1$. 
From the third to the fourth line having used that $\textrm{Pf}(A^c) = \pm \textrm{Pf}(A)$ so equivalent up to a phase due to parity and overall monomial phase. The second term is evaluated by averging the state itself 
\begin{align}
    \mathbb{E}\bigg[  \textrm{Tr}( \rho_i \rho_j  \rho_j  \rho_i )\bigg] &=     \mathbb{E}\bigg[  \textrm{Tr}( \rho_i \rho_j  )\bigg] = \textrm{Tr}\bigg( \mathbb{E}_i[\rho_i]\mathbb{E}_j[\rho_j] \bigg) = \frac{n_{\textrm{fix}}}{2^N}
\end{align}
So that overall the resulting contribution is strongly suppressed in the thermodynamic limit
\begin{equation}
   \boxed{ \beta = \frac{2^{N+1}}{\binom {2N}{N}} - \frac{(n_{\textrm{fix}})^2}{2^N} \sim  
    O\bigg(\frac{ \sqrt N}{2^N}\bigg). }
\end{equation}
The final diagram type is the single particle witness, which is connected to the fourth moment. Due to work done in the preamble this is again an application of the Selberg moment formula~\eqref{eq:sm_selbergid} for $s=2$ and so  
\begin{align}
    \langle W_\Psi(1) \rangle &= \mathbb{E} \bigg[\sum_{P \in \mathbb{P}_N} \textrm{Tr}( \rho_i P \rho_i P \rho_i P \rho_i P) \bigg] 
    = \sum_{P \in \mathbb{P}_N} \mathbb{E} \bigg[  b_i(P)^4 \bigg] = \sum_{k=0}^N \sum_{|A_k| = 2k } \mathbb{E}\bigg[ \prod^k_{i=1} t_i^2 \bigg] \notag \\
     &= \frac{\sqrt{\pi}}{4} \frac{ \Gamma(3+ N)}{\Gamma\bigg( \frac 3 2 + N \bigg)} ,
\end{align}
with the identity $\Gamma\bigg( \frac 1 2 + N\bigg) \Gamma(1+N) = (2N)! / 4^N / \sqrt{\pi}$ a more  obviously combinatorial expression can be written
\begin{equation}
\boxed{
    \langle W_\Psi(1) \rangle =4^N  \frac{ (2+N)!  }{2(1 + 2 N)} \frac{N!}{(2N)!}  \simeq \frac{ \sqrt \pi}{4} N^{3/2} + \frac{21 \sqrt{\pi N}}{32} + \frac{137 \sqrt \pi}{512 \sqrt N}  + O( N^{-3/2}).}
\end{equation}

These case-dependent terms can be combined to determine the total non-stabilizer witness 
\begin{align}
    \langle W_\Psi(m) \rangle &= 
4n_{\textrm{fix}}+\dfrac{1}{m^2}\bigg(\frac{ 3( 2^{2N+1} - 2n_{\textrm{fix}})}{\binom {2N}{N}} + 12\bigg(\frac{2^{N+1}}{\binom {2N}{N}} - \frac{n_{\textrm{fix}}^2}{2^N}\bigg) -9n_{\textrm{fix}} -3 n_\textrm{fix}\bigg)\notag\\
&\phantom{===}+\dfrac{1}{m^3}\bigg(\langle W_\Psi(1)\rangle-\bigg\{\frac{ 3( 2^{2N+1} - 2 n_{\textrm{fix}})}{\binom {2N}{N}}+12\bigg(\frac{2^{N+1}}{\binom {2N}{N}}  - \frac{n_{\textrm{fix}}^2}{2^N}\bigg)-5n_{\textrm{fix}}- 3 n_{\textrm{fix}}\bigg\}\bigg) \notag\\
&\simeq 4 n_{\textrm{fix}} + \frac{6 \sqrt{\pi N} - 12 n_{\textrm{fix}}}{m^2} + \frac{\langle W_\Psi(1)\rangle - 6\sqrt{\pi N} + 8 n_{\textrm{fix}}}{m^3} + O\bigg( N^{-1/2} \bigg).
\end{align}
At leading (orders) in $N$ and $m$ the non-stabilizer witness is given by (with the additional note that $n_{\textrm{fix}} = 2$)
\begin{equation}
\boxed{\langle W_\Psi(m) \rangle
= 4 n_{\textrm{fix}} + \frac{6 \sqrt{\pi N} - 12 n_{\textrm{fix}}}{m^2} + \frac{1}{m^3} \bigg(\frac{\sqrt{ \pi N^3}}{4} + \bigg( \frac {21}{32}- 6\bigg)\sqrt{\pi N}  + 8 n_{\textrm{fix}} \bigg) + O\bigg( N^{-1/2} \bigg).}
\end{equation}

\subsection{Case 3: Random computational-basis states}

The last ensemble we consider is the simplest: each component is drawn independently and uniformly from the $2^N$ computational-basis states $|a\rangle$, $a\in\{0,1\}^N$, and superposed with the random weights and phases used throughout. As every component is a stabilizer state a single component has $\langle W(1)\rangle_\Psi=2^N$ and therefore $M_2(1)=0$ exactly.  The ensemble is however not a $2$-design, and none of the Schur-Weyl results of Case 1 above apply.  Moreover, we will see that for equal weights the initial magic growth is $2\log_2m$, for any $N$, presenting a useful counterexample to the large-$N$ $3\log_2m$ behaviour observed in both Case 1 and Case 2 above. 

Despite its much smaller size, the ensemble average is nevertheless $\mathbb{E}[\rho]=\mathbb{I}/2^N$, and the Schur condition holds with $n_{\textrm{fix}}=1$ and $v=1/D$ for every string. The diagonal moments can be calculated directly with $\langle a|P|a\rangle$ vanishing for any string containing an $X$ or a $Y$,  and equal to the sign $(-1)^{s\cdot a}$ for a $Z$-string $Z_s=\prod_jZ_j^{s_j}$, where $s\in\{0,1\}^N$ marks the sites carrying a $Z$. Hence
\begin{align}
    \mu_2(P)=\begin{cases}1 & P\in\{Z_s\}\\ 0&\textrm{otherwise}\end{cases},
    \qquad
    \sum_{P\in\mathbb{P}_N}\mu_2^2(P)=2^N
    \end{align}
 In the gamma shorthand above this gives
 \begin{align}   
     \gamma_{\mu\mu}=2^N-1,\qquad \alpha=\gamma_{\mu\mu}+n_{\textrm{fix}}=2^N .
     \end{align}

The second moments are as far from flat as they can be. The Parseval sum rule fixes the total weight $\sum_P\mu_2(P)=2^N$. A flat spectrum spreads it evenly, $\mu_2(P)\simeq2^{-N}$ on every non-fixed string, which is the $2$-design value of Case 1 and gives the smallest possible $\sum_P\mu_2^2(P)\simeq2$. Here the same weight sits on only $2^N$ strings, each with $\mu_2=1$, the largest value a single string can take, which gives the largest possible $\sum_P\mu_2^2(P)=2^N$. The ensemble is invariant under every Pauli string ($X$-strings permute the basis states, $Z$-strings only change their phase), and the identity is the only string commuting with all of them, so the closed form for $\beta$ derived above applies with $n_{\textrm{fix}}=1$:
\begin{align}
    \beta=\dfrac{n_{\textrm{fix}}}{D}\sum_{R\in\widetilde{\mathbb{P}}_N}\mu_2^2(R)=\dfrac{2^N-1}{2^N}=1-2^{-N} \approx 1 .
\end{align}
Unlike Cases 1 and 2, where $\sum_P\mu_2^2(P)\ll 2^N$, $\beta$ is not exponentially small here. Although it is a genuine part of the ensemble average,  it is nevertheless negligible in the witness where $12\beta$ enters $b$ alongside $6\sum_P\mu_2^2(P)=6\cdot2^N$, so its relative weight is $O(2^{-N})$, and dropping it changes the growth law only at the order already neglected.

\noindent \textit{Witness and SRE:} Inserting these values into the general form, we get
\begin{align}
    b&=6\bigg\{\sum_{P\in\mathbb{P}_N}\mu_2^2(P)-2(n_{\textrm{fix}}-\beta)\bigg\}=6\cdot2^N-12\cdot2^{-N},
\end{align}
and, grouped by powers of the weights,
\begin{align}
    \langle W_\Psi(m)\rangle&=4+b\,\|p\|_2^4+\Big(2^N-4-b\Big)\|p\|_4^4\notag\\
    &=2^N\Big[6\|p\|_2^4-5\|p\|_4^4\Big]+4-4\|p\|_4^4+O(2^{-N}).
    \label{eq:basis_norms}
\end{align}
For equal weights,
\begin{align}
    \langle W_\Psi(m)\rangle=4+\dfrac{6\cdot2^N}{m^2}-\dfrac{5\cdot2^N+4}{m^3}+O(2^{-N}),
\end{align}
which returns $\langle W_\Psi(1)\rangle=2^N$ at $m=1$, as it should. The $m^{-3}$ coefficient is negative, because the single-component excess $2^N-4$ is smaller than $b\simeq6\cdot2^N$. The $b/m^2$ term therefore dominates from $m=1$ on and there is no $3\log_2m$ regime at any $N$. For $m\ll2^{N/2}$ the floor can be dropped and the SRE climbs as
\begin{align}
    \langle M_2(m)\rangle=N-\log_2\langle W_\Psi(m)\rangle\simeq2\log_2m-\log_2\Big(6-\dfrac5m\Big),
\end{align}
exact at $m=1$ and approaching $2\log_2m-\log_26$, the law quoted in the main text. The Haar ceiling $N-2$ is reached only once $m\sim2^{N/2}$, against $m\sim2^{N/3}$ for random stabilizer states. 

This distinction with Case 1 is worth spelling out. Both random stabilizer states and random basis states share $a=\langle W_\Psi(1)\rangle=2^N$, $n_{\textrm{fix}}=1$ and the floor $c=4$.  However, they differ only in $\sum_P\mu_2^2(P)$, which is $2-2/(2^N+1)$ for the $2$-design and $2^N$ here. The $3\log_2m$ growth of Case 1 is therefore a consequence of the flat second moments guaranteed by the Clifford $2$-design, not of the components being stabilizer states.

\bibliography{main}